\documentclass[opre,nonblindrev]{informs3} 

\OneAndAHalfSpacedXI 

\usepackage{endnotes}
\usepackage{comment}
\usepackage{subcaption}
\let\footnote=\endnote
\let\enotesize=\normalsize
\def\notesname{Endnotes}%
\def\makeenmark{$^{\theenmark}$}
\def\enoteformat{\rightskip0pt\leftskip0pt\parindent=1.75em
  \leavevmode\llap{\theenmark.\enskip}}

\newcolumntype{L}{>{\arraybackslash}m{6cm}}
\newcolumntype{C}{>{\centering\arraybackslash}m{2.5cm}}
\newcommand{\Mod}[1]{\ (\mathrm{mod}\ #1)}

\newcommand{\cA}{{\mathcal A}}
\newcommand{\cP}{{\mathcal P}}
\newcommand{\cN}{{\mathcal N}}
\newcommand{\cC}{{\mathcal C}}
\newcommand{\bR}{{\mathbb R}}
\newcommand{\eps}{\varepsilon}

\usepackage{natbib}
 \bibpunct[, ]{(}{)}{,}{a}{}{,}%
 \def\bibfont{\small}%
\usepackage{enumerate}
\usepackage{hyperref}
\hypersetup{
    colorlinks=true,
    linkcolor=red,
    citecolor = blue,
    filecolor=magenta,      
    urlcolor=cyan,
}

\TheoremsNumberedThrough     
\ECRepeatTheorems

\EquationsNumberedThrough    

\def\tcyc#1{{\bf\color{red}#1}}

\begin{document}


\RUNAUTHOR{Chan, Huang, and Sarhangian}

\RUNTITLE{Dynamic Control of Service Systems with Returns}

\TITLE{Dynamic Control of Service Systems with Returns: Application to Design of Post-Discharge Hospital Readmission Prevention Programs}

\ARTICLEAUTHORS{%
\AUTHOR{Timothy C. Y. Chan, Simon Y. Huang, Vahid Sarhangian}
\AFF{Department of Mechanical and Industrial Engineering, University of Toronto, Toronto ON CANADA \EMAIL{tcychan@mie.utoronto.ca, syx.huang@mail.utoronto.ca, sarhangian@mie.utoronto.ca}} 
} 

\ABSTRACT{%
We study a control problem for queueing systems where customers may return for additional episodes of service after their initial service completion. At each service completion epoch, the decision maker can choose to reduce the probability of return for the departing customer but at a cost that is convex increasing in the amount of \emph{reduction} in the return probability. Other costs are incurred as customers wait in the queue and every time they return for service. Our primary motivation comes from post-discharge Quality Improvement (QI) interventions (e.g., follow up phone-calls, outpatient appointments) frequently used in a variety of healthcare settings to reduce unplanned hospital readmissions. Our objective is to understand how the cost of interventions should be balanced with the reductions in congestion and service costs. To this end, we consider a fluid approximation of the queueing system and characterize the structure of optimal long-run average and \emph{bias-optimal} transient control policies for the fluid model. Our structural results motivate the design of intuitive surge protocols whereby different intensities of interventions (corresponding to different levels of reduction in the return probability) are provided based on the congestion in the system. Through extensive simulation experiments, we study the performance of the fluid policy for the stochastic system and identify parameter regimes where it leads to significant cost savings compared to a fixed long-run average optimal policy that ignores holding costs and a simple policy that uses the highest level of intervention whenever the queue is non-empty. In particular, we find that in a parameter regime relevant to our motivating application, dynamically adjusting the intensity of interventions could result in up to 25.4\% reduction in long-run average cost and 33.7\% in finite-horizon costs compared to the  simple aggressive policy.
}


\maketitle

%


\section{Introduction}
We propose and study a control problem for queueing systems where customers may return for additional episodes of service after their initial service completion. The probability of return for each customer can be reduced through what we refer to as a \emph{post-service intervention}. Post-service interventions are costly but also lead to future savings through eliminating the cost of service for some returning customers, as well as reducing system congestion. Our objective is to understand how the cost of providing post-service interventions should be balanced with the reductions in congestion and service costs.

Our primary motivation comes from Quality Improvement (QI) interventions designed to prevent unplanned hospital readmissions in a variety of healthcare settings. Hospital readmissions are costly and in many cases preventable. QI interventions provided after discharge range from comprehensive transitional care programs (\citealt{stauffer2011}) to multi-tier post-discharge follow-up phone calls for pharmaceutical reconciliations (\citealt{ravnnielsen2018effect}). Systematic reviews in the medical literature suggest that QI interventions can be effective in reducing readmission risk, although to varying degrees (\citealt{leppin2014preventing}). These programs are costly, however, raising the question of whether they are economical for the hospital or more broadly the healthcare system. Findings from economic evaluation studies in the literature range from significant savings to major losses with a systematic review of 21 randomized trials pointing to an overall net loss (\citealt{nuckols2017economic}). The interventions considered in the literature are not usually ``optimized" for cost-effectiveness. More importantly, they do not consider the operational benefits of reduced congestion, and only compare the program cost versus the savings in reduced readmission costs. Intuitively, reducing readmission could reduce congestion at hospitals where capacity strain is a common and growing concern. Nevertheless, given the indirect nature of control (only reducing the risk of readmission) and the relatively long time-scale over which readmissions are defined (usually 30 days), the benefits are not clear a priori. 

While our models and analysis are primarily motivated by the above healthcare application, they are also relevant to other service systems where returns are undesirable and possibly preventable through post-service interventions. For instance, in the case of overcrowded prisons (see, e.g., \citealt{usta2015assessing}) returns correspond to recidivism and post-service interventions correspond to post-incarceration services and resources (e.g., training, education, and healthcare) aimed at reducing the risk of recidivism; see \cite{wallace2015neighborhood} and \cite{badaracco2021effects} for examples.

We study the trade-off between the costs of interventions versus the savings in service and holding costs by introducing a control problem for a multiserver queueing system with returns. At each service completion epoch, the decision maker can choose to reduce the probability of return for the departing customer within a closed interval and at a cost that is convex increasing in the amount of \emph{reduction} in the return probability. That is, in addition to deciding whether to provide the post-service intervention, the decision maker can also adjust the intensity of the intervention. Other costs are incurred as customers wait in the queue and every time they return for service. We are interested in both long-run average cost as well as the transient cost of the system initiated from a congested state. The latter is particularly relevant to healthcare systems, which often experience highly transient dynamics (e.g., due to pandemics, flu season, mass casualty incidents, or regular stochastic fluctuations), and allows us to investigate how the intensity of post-discharge interventions should be adjusted in response to varying levels of congestion in the system.

Queueing systems with returns, also known as Erlang-R, have been studied in the literature, e.g., in support of healthcare staffing decisions (\citealt{yom2014erlang}) and to understand the consequences of speed up (i.e., increased service rate) when it leads to higher probability of return (\citealt{chan2014use}). Dynamic control of the return probability, which is the focus of this paper, has not been studied before. In its most basic form, the Erlang-R queue is a two-node Jackson network where upon service completion at a multiserver node, each customer joins an infinite server node with a certain probability and returns to the first node after some delay in the second node. From a technical standpoint, the problem of controlling the return probability is challenging since the transition rates of the underlying Markov process are unbounded. This prevents one from applying the typical approach of exploiting a uniformized Markov Decision Process (MDP). Although some methods have been proposed in the literature to address this issue (e.g., in \citealt{down2011dynamic}, \citealt{bhulai2014unbdd}, and \citealt{zayas2016dynamic}) they exploit specific structures of their respective problems, which are hard to generalize and do not apply to our setting. In contrast, we rely on analysis of associated fluid control problems which provide structural insights and practical policies. 

Our main results and contributions can be summarized as follows.

\textbf{Queueing model with controllable return probability}: We propose and study a new control problem for the Erlang-R queue, with the key feature that the return probability can be controlled at departure instances. Reducing the return probability incurs a cost that is convex increasing in the amount of reduction. The model captures a tradeoff between the cost of post-service intervention and the savings in holding and return costs.  

\textbf{Long-run average control}: We show that when the fluid model is stable without any intervention (i.e., under the maximum return probability) a fixed return probability, which we refer to as the \emph{equilibrium policy}, minimizes its long-run average cost. In equilibrium the queue is empty and the optimal return probability balances the cost of intervention versus cost of return incurred in equilibrium. This, however, ignores any holding costs incurred starting from an initial condition far from the equilibrium.

\textbf{Bias-optimal transient control}: To minimize the transient cost incurred before reaching the equilibrium, we propose and study a fluid control problem that seeks \emph{bias-optimality}, i.e., minimizes the cost difference with that of the equilibrium policy over an infinite horizon. We show that the infinite-horizon problem is well-defined and is equivalent to a finite-horizon problem that is easier to analyze. We then characterize the structure of the optimal transient policy by exploiting Pontryagin's Minimum Principle. In particular, we show that for states where the queue is non-empty, the optimal policy \emph{under a general convex intervention cost} is characterized by a set of contour lines in the state-space. Points on each line are equivalent with respect to a certain measure of congestion comprised of the number of customers currently in the system, as well as a discounted count of those waiting to return.   When the cost is piecewise-linear with $k$ pieces, $k$ lines ``fan out" in the state-space, dividing it into $k+1$ regions with the same optimal probability of return. This simple \emph{switching control} structure is practically appealing as it motivates the design of ``surge protocols" whereby different levels of the post-service intervention are offered depending on the congestion level of the system. Under the switching control, the fluid dynamics are governed by a system of differential equations with right-hand-side discontinuity, which could exhibit complex dynamics and have multiple solutions. We show that the fluid model has a unique solution under such discontinuous dynamics. This is an important step toward formally justifying the fluid models; see also the discussion in Section \ref{sec:conc}. We further illustrate that a similar structure is optimal when the queue is empty but a ``large" number of customers are waiting to return, except that the contour linear (or boundaries in the case of piecewise linear costs) are nonlinear. Finally, we discuss and illustrate the relevance of our results to systems with time-varying arrivals.


\textbf{Numerical results}: Through extensive simulation experiments, we evaluate the performance of the fluid policy for the stochastic system over both finite and infinite horizons, and in comparison with two benchmark polices: the equilibrium policy which ignores congestion, and a simple policy which always uses the highest level of intervention when all servers are busy. We find that by accounting for holding cost reductions, interventions that may otherwise not be economical could lead to significant cost savings, in particular when the cost of intervention is relatively high. As the holding cost increases, the performance of the fluid policy approaches that of the simple policy. However, in a relevant parameter regime, with high intervention cost and moderate holding cost, dynamically adjusting the intensity of the interventions in response to congestion could result in up to 25.4\% reduction in long-run average cost and 33.7\% in finite-horizon costs compared to the simple policy. Finally, we conduct a case study where we relax some of our modeling assumptions and calibrate the inputs of our model based on efficacy and economic data for a real post-discharge intervention for reducing readmission of heart failure patients. The results illustrate the robustness of our qualitative observations to certain modelling assumptions for practically relevant parameters. 

The rest of this paper is organized as follows. We conclude this section with a brief review of the related literature. In Section \ref{sec:Model} we describe the queueing model and its fluid approximation. We formulate and analyze the associated fluid control problems for long-run average control in Section \ref{sec:longrun} and for transient control in Section \ref{sec:transient}. We discuss the extension and robustness of our results to time-varying arrivals in Section \ref{sec:tv}. Section \ref{sec:numerics} summarizes the results of our numerical experiments and case study. We conclude the paper in Section \ref{sec:conc}. All proofs are presented in the E-Companion. The code to reproduce the numerical experiments is available on GitHub; see Section \ref{ap:supnumerics}.

\subsection{Related Literature}
From a modeling perspective our work is related to previous studies of the Erlang-R queue. \cite{yom2014erlang} proposes and studies the Erlang-R queue under time-varying arrival rates in support of various healthcare staffing problems, including bed capacity planning in wards with readmission, physician staffing in the emergency department, and provider staffing for a mass casualty incident. Several studies have investigated the performance and stability of the Erlang-R model under state-dependent service and return probabilities. \cite{chan2014use} considers a threshold speedup policy where if the queue-length exceeds a fixed threshold, the service rate increases to clear the congestion but at the cost of the return probability also increasing. They utilize a fluid approximation of the model and investigate its stability. \cite{yom2021balancing} extends the analysis to the case where admission control is also used when the number of customers exceeds some threshold. \cite{ingolfsson2020comparison} compare the equilibrium and transient behavior of two fluid models for a system with state-dependent service and return probabilities, differing in terms of when the determination of whether a customer will return to service is made. \cite{barjesteh2021multiclass} study a more general multiclass model where the service rates depend on the workload of the system and the return probabilities on the service rate and investigate the criteria for stability of the system. In contrast to these previous studies which focus on stability and performance analysis, our work is concerned with the dynamic control of the return probabilities at departure instances. We establish the optimality of piecewise-constant (switching) controls when interventions costs are piecewise linear. Under piecewise-constant return probabilities the fluid model is a switching dynamical system with right-hand-side discontinuity. See \cite{perry2011fluid, perry2011ode} for another example of discontinuous control in the context of an overloaded X model. \cite{zhan2014threshold} studies a related routing control problem for call centers with multiple heterogeneous server pools, where customers may immediately call beck with some probability if their issue is not resolved. Server pools differ in terms of their service rate and call resolution probabilities. They study a diffusion control problem with the goal of minimizing average customer waiting time and maximizing the call resolution.

From a methodological perspective, our work relates to the vast literature on queueing control and specifically studies concerned with transient control. Analysis of transient dynamics is typically challenging even under a given control policy. Fluid models approximately capture the transient dynamics of the system, and as such associated fluid control problems have been used in a variety of queueing models to gain insights on the structure of ``good" policies, especially for scheduling and routing problems, e.g., \cite{avram1994stochastic}, \cite{perry2009responding}, \cite{larranaga2013dynamic}, and \cite{hu2020optimal}. See also \cite{zychlinski2023applications} for a recent review of applications of fluid models. For certain fluid control problems, the total cost over an infinite horizon remains finite under appropriate stability conditions, allowing one to obtain stationary optimal policies. This is typically the case for control problems concerned with single-server queues, e.g., a class of scheduling problems studied in \cite{maglaras2000discrete} and a multiclass revenue maximization problem considered in \cite{maglaras2006revenue}, but can also (depending on the cost structure) hold for multiserver problems, e.g., as in \cite{chan2021dynamic}. When costs are also incurred in equilibrium, as in the case of our problem, the infinite horizon cost diverges and hence is not suitable for analysis. Alternatively, one can consider minimizing the total cost until reaching a certain state, e.g., as in \cite{ata2020optimal}, \cite{hu2020optimal}, and \cite{chen2023optimal} which leads to a control problem with state-dependent constraints. For our multiserver problem, the equilibrium is reached asymptotically as time tends to infinity. As such, we instead consider bias-optimal policies. The idea is related to the concept of bias-optimality in MDPs, where a bias-optimal policy is one that minimizes the relative value function among all policies that are long-run average optimal; see, e.g., \cite{lewis2002bias} for a review and \cite{haviv1998bias} for an application to an admission control problem. \cite{larranaga2015asymptotically} also study a bias-optimal fluid control problem, but to derive approximate index policies for a multiclass scheduling problem with abandonment. As such, their problem and approach differ from ours.


With respect to our main motivating application, our work relates to studies aiming to evaluate or propose economical and/or operational interventions to reduce hospital readmissions, see, e.g., \cite{zhang2016hospital, arifouglu2021hospital} and \cite{shi2021timing} among others. More closely related are studies concerned with post-discharge interventions.  \cite{bayati2014data} proposes targeting interventions to high-risk patients as identified using a predictive model. \cite{helm2016reducing} propose a model to optimize the type and timing of checkups to implement post-discharge monitoring plans. \cite{liu2018missed} combines a prediction model with a deterministic optimization model to optimize a post-discharge monitoring schedule and staffing plan to support monitoring patients before they end up readmitting. In a recent study, \cite{chen2022data} studies the dynamic assignment and timing of post-discharge interventions using a model-free Reinforcement Learning (RL) framework. These studies, however, do not consider the potential benefits of reduced congestion at the hospital. In contrast, we focus on understanding how post-discharge interventions can help reduce congestion.

\section{Model Description}\label{sec:Model}
We consider a multiserver queueing network with returns, also known as the Erlang-R model \citep{yom2014erlang}. New customers arrive to the system according to a Poisson process with rate $\lambda$ and have exponentially distributed service requirements with rate $\mu$. There are $N$ identical servers available. If all servers are busy, customers wait in an infinite capacity queue. Upon completing service, and in the absence of a post-service intervention, a customer leaves the system with probability $1-p_u$, and returns to the system after an exponentially distributed time with rate $\nu$ with probability $p_u$. Let $\{X(t);t\geq0\}$ denote the process that keeps track of the number of customers currently in the system (i.e., in service or waiting), and let $\{Y(t);t\geq0\}$ denote the process that keeps track of the number of customers whose return to the system is pending. Following \cite{vericourt2011nurse} and \cite{yom2014erlang}, we refer to the customers currently in the system as \emph{Needy} and those who will return at a later time as \emph{Content}.

The system incurs costs as customers wait, return, and undergo post-service interventions. Needy customers incur a holding cost at rate $h$ as they wait in the queue. For each returning customer, the system incurs a fixed return cost of $r$. Upon service completion, the decision maker has the option of implementing a post-service intervention that can reduce the probability of return for the departing customer. More specifically, we assume that the decision maker can directly control the return probability in the interval $[p_l,p_u]$ where $0<p_l<p_u<1$. The intervention cost of setting the return probability to $p\in[p_l,p_u]$ is denoted by $C(p)$. 

\begin{assumption}\label{assum:cost}
The intervention cost function $C(\cdot)$ is convex, non-negative and decreasing with $C(p_u)=0$. In addition, $C(\cdot)$ has left and right derivatives at the endpoints $p_l$ and $p_u$.
\end{assumption}
Note that by convexity, $C(\cdot)$ must be continuously differentiable at all but countably many points on $(p_l, p_u)$. Moreover, $C(\cdot)$ has non-decreasing left and right derivatives on $(p_l, p_u)$ (and by assumption on $[p_l, p_u]$). We denote the left and right derivatives by $C'_-(\cdot)$ and $C'_+(\cdot)$, respectively.

We consider the set of Markovian policies under which the probability of return is determined based on the state of the system at departure instances of Needy customers. Denote by $p^{\pi}(x,y)$ the probability of return for a Needy customer departing the system at state $(X(t),Y(t))=(x,y)$ under intervention policy $\pi$. Since the dynamics of the process $\{(X(t),Y(t));t\geq 0\}$ depend on the intervention policy we denote the process under policy $\pi$ by $\{(X^\pi(t),Y^\pi(t); t\geq0\}$. Further, denote by $\{R^\pi(t);t\geq 0\}$ the process that keeps track of the cumulative number of returns under policy $\pi$, and by $\{D^\pi(t);t\geq 0\}$ the process that keeps track of the cumulative number of departures from the Needy state. Finally, let $\{t_1,t_2,\ldots\}$ denote the decision epochs, i.e., the sequence of instances when Needy customers depart the system. The expected total cost over the finite horizon of $[0,T]$ and starting from $(X(0),Y(0))$ can be expressed as,
\begin{equation}\label{eq:exp_cost}
    \mathcal{G}^{\pi}(T) = \mathbb{E} \left[ \int_0^T h (X^\pi(t)-N)^+ dt + r R^\pi (T) + \sum_{i=1}^{D^{\pi}(T)} C\left(p^\pi\left(X^\pi(t_i),Y^\pi(t_i)\right)\right)
 \right],
\end{equation}
where $(x)^+ \equiv \max (x,0)$. This cost function is composed of three terms, representing the holding, return, and intervention costs, respectively, over the time horizon $[0,T]$. 

It is natural to consider finding policies that minimize the long-run average cost of the system, that is, 
\begin{equation}
    \limsup_{T\to\infty} \frac{1}{T} \mathcal{G}^{\pi}(T).
\end{equation}
However, if the initial state of the system is far from its steady-state, a long-run average policy could be sub-optimal in the short-term as it ignores the initial transient costs incurred in the system. As such, one may consider searching for a policy that minimizes the finite-horizon cost in \eqref{eq:exp_cost} over an appropriate horizon-length. Due to the complexity of the stochastic control problem, we consider a fluid approximation of the queueing model and formulate associated (deterministic) fluid control problems. The fluid control problem yields approximately optimal policies for the stochastic problem, and provides insights on the structure of optimal policies.
\subsection{The Fluid Model}
The fluid model is obtained by replacing the stochastic arrival, service, and return processes by their deterministic average rates. Denote by $(x(t),y(t))$ the state of the fluid model at time $t\geq 0$ where $x(t)$ and $y(t)$ are the amount of Needy and Content fluid (customers), respectively, in the system at time $t$. Under an intervention policy $\pi$, the fluid dynamics satisfy the following system of ordinary differential equations (ODEs):
\begin{align}
    \dot x(t) &= \lambda + \nu y(t) - \mu (x(t) \wedge N), \label{eq:fluidt_1}\\
    \dot y(t) &= -\nu y(t) + \mu p^\pi (x(t),y(t)) (x(t) \wedge N), \label{eq:fluidt_2}
\end{align}
where $x(t)\wedge N \equiv \min (x(t),N)$ is the number of busy servers at time $t$. The instantaneous rate of change for the fluid trajectories in \eqref{eq:fluidt_1}--\eqref{eq:fluidt_2} is composed of an input and output rate. In \eqref{eq:fluidt_1}, the input rates correspond to the arrival of new and returning customers, whereas the output rate corresponds to service completions of Needy customers. In \eqref{eq:fluidt_2}, the input is the fraction of departing Needy customers who join the Content state, and the output rate is the rate at which current Content customers transition into the Needy state.


To ensure stability, in our analysis of the associated fluid control problems we impose the following assumption on the parameters. 

\begin{assumption}\label{assump:stability}
The maximum return probability satisfies $p_u < 1-\lambda/(\mu N)$.
\end{assumption}
As shown in \cite{chan2014use} for the Erlang-R fluid model with a fixed return probability $p$, if $p<1-\lambda/(\mu N)$, then starting from any initial condition the trajectories converges to the following globally stable equilibrium,
\begin{equation}\label{eq:equil}
    (x_\infty,y_\infty) = \left( \frac\lambda{\mu(1-p)}, \frac{\lambda p}{\nu(1-p)} \right),
\end{equation}
and diverge otherwise. Hence, Assumption \ref{assump:stability} guarantees that in the absence of intervention (i.e., with constant return probability $p_u$) the fluid model remains stable.  

\section{Long-Run Average Control}\label{sec:longrun}
\begin{figure}[]
    \centering
    \includegraphics[scale=0.7]{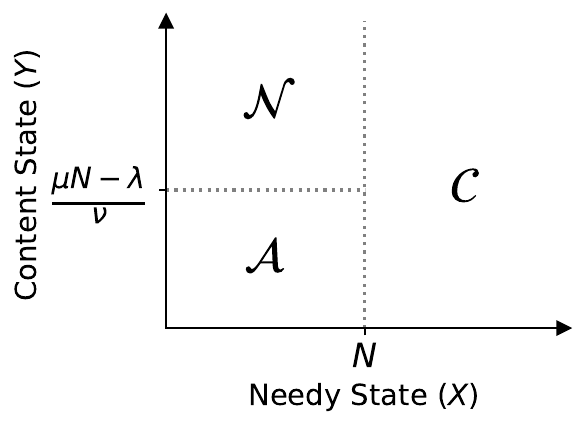}
    \caption{An illustration of the three partitions of the state-space.}
    \label{fig:partition}
\end{figure}

In this section, we formulate the long-run average fluid control problem and characterize the structure of an optimal policy. We assume that fluid waiting in queue incurs a holding cost at rate $h$, returning fluid incurs a cost at rate $r$, and that departing fluid at time $t$ incurs an intervention cost at rate $C(p(t))$. Starting from a given initial condition $(x^0,y^0)$, the problem is to find an admissible intervention policy $p(\cdot)$ that minimizes the long-run average cost:
\begin{equation}
    \min_{p(\cdot) \in \cP} \: \limsup_{T\to\infty}\: \frac1T \int_0^T \left(h(x(t) - N)^+ + r\nu y(t) + C(p(t)) \mu (x(t)\wedge N) \right)\, dt. \label{eq:fluid_1}
\end{equation}
The set $\cP$ denotes the set of admissible policies. A policy $p(\cdot) \in \cP$ is admissible if it jointly satisfies,
\begin{align}
 & \dot x(t) = \lambda + \nu y(t) - \mu (x(t) \wedge N), \quad \forall t \geq 0, \label{eq:fluid_C1}\\
    & \dot y(t) = -\nu y(t) + \mu p(t) (x(t) \wedge N), \quad \forall t \geq 0, \label{eq:fluid_C2}\\
    & p(t)\in [p_l,p_u], \quad \forall t \geq 0, \label{eq:fluid_C3}\\
    & (x(0),y(0)) = (x^0,y^0),\label{eq:fluid_C4}
\end{align}
with an absolutely continuous $(x(t),y(t))$. Before turning into the analysis of the control problem, we partition the state-space $\bR^2_+$ into three regions: $(i)$ $\cC := (N,\infty) \times [0, \infty)$; $(ii)$ $\cA := [0,N] \times [0, \frac{\mu N - \lambda}\nu]$; and $(iii)$ $\cN := [0,N] \times (\frac{\mu N - \lambda}\nu, \infty)$. We refer to $\cC$ as the \emph{congested region} since it represents states where all servers are busy and the queue is non-empty. In contrast, $\cN$ and $\mathcal{A}$ represent states where the queue is empty. Figure \ref{fig:partition} provides an illustration of the three regions. We distinguish between $\mathcal{A}$ and $\mathcal{N}$ because as we establish in Proposition \ref{prop:corner}, region $\mathcal{A}$ is an absorbing region, i.e., trajectories must reach $\cA$ within a bounded time and stay there thereafter. We further note that when $y>(\mu N - \lambda)/\nu$, i.e., as in $\mathcal{N}$ and the upper part of $\mathcal{C}$, we are guaranteed to have $\dot{x}(t)>0$ and $\dot{y}(t)<0$ under Assumption \ref{assump:stability}. In contrast, the signs of $\dot{x}(t),\dot{y}(t)$ could be positive or negative depending on the values of $x,y$ in $\mathcal{A}$ and the lower part of $\cC$.  

\begin{proposition}\label{prop:corner}
$(i)$ For any given initial state $(x^0, y^0) \geq 0$, there exists a time $t_{\cA} < \infty$ such that $(x(t_\cA), y(t_\cA)) \in \cA$ under any admissible policy $p(\cdot)\in\cP$. $(ii)$ If $( x^0, y^0 ) \in \cA$, then $(x(t), y(t)) \in \cA $ for all $t \geq 0$.

\end{proposition}

The proposition states that all trajectories are guaranteed to reach $\cA$ within a common time bound (across all policies) and once they reach $\cA$, they remain in $\cA$ thereafter. Since the queue is empty in $\cA$, we have $(x(t) - N)^+ = 0$ and $x(t) \wedge N = x(t)$ for $t\geq t_{\cA}$. Hence, when considering long-run average performance we may replace the objective in \eqref{eq:fluid_1} with
\begin{equation}
        \min_{p(\cdot)\in\cP} \: \limsup_{T\to\infty}\: \frac1T \int_0^T r\nu y(t) + C(p(t)) \mu x(t) \, dt. \\
\end{equation}

\subsection{Optimizing Amongst Equilibria}\label{subsec:eqopt}
If we restrict our attention to policies that result in convergence to an equilibrium point in the form of \eqref{eq:equil}, minimizing the long-run average cost is equivalent to finding a fixed policy $p_\infty \in \argmin_p J(p)$, where 
\begin{equation}\label{eq:Jp}
    J(p) := \lambda \frac{rp + C(p)}{1-p}.
\end{equation}
We may interpret $J(p)$ as the long-run average cost rate under a constant policy $p(t)=p$ for all $t\geq 0$. We show in Proposition \ref{prop:Jinf} that $J_\infty := J(p_\infty)$ is indeed the optimal long-run average cost and hence focusing on policies under which the trajectories converge to an equilibrium is not restrictive. We note that $J(p)$ only includes the return and intervention costs, as in equilibrium the fluid queue is empty. In contrast, an optimal constant policy for the stochastic system must also account for steady-state holding costs. In Section \ref{ap:exact}, we use numerical examples to illustrate that the sub-optimality of $p_\infty$ is small for practically relevant holding cost rates and system sizes.

Our next result characterizes the structure of the equilibrium cost rate $J(\cdot)$ which, depending on the parameters, may be convex, concave, or neither. However, it is minimized at all of its critical points, or at an endpoint if no critical points exist.

\begin{proposition}\label{equilibrium_policy}
$J(\cdot)$ is unimodal. Further, its minimizer(s) can be characterized as follows:
\begin{enumerate}
    \item[$(i)$] If $(1-p_l)C'_+(p_l) + C(p_l) + r \geq 0$, then $J(\cdot)$ is non-decreasing and hence minimized at $p_l$.
    \item[$(ii)$] If $(1-p_u)C'_-(p_u) + C(p_u) + r \leq 0$, then $J(\cdot)$ is non-increasing and hence minimized at $p_u$. 
    \item[$(iii)$] Otherwise, $J(\cdot)$ has at least one critical point, i.e., $J'_-(p) \leq 0 \leq J'_+(p)$, and its critical points form a closed interval. If this interval has positive length, then they are all minimizers of $J(\cdot)$, and $C(\cdot)$ is linear along this interval. 
\end{enumerate}
\end{proposition}
If the intervention is ``cheap enough", it is optimal to reduce the return probability to its minimum $p_l$, and if it is ``too costly" it is optimal to keep the return probability at $p_u$. Otherwise, the optimal probability takes a value between the two extreme values. In this case, $J(\cdot)$ is roughly U-shaped: it decreases for small $p$ near $p_l$, then attains its minimum, and finally increases for large $p$ near $p_u$. While the minimum is not guaranteed to be unique, the minimizing values are contiguous, forming a closed interval. Any constant policy lying in this interval yields the minimum.

The following corollary specializes the result to the case of a linear intervention cost and follows directly from Proposition \ref{equilibrium_policy}. For brevity, we omit the proof. 
\begin{corollary}\label{cor:lin_eq}
If $C(\cdot)$ is linear, then one of the following two cases holds:
\begin{enumerate}
    \item[$(i)$] If $r(p_u-p_l)/(1-p_u) \leq C(p_l) - C(p_u)$, then $J(\cdot)$ is minimized at $p_l$.
    \item[$(ii)$] If $r(p_u - p_l)/(1-p_u) \geq C(p_l) - C(p_u)$, then $J(\cdot)$ is minimized at $p_u$.
\end{enumerate}
\end{corollary}
Under a linear intervention cost, the optimal return probability is one of the two extreme points of the interval. The condition required for either of the two extreme points to be optimal is intuitive. Under a constant return probability of $p_u$, we incur a total of $rp_u/(1-p_u)$ in readmission costs per patient. If we treat $p_u$ as the baseline policy, the quantity $r(p_u - p_l)/(1-p_u)$ is the reduction in lifetime readmission costs (i.e., considering future returns) that can be achieved via a one-time intervention that reduces the customer's return probability to $p_l$ at their first service completion. This quantity is evaluated against $C(p_l) - C(p_u)$, i.e., the cost of providing such an intervention.


\section{Transient Control}\label{sec:transient}
Under the long-run average cost objective, any policy that eventually enforces an optimal fixed return probability is optimal, even though it may be ``wasteful" in the initial transient period. In particular, starting from an initial state far from the equilibrium, we may incur significant holding costs in the queue, which are ignored under a fixed optimal long-run average policy. In this section, we investigate policies that \emph{also} account for the initial transient costs. 

We begin by formulating the transient control problem and presenting some preliminary results in Section \ref{subsec:formulation}. We next investigate the structure of optimal transient policies starting with a non-empty queue in Section \ref{subsec:nonempty} and starting with an empty queue in Section \ref{subsec:empty}. Finally, we illustrate the results using numerical examples in Section \ref{subsec:examples}.

\subsection{Formulation of the Associated Fluid Control Problem and Preliminaries}\label{subsec:formulation}
We consider a problem that seeks \emph{bias-optimality}, i.e., minimizing the total cost incurred above/below the long-run average cost $J_\infty$:
\begin{align}
    \min_{p(\cdot)\in\cP} \ & \int_0^\infty \left(h(x(t)-N)^+ + r\nu y(t) + C(p(t)) \mu (x(t) \wedge N) - J_\infty \right)\, dt,\label{eq:infinite_horizon}
\end{align}
where an admissible policy $p(\cdot)\in\cP$ must satisfy $\eqref{eq:fluid_C1}$--\eqref{eq:fluid_C4} over an infinite horizon. In the following, we show that this problem is well-defined, i.e., it has a finite optimal cost despite being undiscounted, and confirm that $J_{\infty}$ is the minimum long-run average cost. To this end, we first introduce a finite-horizon problem. We further establish that the finite-horizon problem is equivalent to \eqref{eq:infinite_horizon} for a sufficiently large horizon.

Define \emph{the finite-horizon control problem} as,
\begin{align}
    \min_{p(\cdot)\in\cP} \ & \int_0^T \left( h(x(t)-N)^+ + r\nu y(t) + C(p(t)) \mu (x(t) \wedge N) - J_\infty \right) \, dt + \Psi(x(T), y(T)),\label{eq:finite_horizon}
\end{align}
where,
\begin{equation}
\Psi(x,y) = \frac{(rp_\infty + C(p_\infty)) x + (r+C(p_\infty))y}{1-p_\infty}
.\end{equation}
Observe that the state at time $T$ is assigned the terminal cost $\Psi$, which charges a penalty for each Needy and Content customer in the system at time $T$. This penalty can be interpreted as the future cost of each customer (ignoring congestion), assuming that a constant policy of $p(t)=p_\infty$ is enforced for $t\geq T$. For each Content customer in the system, the penalty includes the return cost $r$ plus the intervention cost $C(p_\infty)$, with each subsequent return incurring the same costs. As the fraction of returning customers equals $p_\infty$, the geometric series summing all the future costs yields $(r+C(p_\infty))/(1-p_\infty)$. Similarly, for Needy customers the penalty includes the intervention cost $C(p_\infty)$, and the future cost of $(r+C(p_\infty))/(1-p_\infty)$ incurred for a fraction $p_\infty$ of customers who transition into the Needy state. Summing together, the future cost of Needy customers is $(r p_\infty+C(p_\infty))/(1-p_\infty)$. Note that this value is equal to $J_\infty/\lambda$ as new customers arriving at a rate of $\lambda$ together contribute an additional cost of $J_\infty$ in equilibrium.


\begin{proposition}\label{tailvalue}
Let $V_T(x,y,t)$ denote the optimal cost-to-go of the finite horizon problem in \eqref{eq:finite_horizon} with horizon $T$, starting from state $(x,y)$ at time $t$. If $(x, y) \in \cA$, then $(i)$ the optimal policy is given by $p(t) = p_\infty$ for all $t \geq 0$, and $(ii)$ the value function is given by $V_T(x,y,t) = \Psi(x, y)$.
\end{proposition}
As expected, $p_\infty$ is an optimal policy when starting from region $\cA$. Furthermore, the value function is independent of time and equals the terminal cost function $\Psi$. This suggests an equivalence to the infinite-horizon problem. The following proposition establishes that the infinite-horizon problem is well-defined.
\begin{proposition}\label{finiteintegral}
Let $V_\infty(x,y)$ denote the optimal cost of the infinite horizon problem in \eqref{eq:infinite_horizon} starting from $(x,y)$. Then, $-\infty < V_\infty(x,y)<\infty$ for all $(x,y)\in\bR^2_{+}$.
\end{proposition}
Proposition \ref{finiteintegral} also implies that $J_\infty$ is indeed the optimal value of the long-run average problem, since otherwise $V_\infty(x,y)$ would be infinite. 

\begin{proposition}\label{prop:Jinf}
$J_\infty$ is the optimal long-run average cost over all policies $p(\cdot) \in \cP$.
\end{proposition}

It is now easy to verify the equivalence of the infinite-horizon problem in \eqref{eq:infinite_horizon} and the finite-horizon problem in \eqref{eq:finite_horizon} for a sufficiently large $T$ (that depends on the initial condition). For $(x^0, y^0) \in \cA$, Proposition \ref{tailvalue} implies that the optimal policy of the infinite-horizon problem agrees with that of the finite-horizon problem on $[0, T]$, and the two problems share the same optimal objective value. The terminal value function $\Psi$ used in the finite-horizon formulation is the value function of the infinite-horizon problem (up to an additive constant). To extend the equivalence to all initial states, we can therefore separate the infinite horizon into a transient period up to a finite horizon $T\geq t_\cA$, and a tail period. By Proposition \ref{prop:corner} we are guaranteed that $(x(T), y(T)) \in \cA$, and hence the tail part must have cost $\Psi(x(T), y(T))$. As such, we can study these two equivalent problems interchangeably.

\subsubsection{Pontryagin's Minimum Principle.} To study the optimal transient policy, we exploit Pontryagin's minimum principle for the finite-horizon formulation with a large horizon $T$. To this end, we define the \emph{Hamiltonian function} as,
\begin{align}
    H(x,y,p,t) &:= h(x - N)^+ + r\nu y + C(p)\mu (x \wedge N) - J_\infty + (\lambda + \nu y - \mu (x \wedge N)) \gamma_1(t) \nonumber\\
    &\qquad + (-\nu y + \mu p (x \wedge N)) \gamma_2(t).
\end{align}
The vector $(\gamma_1,\gamma_2)$ is referred to as the \emph{adjoint}, or \emph{costate} vector, and is analogous to a Lagrange multiplier (see, e.g., \citealt{todorov}). The costate follows the gradient of the value function along the optimal trajectory. It thus captures the marginal cost of each additional Needy and Content patient at each point in time along the optimal trajectory.

Pontryagin's minimum principle provides necessary conditions on optimality. Specifically, an optimal policy $p^*(t)$ must minimize the Hamiltonian at every time $t$, and the Hamiltonian must always equal zero along an optimal trajectory. That is, for all $t\geq0$:
\begin{align}
H(x(t), y(t), p^*(t), t) &= 0,\\
H(x(t), y(t), p^*(t), t) &= \underset{p\in[p_l,p_u]}{\min} H(x(t), y(t), p, t).
\end{align}
Additionally, the costate vector $(\gamma_1, \gamma_2)$ is governed by the following system of ODEs and boundary conditions (see, e.g., \citealt{kirkbook}):
\begin{align}
\dot\gamma_1(t) &= -H_x(x(t), y(t), p(t), t) = \begin{cases}-h, & x(t) > N, \\ \mu(\gamma_1(t) - C(p^*(t)) - p^*(t)\gamma_2(t)), & x(t) \leq N, \end{cases} \label{eq:Gd1}\\
\dot\gamma_2(t) &= -H_y(x(t), y(t), p(t), t) = \nu(- \gamma_1(t) + \gamma_2(t) - r), \label{eq:Gd2}\\
\gamma_1(T) &= \Psi_x(x(T), y(T)) = \frac{rp_\infty + C(p_\infty)}{1-p_\infty}, \label{eq:G1}\\
\gamma_2(T) &= \Psi_y(x(T), y(T)) = \frac{r+C(p_\infty)}{1-p_\infty}.\label{eq:G2}
\end{align}
The system of ODEs describe the evolution of the trajectories $(x, y)$, $(\gamma_1, \gamma_2)$ over time. However, the dynamics differ depending on whether all servers are busy or not. Therefore, we separately analyze the two cases. Before turning to the main analysis, we present the following result which follows directly from the optimality conditions.
\begin{proposition}\label{prop:linear_bangbang}
Under a linear intervention cost function $C(\cdot)$ the optimal transient policy is a bang-bang control policy.
\end{proposition}
With a linear intervention cost function, an optimal policy only switches between the two extreme values of $p_l$ or $p_u$. We further investigate the structure of an optimal transient policy including under a general convex cost function next.

\subsection{Optimal Transient Policy Starting  with a Non-Empty Queue}\label{subsec:nonempty}
In this section, we characterize the structure of the optimal transient policy starting from an initial condition in $\cC$, i.e., when the system is initiated with a non-empty queue. 

From Proposition \ref{prop:corner} we know that trajectories eventually reach $\mathcal{A}$, regardless of the policy used. As noted in Section \ref{sec:longrun}, for $y(t)>(\mu N-\lambda)/\nu$ we have $\dot{x}(t)>0$. Hence, starting in $\cC$, trajectories can only cross into $\cA$ at a point with $y(t)<(\mu N-\lambda)/\nu$ and hence do not enter $\cN$ before doing so. Let $\tau:=\inf\{t\geq0;x(t)=N\}$ denote the time at which the trajectories reach $\cA$ \emph{under an optimal policy}. We can then step backwards from $T$ to $\tau$ at which point the costates are at their constant equilibrium values, the Needy state is $x(\tau) = N$, and the Content state $y(\tau)$ is at some value between $0$ and $(\mu N - \lambda)/\nu$. To step further back in time, we switch to solving a different system which captures the dynamics when the queue is non-empty, using these four values as boundary conditions. Finally, we characterize the policy as a function of the state by exploiting the requirement of Pontryagin's minimum principle that the Hamiltonian must equal zero along the optimal trajectory. 

\subsubsection{Optimal Policy as a Function of Time.} We begin by solving the system in \eqref{eq:Gd1}--\eqref{eq:G2} to determine the costate trajectories. We present the solution in the following lemma. The proof is straightforward and hence omitted.
\begin{lemma}\label{costate_formula}
$(i)$ For $t \in [\tau, T]$, the costates are constant and given by,
\begin{align}
    \gamma_1(t) &= \frac{rp_\infty + C(p_\infty)}{1-p_\infty}, \\
    \gamma_2(t) &= \frac{r+C(p_\infty)}{1-p_\infty}.
\end{align}
$(ii)$ For $t \in [0, \tau)$, the costates are strictly decreasing and are given by:
\begin{align}
    \gamma_1(t) &= h(\tau - t) + \frac{rp_\infty + C(p_\infty)}{1-p_\infty}, \\
    \gamma_2(t) &= \frac{h}\nu (e^{-\nu(\tau-t)} + \nu (\tau - t) - 1) + \frac{r +C(p_\infty)}{1-p_\infty}.
\end{align} 
\end{lemma}

Next, we proceed to characterize the optimal policy $p^*(t)$ as a function of time. Pontryagin's minimum principle states that if a policy $p^*(t)$ is optimal, then $p^*(t)$ must minimize $H(x(t), y(t), p, t)$ for all $t\geq0$. In our case, this condition simplifies to requiring that $p^*(t) \in \argmin_{p} \phi_t(p)$, where $\phi_t(p) := C(p) + \gamma_2(t) p$. Because $C(\cdot)$ is convex, the same must hold for $\phi_t(\cdot)$. As such, minimization occurs at a critical point, characterized by the first-order condition $C'_-(p) \leq -\gamma_2(t) \leq C'_+(p)$, or at an endpoint if no critical points exist. 
\begin{proposition}\label{policyformula}
Any optimal policy must be non-increasing in time during $t \in [0, \tau)$. Specifically, it must be of the form:
\begin{equation}\label{eq:ps_time}
    p^*(t) = \begin{cases}p_u, & \gamma_2(t) < -C'_-(p_u), \\
    p_l, & \gamma_2(t) > -C'_+(p_l) \\
    p \text{ such that }  C'_-(p) \leq -\gamma_2(t) \leq C'_+ (p), & \text{otherwise}, \end{cases}
\end{equation} where $C'_-(\cdot)$ and $C'_+(\cdot)$ are the left and right derivatives of $C(\cdot)$, respectively. Furthermore, the optimal policy $p^*(t)$ is non-decreasing in time and almost everywhere unique on $[0, \tau)$.
\end{proposition}
In the special case where $C(\cdot)$ is continuously differentiable, an interior return probability $p \in (p_l,p_u)$ can only be optimal at an instant where $C'(p) = -\gamma_2(t)$, since $\gamma_2$ is strictly decreasing. In general, if $C'_-(p) \leq -\gamma_2(t) \leq C'_+(p)$ for multiple values of $p$, then all such values are optimal. This can happen for at most countably many $t$ and hence policies satisfying \eqref{eq:ps_time} must agree almost everywhere (with respect to time). These differences are inconsequential as they do not affect the state trajectory nor the objective value. Hence, the optimal trajectory is almost everywhere unique until time $\tau$. 



\subsubsection{Optimal Policy as a Function of State.} So far, we have characterized the optimal policy as a function of time and in terms of the value $\tau$. If we know the value of $\tau$ associated with the initial state $(x^0, y^0) \in\cC$ under the optimal policy, we can then simply apply Proposition \ref{policyformula} starting from time $t=0$ until we reach $\cA$, after which we switch to the equilibrium policy. However, the value of $\tau$ is not known, making the result not directly applicable. Instead, suppose that we start with a particular value of $\tau$. Then $p^*(0), \gamma_1(0)$, and $\gamma_2(0)$ can all be computed for this particular $\tau$. The requirement that $H(x^0, y^0, p^*(0), t) = 0$ then becomes a constraint on the corresponding state $(x^0,y^0)$. In other words, all states which are exactly $\tau$ time away from $\cA$ must satisfy what turns out to be a linear constraint, allowing for characterization of the optimal policy as a function of the state. As we discuss below further, this characterization provides structural insights but also allows for the computation of the optimal policy for all states in $\cC$. 




\begin{theorem}\label{policystructure}
Suppose $(x,y)\in\cC$. Under the optimal policy, there exists a unique $\tau > 0$ such that $(x,y)$ lies on the line:
\begin{equation}\label{lineequation}
    h(x-N) + h(1-e^{-\nu\tau})y - J_\infty + (\lambda - \mu N)\gamma_1(\tau) + \mu N \left( C(p^*) + \gamma_2(\tau) p^* \right) = 0,
\end{equation}
where,
\begin{align}
    \gamma_1(\tau) &= h\tau + \frac{rp_\infty + C(p_\infty)}{1-p_\infty},\\
    \gamma_2(\tau) & = \frac{h}\nu (e^{-\nu\tau}+\nu\tau-1) + \frac{r+C(p_\infty)}{1-p_\infty},\\
    p^* & = \argmin_{p\in[p_l,p_u]} \{C(p) + \gamma_2(\tau) p\}.\label{eq:p*}
\end{align}

\end{theorem}

Theorem \ref{policystructure} characterizes the optimal policy by mapping each state to the optimal return probability $p^*$ and the corresponding clearing time $\tau$ through \eqref{lineequation}--\eqref{eq:p*}. To interpret the characterization, it is instructive to view the relationship in the opposite direction: for each $\tau>0$, there exists a set of initial conditions in $\cC$, characterized by \eqref{lineequation}, which share the same optimal return probability $p^*$ and whose optimal queue clearing time is exactly $\tau$. We show (see Lemma \ref{increasingintercept} in proof of Theorem \ref{policystructure}) that these lines have increasing slopes and $(x=N)$--intercepts with respect to $\tau$, i.e., they do not intersect in the congested region, ``fanning out'' from the boundary $x(\tau)=N$. It follows that every state lies on exactly one such line, and every point on the line associated with $\tau$ must be exactly $\tau$ time away from reaching $\cA$. 

The identity \eqref{lineequation} implies that the set of states sharing the same optimal return probability is characterized by a line of the form $x + (1-e^{-\nu\tau})y = a$ where $a$ is a constant that depends on the system and cost parameters as well as $\tau$. At any point along the line, we are trading off between Needy and Content customers without changing the intervention policy. Note that the Content state is discounted by a factor of $1-e^{-\nu\tau}$, reflecting the fact that the customers readmit in the future, when we will have made more progress in reducing congestion. The simple structure also provides insights on the impact of return time. Indeed, if $\nu$ is very large (i.e., customers return soon after discharge, immediately contributing to the congestion) then the discount is very small and the returning customers are treated on par with Needy customers. On the other hand, if $\nu$ is very small (i.e., patients readmit ``much later", by which time the congestion will be completely cleared) the Content customers are entirely ignored. 



The theorem further supports the computation of the optimal policy for states in $\cC$. Instead of treating the policy as a function that varies through time as we move along a specific trajectory, Theorem \ref{policystructure} allows us to relate the policy to all states in $\cC$. It allows us to find all states corresponding to a given value of the congestion-clearing time $\tau$. By computing these lines for a grid of $\tau$ values, we can simultaneously trace out the path of \emph{all} trajectories moving through $\cC$. We illustrate the structure of optimal policies for different intervention costs using examples in Section \ref{subsec:examples}.

\subsubsection{Piecewise Linear Intervention Cost and Switching Control.}
We next consider the special case where $C(\cdot)$ is piecewise linear. The following corollary follows from Theorem \ref{policystructure}. 

\begin{corollary}\label{cor:piecewise}
Let $C(\cdot)$ be a piecewise linear function with $k\geq 1$ pieces. Let $p_1, \ldots, p_{k-1}$ be the ``breakpoints" of $C(\cdot)$, and $p_0 := p_l, p_{k+1} := p_u$. An optimal policy divides $\mathcal{C}$ into (up to) $k+1$ disjoint regions $\{\Omega_0,\ldots,\Omega_{k}\}$ where constant return probabilities $p_0<\ldots<p_k$ are used in the corresponding regions. The boundaries of the regions are line segments of the form $x+(1-e^{-\nu \tau_i})y-a_i=0$.
\end{corollary}

Consider the special case of a linear cost, i.e., $C(p)= M (p_u - p)/(p_u-p_l)$ where $M$ is the cost of intervention at its highest level. We know from Proposition \ref{prop:linear_bangbang} that the optimal control policy is bang-bang (at all times, over the entire state-space). In this case, the state-space can be divided into two regions: one where no intervention is provided (with return probability $p_u$), and another where the highest level of intervention is provided (with return probability $p_l$). The part of the boundary in $\cC$ corresponds to the points whose clearing time $\tau$ satisfies, 
\begin{equation}
    \frac{M}{p_u-p_l} = \gamma_2(\tau)= \frac{h}{\nu}(e^{-\nu\tau}+\nu\tau - 1) + \frac{r+C(p_\infty)}{1-p_\infty},
\end{equation}
i.e., a single line which divides the two regions, and it represents the states with a queue clearing in $\tau$ time units. This single line of the form $x + (1-e^{-\nu\tau})y = a$ completely characterizes the optimal policy for states in the congested region $\cC$. Similarly, for a piecewise linear cost function, the optimal policy for states in $\cC$ takes the form of multiple lines dividing the state-space into regions where the optimal return probability is constant within each region and switches when crossing the boundaries of the regions. 

In Corollary \ref{cor:piecewise} we establish the optimality of a \emph{switching control} in the case of piecewise linear intervention costs, i.e., where the optimal return probability switches from one value to another in different regions of the state-space. In this case, the fluid model is a \emph{switching dynamical system} with righ-hand-side discontinuity. It is well-know in the dynamical system theory literature that discontinuity in control could lead to complex dynamics including sliding modes and chattering along the switching boundary, see, e.g., \cite{perry2016chattering} for an example of chattering in a queueing system under poorly chosen control parameters. In the following, we examine the existence and uniqueness of the solution to the ODE \eqref{eq:fluidt_1}--\eqref{eq:fluidt_2} under a switching control of the form established in Corollary \ref{cor:piecewise}. More specifically, consider a policy that divides the state-space into $k+1$ disjoint regions where $k$ is the number of pieces in the intervention cost function, and (with a slight abuse of notation) denote the regions by $\Omega_i$ such that their closures denoted by $\bar{\Omega}_i$ covers $\mathbb{R}^2_+$, i.e, $\cup_{i=1}^{k+1} \bar{\Omega}_i = \mathbb{R}^2_+$. Denote the boundaries by $\partial \Omega_i$ such that the points of discontinuity belong to two boundaries where different levels of interventions are provided below and above the boundary. We then have the following ODE with righ-hand-side discontinuity:
\begin{align}
    \dot x(t) &= \lambda - \mu (x(t)\wedge N) + \nu y(t),\label{eq:switchedDS1}\\ 
    \dot y(t) &= -\nu y(t) + \mu p_i (x(t) \wedge N),\quad \mbox{if } (x(t),y(t))\in\Omega_i.\label{eq:switchedDS2}
\end{align}
Given the discontinuous right-hand-side, we consider Filippov solutions \citep{filippov2013differential}. According to Filippov's notion of solutions, once enriches the set of time-derivatives for points on the boundary by considering all convex combinations of the vectors in the two regions. An absolutely continuous function is then a solution of the switched system in the sense of Filippov if it satisfies the resulting \emph{differential inclusion}; see Section \ref{ap:filippov} for details. The following theorem establishes the existence and uniqueness of Filippov solutions starting with an initial condition in $\cC$ under Assumption \ref{assump:stability}.
\begin{theorem}\label{thm:unique}
    There exists a unique Filippov solution of \eqref{eq:switchedDS1}--\eqref{eq:switchedDS2} starting from each initial condition in $\cC$ when boundaries in $\cC$ are line segments of the form $x+(1-e^{-\nu \tau_i})y-a_i=0$.
\end{theorem}
We note that the result does not require the boundaries to use optimal parameters $\tau_i,a_i$, implying the robustness of the structure to errors in determining the control parameters. Theorem \ref{thm:unique} rules out unintended dynamics (e.g., chattering) under the specified switching control. In particular, we show in the proof that trajectories that hit the boundary either cross into the next region, or if they slide on the boundary this must happen temporarily. The proof relies on the smoothness of the boundary and its specific form characterized in Theorem \ref{policystructure} and Corollary \ref{cor:piecewise}.

\subsection{Optimal Transient Policy Starting with an Empty Queue}\label{subsec:empty}
We next turn to the case where the system is initiated with an empty queue. When the number of Content customers is also ``small", i.e., starting in $\cA$, we already know the optimal policy coincides with the equilibrium policy. With a ``large" number of Content customers, i.e., starting from $\cN$ where $y > (\mu N - \lambda)/\nu$, it could still be beneficial to reduce the return probability of the current Needy customers to avoid higher congestion levels in future when the Content customers return. 

An analytical characterization of the optimal policy for this part of the state-space is, however, much more challenging. When studying the optimal policy in $\cC$, we benefited from two features of the system of ODEs \eqref{eq:Gd1}--\eqref{eq:G2} that characterized the costates. First, the equation $\dot\gamma_1(t) = -h$ could be solved explicitly without reference to $\gamma_2$. The resulting linear expression for $\gamma_1$ could then be used in solving for $\gamma_2$. Second, there was a fixed boundary condition for $\gamma_1$, $\gamma_2$. In particular, optimal trajectories are guaranteed to enter $\cA$ from $\cC$. Once they reach $\cA$, the marginal cost of each customer in the system does not depend on the specific state. In contrast, in $\cN$, the dynamics of $\gamma_1,\gamma_2$ depend on each other. We are thus forced to consider $\gamma_1$, $\gamma_2$, and $p$ simultaneously, facing a three-dimensional differential equation involving a minimization (due to the presence of $p$). Moreover, the boundary conditions are more complex. Starting from a point in $\cN$, we may transition to the congested region. As such, the boundary values of $\gamma_1$ and $\gamma_2$ can vary depending on where the line $x=N$ is crossed.


Given these complexities, we numerically investigate the optimal policy in region $\cN$, and provide analytical support by establishing certain properties of the optimal trajectories. Taken together, the fluid dynamics \eqref{eq:fluid_C1}--\eqref{eq:fluid_C2} and costate dynamics \eqref{eq:Gd1}--\eqref{eq:G2} form a full-rank system of ODEs. For any state in $\cA$, we know that the corresponding costates must be equal to their terminal values as specified in Lemma $\ref{costate_formula}$, thus providing a complete set of boundary conditions. We can then solve for the optimal trajectory passing through that state, and obtain the optimal policy for every point along the trajectory. By repeating this procedure for a grid of closely-spaced points along the boundary of $\cA$, we obtain a grid of trajectories fanning outward from $\cA$. We can then interpolate to obtain the policy for any point in the state-space. While this is an approximate solution, we can increase the precision of the solution by increasing the density of the starting points. 

As we illustrate and further discuss in Section \ref{subsec:examples}, we observe that the optimal policy in $\cN$ has a similar structure to that in $\cC$, except that the contour lines are nonlinear. Specifically, they are extensions of the linear contours in $\cC$ and similarly monotone in state. In the case of a piecewise linear intervention cost, the state-space is divided into different regions where different intensities of interventions are provided, but with nonlinear boundaries. In addition, we conjecture that a similar uniqueness result to Theorem \ref{thm:unique} also holds for initial conditions in $\cN$; see Section \ref{ap:filippov} for additional details.

We conclude this section with two supporting results. First, we show that starting in $\cN$, if the optimal policy ever deviates from $p_\infty$, we must enter the $\cC$ region.
\begin{proposition}\label{prop:noaggression}
Starting from a state in $\cN$, it is never optimal to reduce the return probability to avoid entering $\cC$.
\end{proposition}
Second, in the special case of a linear intervention cost, we show that the optimal trajectory switches between intervention and no-intervention at most twice before entering $\cC$, hence supporting our numerical observations.
\begin{proposition}\label{prop:twoswitches}
    Assume that the intervention cost $C(\cdot)$ is linear. Starting from an initial condition in $\cN$, an optimal trajectory makes at most two policy changes before entering $\cC$. 
\end{proposition}
The proposition implies that starting in $\cN$, either we never intervene, or if we intervene and stop, we never start intervening again.

\subsection{Illustrative Examples}\label{subsec:examples}
In this section, we use numerical examples to illustrate the structure of optimal policies under different cost structures. We further discuss and illustrate the properties of optimal trajectories, focusing in particular on initial conditions in $\cN$. 

\subsubsection{Optimal Policy Structure.} We present three examples with common system parameters $\lambda=9.5,\mu=1/4,\nu=1/15,N=50$. The return cost is normalized at $r=1.0$ and the holding cost rate is $h=0.25$. We assume that the baseline return probability is $p_u=0.2$ which can be reduced to as low as $p_l=0.1$, i.e., $[p_l, p_u]=[0.1,0.2]$. The cost functions share the same maximum and minimum intervention costs of $C(0.2)=0$ and $C(0.1)=0.5$, but differ in the interior of $[p_l, p_u]$.

\begin{figure}[]
\center{\includegraphics[width=\textwidth]{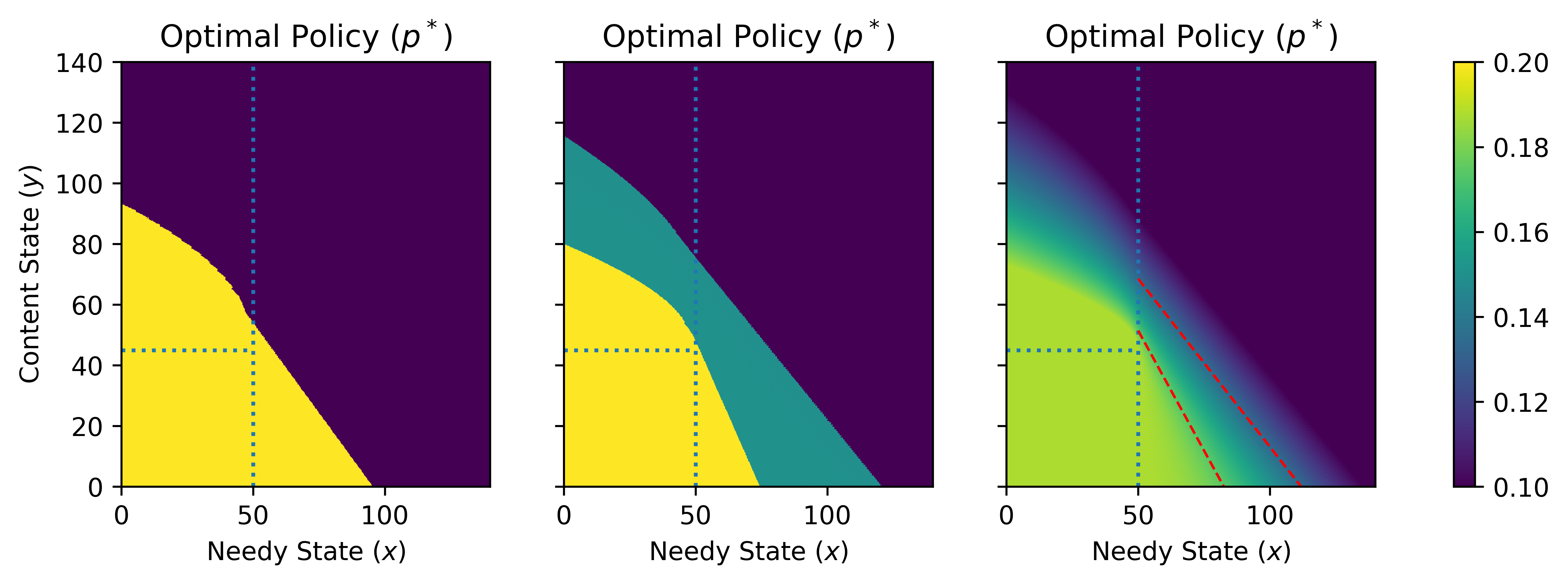}}
\caption{Optimal transient policies under three intervention cost structures: linear $C(p) = 5(0.2-p)$ (left), piecewise linear $C(p) = 2(0.2-p) + 6(0.15-p)^+$ (middle), and quadratic $C(p) = 50(0.2-p)^2$ (right). Other system parameters are $N=50, \lambda=9.5, \mu=1/4, \nu=1/15, p_l=0.1, p_u=0.2, r=1.0, h=0.25$. Two (isochromatic) contour lines in the $\cC$ region for the quadratic intervention cost are highlighted in red.}
\label{fig:policy_examples}
\end{figure}
First, we consider a linear intervention cost function $C(p)=5(0.2-p)$. Under this cost structure, no intervention is provided in equilibrium. The optimal transient policy is presented in Figure \ref{fig:policy_examples} (left). There is a single boundary separating the $p_l$ region and the $p_u$ region, and the portion of this boundary lying in $\cC$ is a line. For states in $\cN$ the boundary takes the form of a nonlinear curve. Interestingly, at lower Needy states, the policy switches from no intervention to full intervention at lower Content states compared to an extension of the line in $\cC$. This can be explained noting that the service completion rate is $x(t)\mu$ in this region and hence with a lower number of Content customers the system experiences a slow-down in clearing the customers from the system. As such, the intervention is triggered with a smaller number of Content customers with pending return, in order to compensate for the slow-down.

Next, we consider a piecewise linear cost $C(p) = 2(0.2-p) + 6(0.15-p)^+$ which has a breakpoint at $C(0.15) = 0.1$. Compared to the linear intervention cost which has derivative $C'(p) = -0.5$, the partial intervention of $p=0.15$ comes at a lower marginal cost with $C'(p) = -0.2$, but moving from partial to full intervention comes at a higher marginal cost with $C'(p) = -0.8$. The optimal transient policy is presented in Figure \ref{fig:policy_examples} (middle). The policy structure is similar to that of the linear cost. We once again apply no intervention ($p_h$) in the bottom-left of the state-space, and apply full intervention ($p_l$) in the upper-right. However, there is now also a strip in the middle which corresponds to mid-level intervention ($p=0.15$). The boundary between the partial and full intervention regions is less steep than the one between the partial and no-intervention regions. 

Finally, we consider a strictly convex intervention cost $C(p) = 50(0.2-p)^2$. In this case, using Proposition \ref{equilibrium_policy} and minimizing $J(p)$ in \eqref{eq:Jp}, we find that the equilibrium optimal policy is to apply a modest level of intervention, yielding a return probability of $p_\infty = 0.1876$. Figure \ref{fig:policy_examples} (right) presents the optimal transient policy. We observe that the policy is monotone with respect to both $x$ and $y$, decreasing from $p_\infty$ near the bottom-left to $p_l$ in the upper-right. Moreover, this happens continuously such that the policy contours form lines in $\cC$ as described in Theorem \ref{policystructure}. Two such lines are highlighted on the figure as examples. As established in Theorem \ref{policystructure}, we observe from the slopes of these two lines that the contours become less steep as $p^*$ decreases, fanning out as we move toward the upper-right. As the system becomes more congested, the Content customers are thus discounted at a smaller rate relative to the Needy customers in determining the level of intervention. This is intuitive since if the queue takes longer to clear, then Content customers contribute more to future congestion costs whereas with a small queue-length, the congestion will already have cleared by the time most of the Content customers return. In addition, we observe that a similar structure holds in $\cN$ except that the contours are nonlinear.
\subsubsection{Optimal Trajectories.} 
\begin{figure}[]
    \centering
    \includegraphics[width=0.4\linewidth]{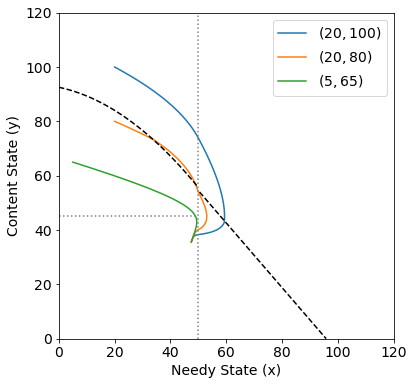}
    \includegraphics[width=0.4\linewidth]{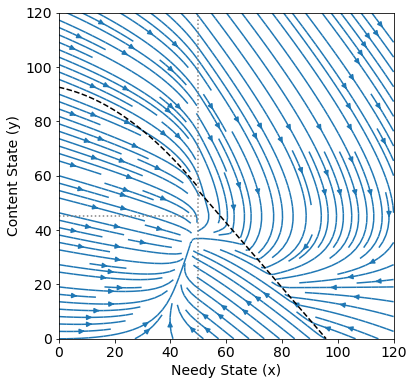}
    \caption{Optimal trajectories starting from three initial conditions in $\mathcal{N}$ (left) and stream plot under the optimal policy (right). System and cost parameters are $N=50,\bar{\lambda}=0.95,\mu=1/4, \nu=1/15$; and $r=1, M=0.5, h=1/4$ with $C(p) = 5(0.2-p)$. The dashed line corresponds to the intervention boundary.}
\label{fig:opt_traj}
\end{figure}

Figure \ref{fig:opt_traj} presents an stream plot of the ODE under the optimal policy with the linear intervention cost function, as well as selected optimal trajectories starting in $\cN$. The trajectories illustrate different paths starting from $\cN$: they may directly enter $\cA$ or first enter $\cC$. In the latter case, as shown in Proposition \ref{prop:twoswitches}, we can have at most two switches. Note that this implies that the optimal policy is not necessarily monotone (with respect to time)  starting in $\cN$ and in contrast to starting in $\cA$ and $\cC$.


\section{Time-Varying Arrivals}\label{sec:tv}

So far, we have assumed a stationary arrival rate in our analysis. In many service systems, including our motivating healthcare setting, arrival rates are subject to significant temporal variation; see, e.g., \cite{green2007coping} and \cite{yom2014erlang}. Hence, in the following, we discuss the application and robustness of the results and policies developed for the stationary system to the case with time-varying arrival rates. Specifically, we assume that the arrival process follows a non-stationary Poisson process with a periodic rate $\lambda(t)$ with period $f$, i.e., $\lambda(t+f)=\lambda(t)$ for all $t\geq0$, and average (over a period) equal to $\bar{\lambda}$. In addition, we extend the set of admissible policies to allow for time-dependent policies. We continue to rely on fluid approximations to develop insights. 

Under time-varying arrivals, the trajectories $(x(t),y(t))$ are governed by
\begin{align}
 & \dot x(t) = \lambda(t) + \nu y(t) - \mu (x(t) \wedge N), \label{eq:fluidtv_C1}\\
    & \dot y(t) = -\nu y(t) + \mu p(x(t),y(t),t) (x(t) \wedge N), \label{eq:fluidtv_C2}
\end{align}
for $t\geq 0$. Under a fixed return probability $p<1-\bar{\lambda}/(\mu N)$, we expect the trajectories to converge to a periodic equilibrium (i.e., a cyclic orbit in the state-space) with the same period $f$ as the arrival rate. Paralleling our stationary analysis, if we restrict our attention to policies that converge to a periodic equilibrium, we may interpret $J_{\infty}$ as the optimal long-run average cost. Note that with time-varying arrivals the equilibrium may involve a queue that builds up and empties during the period, and as such $J_{\infty}$ we may also include holding costs. Therefore, the optimal policy may not be constant, and could vary depending on the congestion level. 

Consider minimizing the total cost over a long but finite time horizon $T$, plus some time-dependent terminal cost function. While we cannot specify the terminal cost function $\Psi_T$, we are still able to gain insights by examining the necessary optimality conditions of a control problem of the form,
\begin{equation}
    \min_{p(\cdot)\in\cP} \int_0^T h(x(t)-N)^+ + r\nu y(t) + C(p(t))\mu(x(t) \wedge N) \,dt + \Psi_T(x(T), y(T)),
\end{equation}
where an admissible policy must satisfy the fluid dynamics in \eqref{eq:fluidtv_C1}--\eqref{eq:fluidtv_C2} in addition to \eqref{eq:fluid_C3}--\eqref{eq:fluid_C4}. Intuitively, if $T$ is sufficiently large, this should approximate the bias-optimal objective with an optimal policy achieving both the optimal long-run average and transient costs. 

Applying Pontryagin's minimum principle to this problem, we observe that as in the stationary case the optimal policy is given by $p^*(t) = \argmin_p \{ C(p) + \gamma_2(t) p \}$. We thus pay particular attention to the trajectory of the costate $\gamma_2(t)$, since its value determines the optimal policy. The costate trajectories are governed by the same equations \eqref{eq:Gd1}-\eqref{eq:Gd2}.
The time-varying arrival rate $\lambda(t)$ affects the timing of the transitions when $x(t)$ crosses above and below the capacity $N$. Characterization of an optimal policy hence becomes much more difficult, similar to the case of region $\cN$ under stationary arrivals. In particular, each entry into or exit from $\cC$ marks a change in the system of ODEs governing the dynamics, which requires computation of new boundary conditions. As such, we focus on developing an understanding of the parameter regimes where the time-variation has a significant impact on the performance of a policy.  

In the case of a fixed probability of return, \cite{yom2014erlang} find that when the service times are ``much longer'' than the scale of time variability (e.g., hourly variation in arrival rate and service times in the order of days) the impact of time-variation is small. Similar observations are made in \cite{chan2014use} in a system subject to speedup under a threshold policy. We study the effect of the time-scale of the parameters for our problem by examining the costate trajectories. The higher the transition rates $\mu$ and $\nu$ (i.e. the shorter the timescales), the faster the costates change. If we fix the period $f$ while increasing the service and time-to-return times (i.e., decrease $\mu$ and $\nu$), we expect to see less variation in $\gamma_2$ and hence less variation in the policy. Conversely, holding $\mu$ and $\nu$ unchanged and varying the arrival rate faster would achieve the same. Intuitively, shorter-term fluctuations have less impact on the policy, since they are smoothed out when considering the timescale of the service and return. 

\begin{figure}[]
    \centering
    \begin{subfigure}{0.38\textwidth}\centering
    \includegraphics[width=\linewidth]{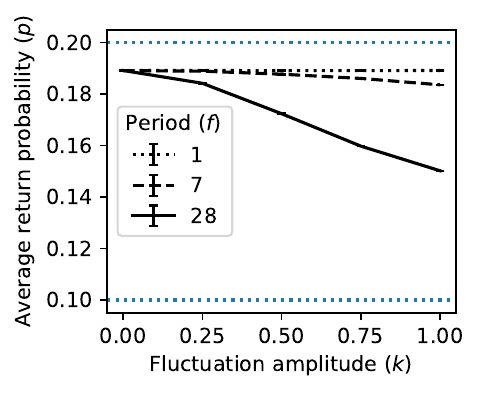}
    \caption{Linear Intervention Cost}
    \end{subfigure}
    \begin{subfigure}{0.38\textwidth}\centering
    \includegraphics[width=\linewidth]{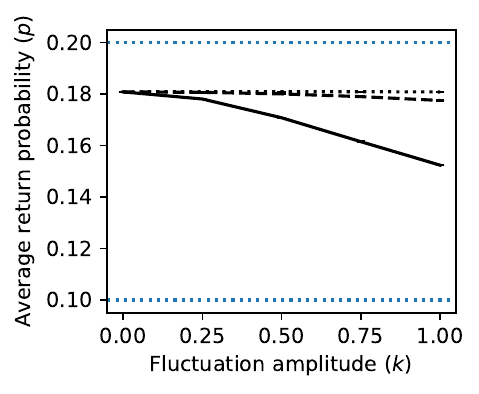}
    \caption{Quadratic Intervention Cost}
    \end{subfigure}
    \caption{Long-run average probability of return under the fluid policy for three periods and varying amplitudes of the arrival rate. Other system and cost parameters are $N=50,\bar{\lambda}=0.95,\mu=1/4, \nu=1/15$; and $r=1, M=0.5, h=1/4$ with $C(p) = 5(0.2-p)$ (left) and $C(p) = 50(0.2-p)^2$ (right).}
\label{fig:varying_intervention}
\end{figure}

We illustrate these effects using simulation experiments. We consider sinusoidal arrival rate functions of the form,
\begin{equation}\label{eq:sinarrival}
    \lambda(t) = \bar \lambda( 1+k\sin(2\pi t/f)),
\end{equation}
and vary the fluctuation amplitude $k \in \{0, 0.25, 0.5, 0.75, 1.0\}$ as well as the period $f \in \{1, 7, 28\}$ to capture daily, weekly, and monthly timescales. We fix the other system parameters to $N=50,\bar{\lambda}=0.95,\mu=1/4, \nu=1/15$ and the cost parameters to $r=1, M=0.5, h=1/4$ and consider the same linear and quadratic intervention cost functions from Section \ref{subsec:examples}. We compute the fluid policy using the stationary arrival rate $\bar{\lambda}$ and apply it to the stochastic system (see Section \ref{sec:numerics}). 

Figure \ref{fig:varying_intervention} presents the long-run average probability of return under the fluid policy for the three periods and varying relative amplitudes. We observe that hourly variations in the arrival rate have effectively no impact. Under strong intra-week fluctuation, we begin to see more frequent interventions, but the impact is still modest. It is only when we move to the extreme case of month-long variation, which has the same order as the service time and time-to-return, that we observe a more sizable impact. We find consistent results with respect to other performance measures, e.g., the long-run average proportion of time all servers are busy; see Appendix \ref{ap:supnumerics}.

Our experiments suggest that in a parameter regime relevant to our motivating application, namely significant hourly arrival rate variations and service and return times that are in the order of days, our fluid policy that uses the average arrival rate $\bar{\lambda}$ should continue to preform well. We demonstrate the performance of the policy for systems with time-varying arrivals in the numerical study of Section \ref{sec:numerics}.

\section{Numerical Experiments}\label{sec:numerics}
In this section, we first conduct a simulation study to: (1) investigate the performance of the fluid-based control policies for the original stochastic system (with both stationary and time-varying arrivals), and (2) identify parameter regimes where the post-service intervention can lead to significant cost savings. Next, in Section \ref{sec:case}, we conduct a case study where we examine the robustness and application of the results to the hospital readmission application. 

We directly translate the fluid policy to an admissible policy for the stochastic system. At each departure epoch $t_k,k\geq0$ the return probability is determined through $p^{\tilde{\pi}}(X(t_k),Y(t_k))$ where $p^{\tilde{\pi}}(\cdot,\cdot)$ is the optimal solution of the transient fluid control problem starting from state $(X(t_k),Y(t_k))$. Note that the translated policy is determined based on the state of the system at each departure epoch, which may deviate from the predicted fluid trajectory due to stochastic fluctuations. We refer to this policy as as the \emph{fluid policy} and evaluate its performance based on both the long-run expected cost of the system as well as the expected cost over a finite-horizon of 90 days and starting from different initial states. We compare the performance against two benchmark policies:

\begin{enumerate}
    \item The \emph{equilibrium policy} which uses the fixed optimal (fluid) long-run average return probability $p_\infty$ at all times. This policy ignores the congestion and its associated costs, only considering the direct trade-off between readmission and intervention costs. As noted in Section \ref{sec:longrun}, the optimal fixed return probability for the stochastic system that also accounts for steady-state holding costs can be computed numerically; see Section \ref{ap:exact}. Here, we use the fluid optimal since it can be viewed as an optimistic version of the status-quo practice for the hospital readmission application, where only readmission and intervention costs are considered. By comparing the difference in costs under the fluid policy and the equilibrium policy, we gain an understanding of the value of congestion-awareness. 
    \item The \emph{simple policy} which uses the return probability $p_\infty$ whenever $X(t_k) \leq N$, and the lowest possible return probability $p_l$ (highest level of intervention) whenever the queue is non-empty. This policy considers congestion, but not the specific level of congestion. By comparing the performance of the fluid policy with that of the simple policy, we thus quantify the value of dynamically adjusting the intervention level in response to varying levels of congestion.
\end{enumerate}

We compare the fluid policy against these benchmarks by estimating the absolute and relative reduction in the expected cost of the system achieved by the policy. The relative reduction is defined as one minus the ratio between the expected cost incurred under the fluid policy and that incurred under the benchmark policy. We construct a confidence interval for this estimate using Fieller's method (\citealt{fieller1932}).

\textbf{System parameters.} Throughout our experiments, we fix the number of servers at $N = 50$ and the service rate at $\mu = 1/4$, representing an average service time of 4 days. We vary the return rate in our experiments in $\nu \in \{1/10, 1/15, 1/20\}$ representing an average time-to-return between 10 and 20 days. We consider the interval of return probabilities $[p_l, p_u] = [0.1, 0.2]$. The 20\% return probability without intervention is representative of higher-risk patient cohorts for whom one would be interested in intervening. Finally, we vary the exogenous arrival rate $\lambda \in \{9.0,9.5,9.8\}$. Note that in the absence of an intervention (i.e., under a constant return probability $p=0.2$), the arrival rates corresponds to a utilization of 90\%-98\%.

\textbf{Cost parameters.} We fix the return cost in our experiments at $r=1$, representing a reference value against which the other costs are measured. Note that this cost is charged once for the entire service. Since the average service time is 4 days, the return cost is $0.25$ per day for the average service time. Denote by $M$ the maximum intervention cost corresponding to the lowest return probability. We consider two functional forms for the intervention cost: (1) linear with $C(p) = 10M(0.2-p)$ and (2) quadratic with $C(p) = 100M(0.2-p)^2$. While $C(p_l)$ and $C(p_u)$ are identical under both forms, the strict convexity of the quadratic cost offers the ability to partially reduce the readmission probability for a lower marginal cost. As a baseline, we set $M = 0.5$, which is half the return cost, but vary $M\in \{0.2,0.5,1.0\}$ in our experiments. Finally, we consider a baseline holding cost rate of $h=0.25$ per day for each customer waiting in queue, equal to the average per-day cost of return. We vary $h\in\{0.05, 0.1, 0.25,0.5,1.0\}$ in our experiments. 

\subsection{Stationary Arrivals}\label{subsec:stat_exp}
We present two sets of experiments. In the first set, we fix the system parameters at their nominal values and vary the cost parameters. In the second set, we fix the cost parameters at their nominal values and vary the arrival and return rates. We present the results under both finite-horizon and infinite-horizon (long-run) settings. For the finite-horizon experiments, we consider three initial state values $(X(0),Y(0))\in\{(25,65),(65,25),(65,65)\}$ and examine the performance of the policies over a 90-day horizon. 

\begin{figure}[]
    \centering
    \begin{subfigure}{0.32\textwidth}\centering
    \includegraphics[width=\linewidth]{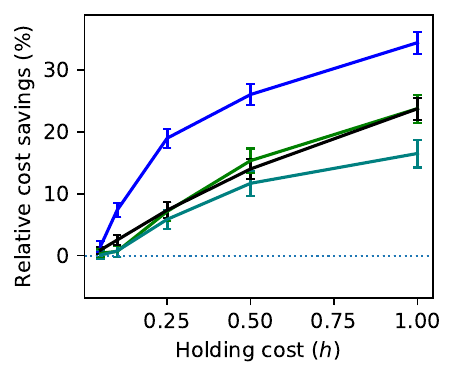}
    \caption{$M=0.2$}
    \end{subfigure}
    \begin{subfigure}{0.32\textwidth}\centering
    \includegraphics[width=\linewidth]{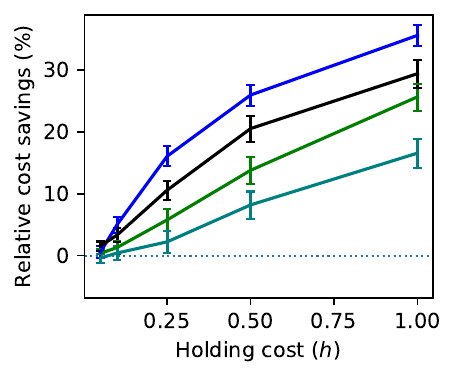}
    \caption{$M=0.5$}
    \end{subfigure}
    \begin{subfigure}{0.32\textwidth}\centering
    \includegraphics[width=\linewidth]{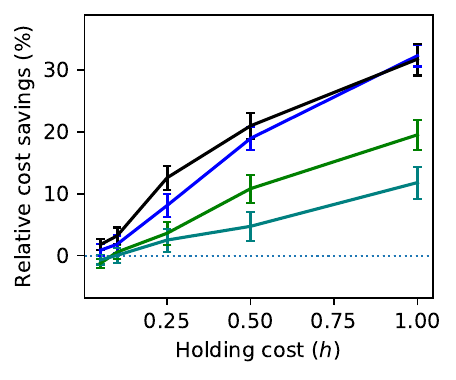}
    \caption{$M=1.0$}
    \end{subfigure}
    \begin{subfigure}{0.32\textwidth}\centering
    \includegraphics[width=\linewidth]{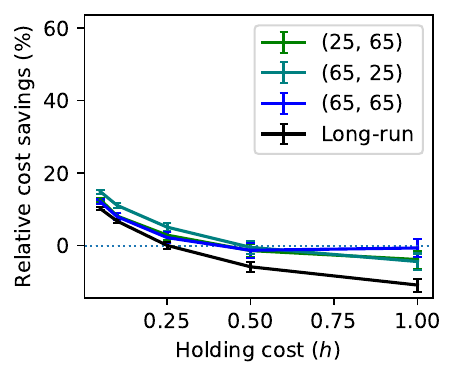}
    \caption{$M=0.2$}
    \end{subfigure}
    \begin{subfigure}{0.32\textwidth}\centering
    \includegraphics[width=\linewidth]{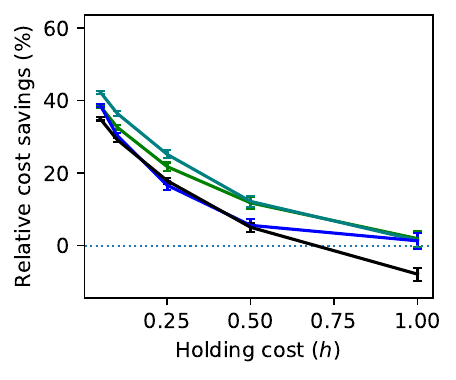}
    \caption{$M=0.5$}
    \end{subfigure}
    \begin{subfigure}{0.32\textwidth}\centering
    \includegraphics[width=\linewidth]{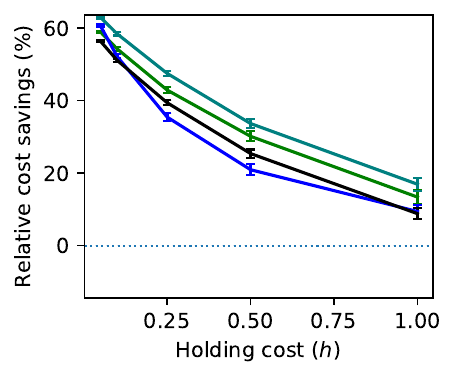}
    \caption{$M=1.0$}
    \end{subfigure}
    \caption{Relative reduction in expected cost for both finite-horizon and long-run experiments under the fluid policy with respect to the equilibrium policy (top row) and simple policy (bottom row) under the quadratic intervention cost and for different values of cost parameters. System parameters are fixed at $N=50,\lambda=0.95,\mu=1/4,\nu=1/15$ and and the return cost is normalized at $r=1$.}
    \label{fig:cost_exp_quad}
\end{figure}
We begin by examining the performance of the fluid policy for different cost parameters. Figure \ref{fig:cost_exp_quad} presents the relative reduction in expected cost under the fluid policy compared to the equilibrium (top row) and simple (bottom row) policies for the quadratic intervention cost function and under varying maximum intervention cost $M$ and holding cost $h$. We observe that when the intervention is relatively expensive, i.e., comparable to the return cost, and the holding cost is moderate, i.e., $h=0.5$, the fluid policy outperforms the benchmarks with respect to both the long-run average and finite-horizon costs. The cost savings are quite significant, e.g., for $h=0.5$ and $M=1$ we observe a 21.0\% reduction in the expected long-run average cost compared to the equilibrium (no intervention) policy and a 25.4\% reduction compared to the simple policy. If the holding cost is very small and the intervention is expensive, the equilibrium policy could lead to the same or slightly lower cost. If the intervention is cheap and the holding cost is relatively much larger, the simple policy can outperform the fluid policy. The reason is that the fluid policy (which ignores stochastic fluctuations) could be overly optimistic about congestion costs in particular in the long-run. Nevertheless, the fluid policy never performs worse than both benchmarks simultaneously (i.e., for the same set of parameters). In addition, we observe that the absolute differences are very small in these cases. In contrast, when the fluid policy outperforms the benchmarks, both absolute and relative savings are significant; see Appendix \ref{ap:supnumerics} for details. 
 

\begin{figure}[]\label{fig:rate_exp}
    \centering
    \begin{subfigure}{0.32\textwidth}\centering
    \includegraphics[width=\linewidth]{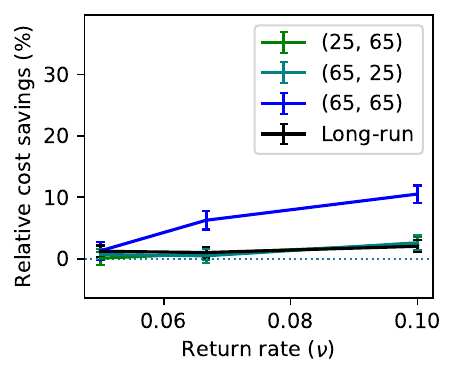}
    \caption{$\lambda=9.0$}
    \end{subfigure}
    \begin{subfigure}{0.32\textwidth}\centering
    \includegraphics[width=\linewidth]{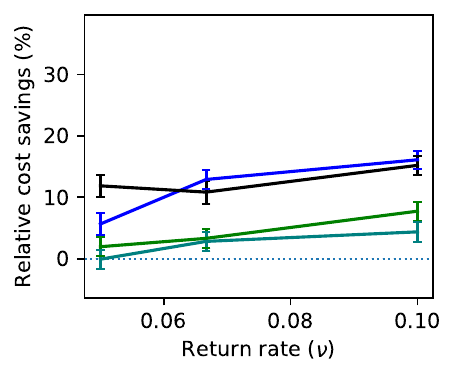}
    \caption{$\lambda=9.5$}
    \end{subfigure}
    \begin{subfigure}{0.32\textwidth}\centering
    \includegraphics[width=\linewidth]{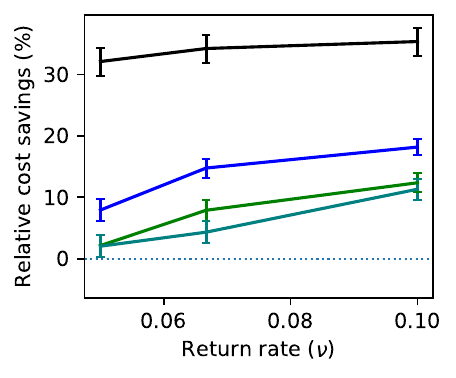}
    \caption{$\lambda=9.8$}
    \end{subfigure}
    \begin{subfigure}{0.32\textwidth}\centering
    \includegraphics[width=\linewidth]{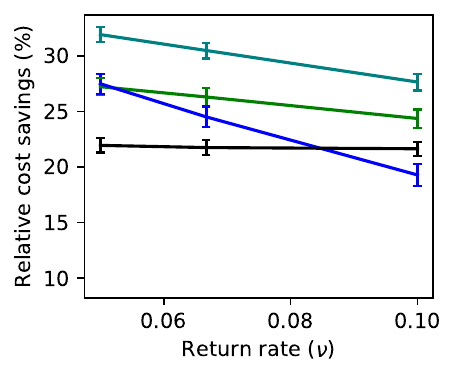}
    \caption{$\lambda=9.0$}
    \end{subfigure}
    \begin{subfigure}{0.32\textwidth}\centering
    \includegraphics[width=\linewidth]{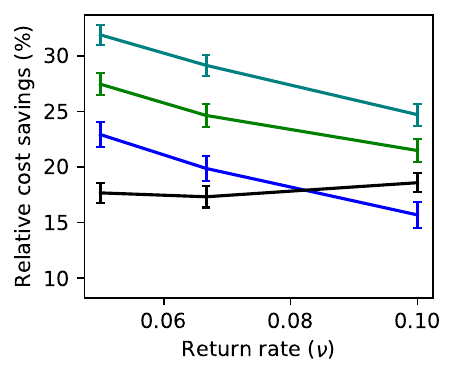}
    \caption{$\lambda=9.5$}
    \end{subfigure}
    \begin{subfigure}{0.32\textwidth}\centering
    \includegraphics[width=\linewidth]{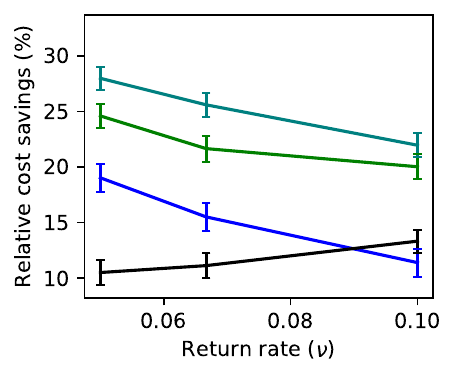}
    \caption{$\lambda=9.8$}
    \end{subfigure}
    \caption{Relative reduction in expected cost for both finite-horizon and long-run experiments under the fluid policy with respect to the equilibrium policy (top row) and simple policy (bottom row) under the quadratic intervention cost and for different values of arrival $\lambda$ and return $\nu$ rates. Other cost and system parameters are fixed at $N=50,\mu=1/4, r=1, M=0.5, h=1/4, C(p) = 50(0.2-p)^2$.}
    \label{fig:rate_exp_quad}
\end{figure}

Next, we examine the performance of the fluid policy for different system loads and return rates. Figure \ref{fig:rate_exp_quad} presents the relative reduction in expected cost under the fluid policy compared to the equilibrium (top row) and simple (bottom row) policies for the quadratic intervention cost function and under varying arrival rates $\lambda$ and return rates $\nu$, while fixing the cost parameters at their nominal values, i.e., $h=1/4, M=0.5$ and $r=1$. We first discuss the impact of return rate. We observe that the return rate can have a significant effect on the transient cost savings, whereas the long-run average savings are less sensitive. As the return rate increases, the value of adjusting the intervention in response to congestion increases and hence the cost savings compared to the equilibrium policy increase. At the same time, the cost savings compared to the simple policy decrease but remain positive. Next, we consider the impact of the arrival rate. As the arrival rate increases, the system load and hence the congestion increases, making it more valuable to increase the intensity of the intervention in response to congestion. Therefore, we observe a significant increase in cost savings compared to the equilibrium policy both for the long-run and transient experiments. As expected, when compared to the simple policy, the cost savings decrease with higher arrival rates but remain positive.  

We conduct the same experiments under the linear intervention cost function. Overall, the observations remain similar but the cost savings compared to the equilibrium (no intervention) policy can be higher when positive. The detailed results can be found in Appendix \ref{ap:supnumerics}.

The above numerical results lead to \emph{two key observations}: First, we observe significant cost savings compared to the equilibrium policy which is agnostic to congestion, demonstrating that accounting for the congestion cost can make it optimal to intervene in cases where otherwise it may not seem economical to do so. Under none of the simulated parameter regimes did the congestion-agnostic equilibrium policy $p_\infty$ equal the lowest value $p_u$. The additional intervention applied by the fluid policy became economical only due to the consideration of holding costs, especially when the starting state was far from the equilibrium. Moreover, when the holding cost was moderate or high, these savings were seen even in the long-run. From time to time, stochastic fluctuations bring the system away from the equilibrium point and into $\cC$, so the fluid policy's ability to efficiently steer the system back toward the equilibrium remains valuable.

Second, we find that in a relevant parameter regime to our hospital readmission application, namely costly interventions and moderate holding costs (relative to cost of return), dynamically adjusting the intervention level based on the observed congestion can lead to significant cost savings. 


\subsection{Time-Varying Arrivals}\label{subsec:varyingnumerics}
In this section, we examine the robustness of the observations made in the previous section for systems with time-varying arrivals. We use the same sinusoidal arrival rate as in \eqref{eq:sinarrival} in our simulation experiments and evaluate the performance of the fluid policy (which uses the average arrival rate $\bar{\lambda}$) against the two benchmarks for varying periods and amplitudes of the arrival rate. 

\begin{figure}[]
    \centering
    \begin{subfigure}{0.32\textwidth}\centering
    \includegraphics[width=\linewidth]{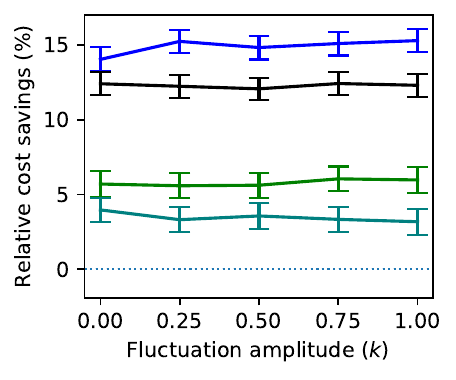}
    \caption{$f=1$}
    \end{subfigure}
    \begin{subfigure}{0.32\textwidth}\centering
    \includegraphics[width=\linewidth]{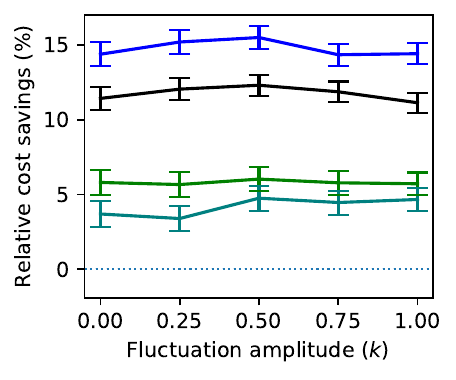}
    \caption{$f=7$}
    \end{subfigure}
    \begin{subfigure}{0.32\textwidth}\centering
    \includegraphics[width=\linewidth]{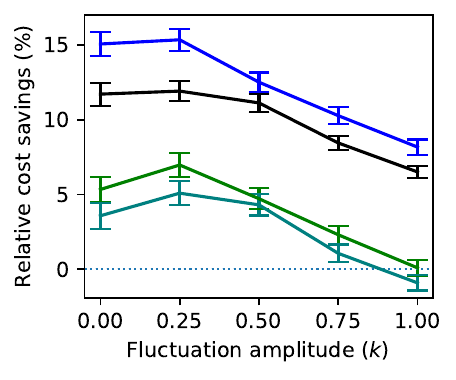}
    \caption{$f=28$}
    \end{subfigure}
    \begin{subfigure}{0.32\textwidth}\centering
    \includegraphics[width=\linewidth]{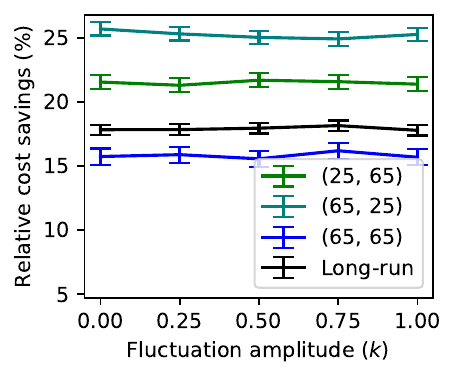}
    \caption{$f=1$}
    \end{subfigure}
    \begin{subfigure}{0.32\textwidth}\centering
    \includegraphics[width=\linewidth]{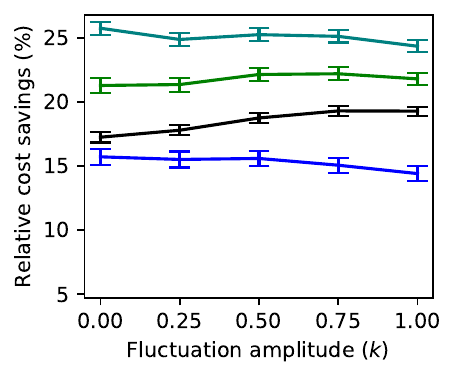}
    \caption{$f=7$}
    \end{subfigure}
    \begin{subfigure}{0.32\textwidth}\centering
    \includegraphics[width=\linewidth]{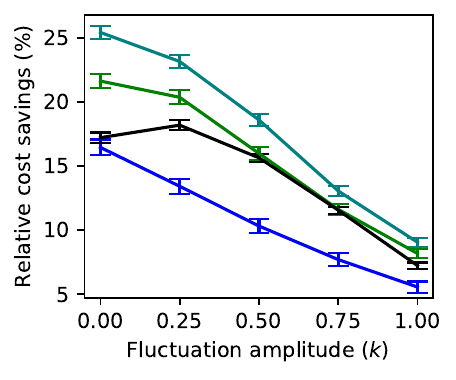}
    \caption{$f=28$}
    \end{subfigure}
    \caption{Relative reduction in expected cost under the fluid policy with respect to the equilibrium policy (top row) and simple policy (bottom row) under different time-varying arrival rate functions. System and cost parameters are fixed at $N=50,\bar\lambda=0.95,\mu=1/4,\nu=1/15, r=1, h=1/4, C(p)=50(0.2-p)^2$.}
    \label{fig:varying_exp_quadratic}
\end{figure}

We observe that as long as the scale of service and time-to-return times are large relative to the scale of time-variation, the fluid policy (which ignores time-variation) still performs well even when the amplitude of the variation is high. We illustrate this by considering the quadratic cost function and baseline system parameters where  the average service time and time-to-return are $1/\mu=4$ and $1/\nu=15$ days, respectively. Figure \ref{fig:varying_exp_quadratic} presents the relative reduction in expected cost achieved under the fluid policy in contrast to the two benchmark policies and for different time-variation scales and amplitudes. In the cases of intra-day and intra-week variation, the cost savings remain fairly consistent even as we increase the amplitude of the fluctuation. In these cases, the fluid policy consistently outperforms the two benchmark policies. It is only when the period is lengthened to 28 where we observe a reduction in the cost savings for larger amplitude values. We observe consistent results for other system parameters and cost functions; see Appendix \ref{ap:supnumerics} for the same experiments under the linear cost function.

\subsection{Case Study: Hospital Readmission}\label{sec:case}
In this section, we conduct a numerical study using parameters that are calibrated based on a real hospital setting and post-discharge readmission intervention plan reported in the literature. In doing so, we further relax some of our modeling assumptions to examine the robustness of our observations.

\textbf{Cost and intervention parameters}. Our intervention and cost parameters are modeled after a transitional care program for heart failure patients studied in \cite{stauffer2011}. The intervention involves multiple home visits after discharge by an Advanced Practice Nurse (APN) who is also available for telephone calls. The cost of intervention per patient is $M=1110$ and the cost of a readmission is $r=5000$. Estimated readmission probabilities with and without the intervention are $p_l=0.081$ and $p_u=0.141$, respectively, i.e., the intervention has a 42\% efficacy. Since the intervention does not have multiple levels, we construct the intervention cost function $C(\cdot)$ by interpolating the cost of the two extremes on the interval $[p_l, p_u]$. However, we know by Proposition \ref{prop:linear_bangbang} that the resulting fluid policy will be bang-bang, always choosing either $p_u$ (no intervention) or $p_l$ (intervention). We vary the holding cost $h \in \{500, 1000, 1500, 2000, 2500, 3000\}$ in our experiments.

From Corollary \ref{cor:lin_eq}, we know that the equilibrium policy under a linear intervention cost is determined by comparing the expected total cost savings associated with intervention against the intervention cost. Under the costs above, the expected lifetime savings of $r(p_u-p_l)/(1-p_u) = 349.2$ are $0.31$ times the intervention cost $M=1110$. This is somewhat higher than the corresponding value of $0.25$ under the baseline parameters of the numerical experiments in Section \ref{subsec:stat_exp}. We therefore have a higher incentive to intervene, and expect the fluid policy to be closer to the simple policy.

\textbf{System parameters}. We calibrate the parameters of our model based on those of the Internal Ward A of Ramban Hospital (referred to as Ward A hereafter) described in \cite{armony2015on}. We set the number of servers to the number of beds of Ward A, i.e., $N=45$ and use the Log-Normal distribution fitted to Length-of-Stay (LOS) data, which has log-mean $1.38$, log-standard deviation $0.83$, and mean $1/\mu=5.6$. The arrival rate follows a non-stationary Poisson process with rate, 
\begin{equation}
    \lambda(t) = \begin{cases}6.14(1-0.8 \sin(2\pi t)), & t\Mod{7} \leq 5 , \\ 5.32(1-0.8 \sin(2\pi t)), & t \Mod{7} > 5.  \end{cases}
\end{equation}
The overall average arrival rate of $\bar{\lambda}=5.91$ is based on Ward A, which receives $206.3$ patients per month. We adjust this value by a factor of $1-p_u$, removing readmissions to obtain the net arrival rate of new patients for our case study. In addition, we add hourly fluctuations with relative amplitude of $0.8$, and reduce weekend arrival rates compared to weekdays. The return times follow a distribution with density $0.057 \exp(-x/25), x \in (0,30)$ with mean $\nu=1/12.07$, equivalent to an exponential distribution with mean $25$, conditional on being less than 30. These parameters yield an average load of $\bar\lambda/\mu(1-p_h) = 38.6$ in the absence of intervention, corresponding to a $\rho=0.86$ utilization.   

\begin{figure}[]
    \centering
    \includegraphics[width=0.4\textwidth]{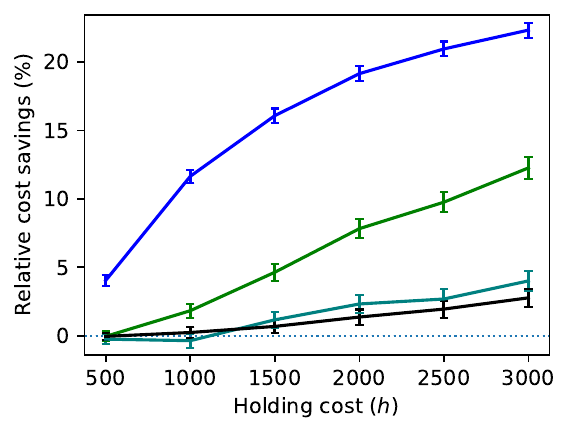}
    \includegraphics[width=0.4\textwidth]{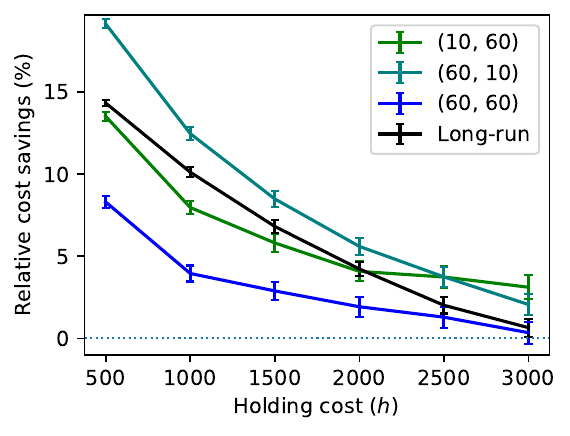}
    \caption{Relative reduction in expected cost in the case study as measured against the equilibrium policy (left) and the simple policy (right) for different holding cost values.}
    \label{fig:case_results}
\end{figure}

\textbf{Results.} Figure \ref{fig:case_results} presents the relative reduction in expected cost savings under the fluid policy compared to the two benchmarks. As in Section \ref{sec:numerics} we consider both the long-run average cost, and transient cost over a 90-day horizon starting with three different initial states. Despite the relaxations to the model assumptions and lower utilization, the fluid policy performs the same or better than both benchmark policies. In addition, our qualitative observations in Section \ref{subsec:stat_exp} continue to hold.

As reported in \cite{stauffer2011} the intervention is not economical when only considering the costs of readmission versus intervention. However, even for relatively small holding costs, the intervention could be economical for a system starting from a state far from the equilibrium. For larger holding cost values, we observe considerable savings through the intervention specially over a finite horizon. The savings are also considerably larger compared to the simple policy, as long as the holding cost is not too high.

\begin{figure}[]
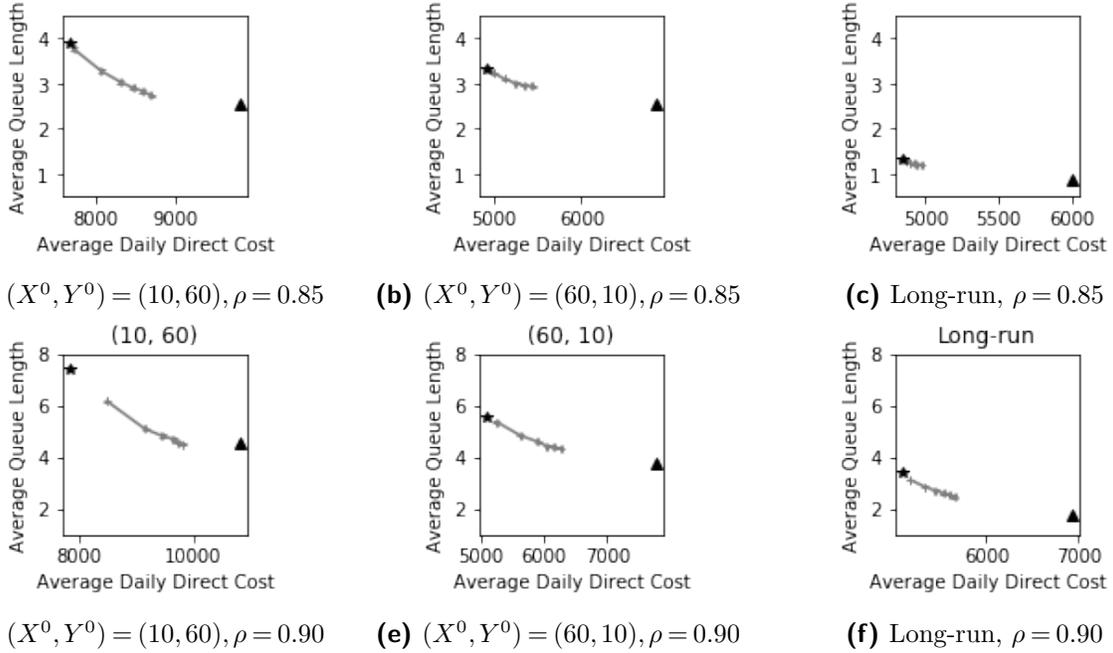

\begin{subfigure}{0.33\textwidth}
    \centering
    \includegraphics[width=0.7\textwidth]{case_figs/Tradeoff-1-1.pdf}
    \caption{$(X^0, Y^0) = (10, 60), \rho=0.85$}
\end{subfigure}
\begin{subfigure}{0.33\textwidth}
    \centering
    \includegraphics[width=0.7\textwidth]{case_figs/Tradeoff-1-2.pdf}
    \caption{$(X^0, Y^0) = (60, 10), \rho=0.85$}
\end{subfigure}
\begin{subfigure}{0.33\textwidth}
    \centering
    \includegraphics[width=0.7\textwidth]{case_figs/Tradeoff-1-3.pdf}
    \caption{Long-run, $\rho=0.85$}
\end{subfigure}

\begin{subfigure}{0.33\textwidth}
    \centering
    \includegraphics[width=0.7\textwidth]{case_figs/Tradeoff-1-4.pdf}
    \caption{$(X^0, Y^0) = (10, 60), \rho=0.90$}
\end{subfigure}
\begin{subfigure}{0.33\textwidth}
    \centering
    \includegraphics[width=0.7\textwidth]{case_figs/Tradeoff-1-5.pdf}
    \caption{$(X^0, Y^0) = (60, 10), \rho=0.90$}
\end{subfigure}
\begin{subfigure}{0.33\textwidth}
    \centering
    \includegraphics[width=0.7\textwidth]{case_figs/Tradeoff-1-6.pdf}
    \caption{Long-run, $\rho=0.90$}
\end{subfigure}
\caption{Tradeoff between direct (readmission and intervention) costs and queue-length for long-run average and finite-horizon costs with different initial conditions. The star and triangle symbols respectively correspond to the equilibrium and simple policy. }\label{fig:qlength-case}
\end{figure}

Finally, we examine the tradeoff between direct costs of returns and intervention, versus the average queue-length under the fluid policy. In doing so, we treat the holding cost as a parameter and vary it in the same range as before to obtain different pairs of average queue-length and costs. In addition to the case with 86\% utilization, we also present the results for 90\% utilization. Figure \ref{fig:qlength-case} presents the tradeoff curves for the long-run average and two finite-horizon cases. Note that each point on the curves corresponds to a certain holding cost, and the end-points marked with a star and triangle correspond to the the equilibrium and simple policy, respectively. Increasing the holding cost leads to higher intervention costs, but lower return costs and queue-length. The tradeoff curves appear to be convex, that is the average queue-length can be reduced in return for a small increase in direct costs. We note that the smaller values of queue-length in the long-run average are due to the time-varying arrival rate, and the peak congestion over each period (and the corresponding reduction) is larger.  

\section{Discussion and Future Work}\label{sec:conc}
In this work, we propose and study a new control problem for queueing systems with returns. Our model captures the tradeoff between the cost of post-service interventions and the benefits through reduced readmissions including reduced congestion costs. We study associated fluid control problems and characterize the structure of the long-run average and transient optimal policies under a general convex intervention cost. Through our analysis of the fluid control problems, we obtain approximately optimal policies for the stochastic system and gain insights into the structure of congestion-aware intervention policies. In particular, under a piecewise linear cost function, a policy divides the state-space into as many regions as there are pieces in the cost function, and prescribes a different level of intervention intensity (probability of return) in each region corresponding to the breakpoints of the cost function. The level of intervention is determined based on an interpretable measure of congestion risk comprised of the number of Needy as well as the discounted number of Content customers. In the context of our motivating hospital readmission application, the structure motivates the design of practical surge protocols that economically use post-discharge interventions to recover from highly congested states. In the following, we discuss some future directions of our work along both practical and technical dimensions. 

\subsection{Model Extensions}
We assume that the intensity of intervention for each customer is determined at the time of service completion, and that receiving an intervention only impacts the return probability. In the context of our motivating post-discharge readmission prevention programs, there are several examples that fit this simple structure. At the same time, interventions and their impacts could be more complex. For instance, the intervention could also increase the return time, e.g., for geriatric patients \citep{morkisch2020components}. In this case, the same framework developed here can be used to analyze the optimal allocation of interventions; see Section \ref{ap:timetoreturn} for preliminary analysis. Further, in some cases patients are monitored over a certain time-horizon after discharge and the intervention level can be adjusted at these epochs. This is studied in \cite{chen2022data} for a simple intervention with binary states and using an RL framework in the general case, but without considering congestion costs. Examining the impact of hospital congestion on the optimal timing and intensity of interventions would be an interesting direction for future work. 

Our structural results provide insights on the impact of both Content and Needy states on the optimal level of intervention. In practice, the Content state is often not directly observable. However, the value of the Content state can be estimated through predictive models and then used to compute the intervention level. When the intervention cost is piecewise, the intervention level remains relatively insensitive to the precise value of the Content state. As such, small prediction errors are likely to not impact the level of intervention. We illustrate this for our case study and using a naive estimation of the Content state in Section \ref{ap:noisy}. Moreover, we have utilized a parsimonious model to understand the interaction between congestion costs and costly post-service interventions and gain insights on optimal policy structure. An interesting area for future work would be the integration of our policies with personalized predictions of readmission risk to develop practical policies. In this context, one can expand the state-space to keep track of the Content and Needy states for different types of patients and develop congestion risk scores that depend on the multi-dimensional vector. One can then leverage the structure established here and tune the control parameters using, e.g., an approximate dynamic programming approach and a high-fidelity simulation model. See \cite{bravo2023interpretable} for another example of congestion risk scores.

\subsection{Technical Extensions}
We used a heuristic derivation of the fluid model in our analysis. When the policy is Lipschitz, the convergence to the fluid model can be justified using a Strong Functional Law of Large Number (FSLL) established in \cite{mandelbaum1998strong}. In the case of switching policies that we show are optimal under piecewise linear costs, establishing convergence requires additional technical developments. The uniqueness of solution to the system of ODEs shown here is a first step to this end. We suspect that convergence holds (see Section \ref{ap:weak} for numerical evidence) and can be proved using the framework of \cite{pang2007martingale} when the the fluid model has a unique solution; see, e.g., \cite{perry2013fluid} for a proof of weak convergence in the presence of discontinuous policies. When Assumption \ref{assump:stability} is relaxed and the system requires some level of intervention to stabilize, the analysis could be more complicated as the system may exhibit more complex dynamics, e.g., chattering around the intervention boundary, and may not have a unique solution. We leave an analysis of the stability of the system with controllable return probabilities for future work.

Our characterization of the optimal transient policy relies on the notion of bias-optimality for fluid control problems. The proposed cost criteria can be utilized to study transient policies for other multiserver queueing control problems where the total cost of the system does not remain finite over an infinite horizon. Compared to using state-dependent constraints, this approach allows for an easier application of Pontryagin's Minimum Principle. We believe this approach is applicable to other complex queueing control problems. In addition to applying it to other problems, establishing a formal connection between the bias-optimal value function and that of the corresponding MDP, and establishing asymptotic optimality of the resulting policies would be an interesting direction for future work.
\ACKNOWLEDGMENT{The authors are grateful to the area editor Ramandeep Randhawa, an anonymous associate editor, and two anonymous referees for their valuable comments and suggestions that helped improve the paper. Vahid Sarhangian thanks Ohad Perry for helpful discussions related to discontinuous control.}

\bibliographystyle{abbrvnat}
\bibliography{bibliography}

\ECSwitch
\renewcommand*{\theHsection}{\thesection}
\renewcommand*{\theHsubsection}{\thesubsection}

\ECHead{Proofs and Supplementary Analysis}

\section{Proofs of Lemmas and Propositions}

\proof{Proof of Proposition \ref{prop:corner}.}

$(i)$ To prove the first claim, we bound the time at which the fluid state reaches region $\cA$ starting from a given initial state. Consider $\hat{y}$ whose dynamics are governed by:
\begin{equation}
    \dot{\hat y}(t) = \mu N p_u -\nu \hat{y}(t),
\end{equation}
for $t\geq0$ and starting from $\hat y(0) = y^0$. Since $\dot{y}(t) - \dot{\hat y}(t) \leq -\nu(y(t)-\hat{y}(t))$ and $y(0) - \hat y(0) = 0$, Gronwall's inequality for absolutely continuous functions (see, e.g., \citealt{marx2022path}) yields $y(t) - \hat y(t) \leq 0$ for all $t \geq 0$. Therefore, $\hat{y}$ is an upper bound for $y$.

Next, note that since $\lim_{t\to\infty} \hat y(t) = (\mu N p_u)/\nu < (\mu N - \lambda)/\nu$, there exists a time $t_1 < \infty$, such that $y(t) \leq \hat y(t) \leq (\mu N - \lambda)/\nu$ for all $t \geq t_1$ and under all policies $p(\cdot) \in \cP$.

Define $t_\cA := t_1 + \frac{x_0+y_0+\lambda t_1}{\mu N (1-p_u)-\lambda} + 1$. Under a given policy $p \in \cP$, if $x(t_1) \leq N$, then $(x(t_1), y(t_1)) \in \cA$. If $x(t_1) > N$, note that $x(t_1)+y(t_1)\leq x^0+y^0+\lambda t_1$, since the total number of customers (both Needy and Content) always grows at a rate of at most $\lambda$. As long as $x(t) > N$, all servers are busy and the total number of customers in the system decreases at a rate of at least $\mu N (1-p_u) - \lambda $ (which is positive by Assumption $\ref{assump:stability}$). The queue therefore must empty at some time before $t_\cA$, else there would be a negative number of customers remaining in the system, an impossibility.

Therefore under any policy $p(\cdot) \in \cP$, there is a $t \leq t_\cA$ such that $(x(t), y(t)) \in \cA$. Applying the result of part (ii) then gives us $(x(t_\cA), y(t_\cA)) \in \cA$.


$(ii)$ Since the trajectory $(x(t),y(t))$ is (almost everywhere) differentiable, it suffices to show that its gradient points inward along the boundaries of $\cA$ under any admissible policy. Along the vertical boundary line $x = N$, the gradient points leftward:
    \begin{align}
        \dot{x}(t) &= \lambda + \nu y(t) - \mu (x(t)\wedge N)\\
         &= \lambda + \nu y(t) - \mu N \\
         & \leq \lambda + (\mu N - \lambda) - \mu N = 0 \label{eq:xN},
    \end{align}
where the inequality in \eqref{eq:xN} is strict for $y(t) < (\mu N - \lambda)/\nu$. Along the horizontal boundary line $y = (\mu N - \lambda)/\nu$, the gradient points downward:
    \begin{align}
    \dot y(t) &= -\nu y(t) + \mu p(t) (x(t) \wedge N) \\
    &= -(\mu N - \lambda) + \mu p(t) (x(t) \wedge N) \\
    &\leq \lambda - \mu N + \mu p_u N \label{eq:ymuN}\\
    &= \lambda - \mu N(1-p_u) < 0.
\end{align}
Finally, at the top-right corner $(N, (\mu N - \lambda)/\nu)$ the gradient points straight downwards with $\dot {x}(t)= 0$ and $\dot{y}(t) < 0$. Note that since $\dot y < 0$ and $\dot x = 0$, we can see that $\dot x = \lambda - \nu y - \mu(x \wedge N)$ is decreasing. The trajectory will thus curve inwards and the state will remain in $\cA$.
 \Halmos\endproof

\proof{Proof of Proposition \ref{equilibrium_policy}.}
Recall that $C'_-(\cdot)$ and $C'_+(\cdot)$ denote the left and right derivatives of $C(p)$. While $C(\cdot)$ is not guaranteed to be differentiable everywhere, we know by convexity that $C'_-(\cdot)$ and $C'_+(\cdot)$ exist and are non-decreasing with $C'_-(\cdot) \leq C'_+(\cdot)$ on all of $[p_l, p_u]$. We have:
\begin{equation} 
    J'_- (p) = \frac{(1-p)C'_-(p) + C(p) + r}{1-p}, \qquad J'_+(p) = \frac{(1-p)C'_+(p) + C(p) + r}{1-p}. \label{eq:J'}
\end{equation}
$J(\cdot)$ is maximized at points where $C'_-(p) \leq 0 \leq C'_+(p)$ (which we will call critical points with a slight abuse of terminology). Denote the numerators of \eqref{eq:J'} by $q_{-}(p) $ and $q_+(p)$, respectively. The denominator is positive on $[p_l, p_u]$, so the signs of $q_{-}$ and $q_+$ determine the signs of $J'_-$ and $J'_+$, respectively.

Let $p_1 < p_2$. We use the convexity of $C(\cdot)$ to show that $q_+(p_1) \leq q_+(p_2)$. We know that $C(p_1) + (p_2 - p_1) C'_+(p_1) \leq C(p_2)$ and that $C(p_2) - (p_2 - p_1) C'_+(p_2) \leq C(p_1)$. It follows that, $$q(p_1) \leq \frac{(1-p_1)C(p_2) - (1-p_2)C(p_1)}{p_2 - p_1} \leq q(p_2).$$ Thus, $q_+$ is increasing, and hence the sign of $J'_+(p)$ is non-decreasing. The same argument applied to $q_-$ implies that the sign of $J'_-(p)$ is also non-decreasing. Therefore, $J(p)$ is either U-shaped or monotone, and its critical points (if any) form a closed interval. Since $q_-$ and $q_+$ are non-decreasing, they are minimized at $p_l$ and maximized at $p_u$. If $q_-(p_l) \geq 0$, then $J'_-(p)$ is non-negative (and similarly with $J'_+(p)$) and thus $J(p)$ is minimized at $p_l$. Similarly, if $q_+(p_u) \leq 0$, then $J'_-(p)$ is non-positive (and similarly with $J'_+$) and $J(p)$ is minimized at $p_u$. Otherwise, $q_-(p_l) < 0 < q_+(p_u)$, implying the existence of a critical point of $J(p)$ in the interior of $[p_l, p_u]$ (where $q_-(p) \leq 0 \leq q_+(p)$). The proof is complete.
\Halmos
\endproof

%

\proof{Proof of Proposition \ref{tailvalue}.}
Since $(x^0,y^0)\in\cA$, by Proposition \ref{prop:corner} the queue will remain empty for all $t \geq 0$. We verify the correctness of the value function using the Hamilton-Jacobi-Bellman (HJB) equations. Here we suppress the reference to the specific horizon length $T$ since the proof does not depend on its specific value.
\begin{align}
    & \dot V(x,y,t) + \min_{p} \left\{ \dot x(t) V_x(x,y,t) + \dot y(t) V_y(x,y,t) + r\nu y + C(p)\mu x - J_\infty \right\}  \nonumber\\
    &\quad= 0 + \min_{p\in\cP} \left\{ (\lambda + \nu y - \mu x) \frac{ r p_\infty + C(p_\infty)}{1-p_\infty} + (-\nu y + \mu x p)\frac{r + C(p_\infty)}{1-p_\infty} + r\nu y + C(p)\mu x - J_\infty \right\} \nonumber\\
    &\quad= \mu x \left( - \frac{ r p_\infty + C(p_\infty)}{1-p_\infty} + \min_{p\in\cP} \left\{ \frac{r + C(p_\infty)}{1-p_\infty} p + C(p) \right\} \right) \label{hjb_min} \\
    &\quad= \mu x \left( - \frac{ r p_\infty + C(p_\infty)}{1-p_\infty} + \frac{r + C(p_\infty)}{1-p_\infty} p_\infty + C(p_\infty) \right) \nonumber\\
    &\quad= 0.\nonumber
\end{align}
It is easy to verify that the quantity minimized in \eqref{hjb_min} is convex with critical point $p_\infty$, and hence minimized at $p_\infty$. Additionally, we have $V(x,y,T) = \Psi(x,y)$ by definition. The HJB equations are thus satisfied. It follows that $V(x,y,t)$ is the value function for $(x,y) \in \cA$, and the optimal policy is given by $p(t) = p_\infty$ for all $t \geq 0$.
\Halmos\endproof

\proof{Proof of Proposition \ref{finiteintegral}.} We prove the proposition by obtaining finite lower and upper bounds for the value function. By Proposition \ref{prop:corner}, there is a bound $t_\cA$ such that $t_\cA \in \cA$ under any policy $p(t)$, and that $(x(t), y(t)) \in \cA$ for all $t \geq t_\cA$ as well.

We will bound the integral up to a large finite time $T \geq t_\cA$. We can split the objective into an integral up to time $t_\cA$ and another integral from $t_\cA$ onwards, and obtain a lower bound by ignoring the costs incurred before $t_\cA$:
\begin{align}
&\min_{p(\cdot)} \int_0^T h(x(t)-N)^+ + r\nu y(t) + C(p(t))\mu(x(t) \wedge N) - J_\infty \,dt \\
&= \min_{p(\cdot)} \bigg\{ \int_{t_\cA}^T h(x(t)-N)^+ + r\nu y(t) + C(p(t))\mu(x(t) \wedge N) - J_\infty \,dt \\ &\qquad + \int_0^{t_\cA} h(x(t)-N)^+ + r\nu y(t) + C(p(t))\mu(x(t) \wedge N) - J_\infty \,dt \bigg\} \\
&\geq \min_{p(\cdot)} \left\{ \int_{t_\cA}^T h(x(t)-N)^+ + r\nu y(t) + C(p(t))\mu(x(t) \wedge N) - J_\infty \,dt \right\} -J_\infty t_\cA
\end{align}

In addition, because $T \geq t_\cA$, we are guaranteed that $(x(T), y(T)) \in \cA$ and therefore that $\Psi(x(T), y(T)) \leq \Psi(N, (\mu N - \lambda)/\nu)$. Adding a term $\Psi(x(T), y(T))$ to the objective yields the value function from Proposition \ref{tailvalue}, and thus increases the minimum by at most $\Psi(N, (\mu N - \lambda)/\nu)$. This yields,
\begin{align}
& \min_{p(\cdot)} \left\{ \int_{t_\cA}^T h(x(t)-N)^+ + r\nu y(t) + C(p(t))\mu(x(t) \wedge N) - J_\infty \,dt \right\} -J_\infty t_\cA \\
&\geq \min_{p(\cdot)} \left\{ \int_{t_\cA}^T h(x(t)-N)^+ + r\nu y(t) + C(p(t))\mu(x(t) \wedge N) - J_\infty \,dt + \Psi(x(T), y(T)) \right\} \\ &\qquad - \Psi\left(N,\frac{\mu N - \lambda}\nu\right)  -J_\infty t_\cA \\
&= \Psi(x(t_\cA), y(t_\cA)) - \Psi\left(N,\frac{\mu N - \lambda}\nu\right) -J_\infty t_\cA > -\infty.
\end{align}
We thus have a finite lower bound independent of $T$. Taking $T\to\infty$ makes this a lower bound on the infinite-horizon problem.

We obtain an upper bound in a similar fashion. Since the total number of customers (both Needy and Content) grows at a rate of at most $\lambda$, we know that $K := x^0 + y^0 + \lambda t_\cA < \infty$ is an upper bound on $x(t) + y(t)$ (as well as $x(t)$ and $y(t)$ individually) for all $t \leq t_\cA$. We can thus place a finite bound on the costs up to time $t_\cA$:
\begin{equation}
    \int_0^{t_\cA} h(x(t)-N)^+ + r\nu y(t) + C(p(t))\mu(x(t) \wedge N) - J_\infty \,dt \leq (hK + r\nu K + C(p_l)N - J_\infty)t_\cA.
\end{equation}
Since $\Psi(x(T), y(T)) \geq 0$, adding the terminal cost to the objective yields an upper bound. We thus have the following inequality:
\begin{align}
&\min_{p(\cdot)} \int_0^T h(x(t)-N)^+ + r\nu y(t) + C(p(t))\mu(x(t) \wedge N) - J_\infty \,dt \\
&\leq \min_{p(\cdot)} \left\{ \int_{t_\cA}^T h(x(t)-N)^+ + r\nu y(t) + C(p(t))\mu(x(t) \wedge N) - J_\infty \,dt \right\} \\
& \qquad + (hK + r\nu K + C(p_l)N - J_\infty)t_\cA \\
&\leq \min_{p(\cdot)} \left\{ \int_{t_\cA}^T h(x(t)-N)^+ + r\nu y(t) + C(p(t))\mu(x(t) \wedge N) - J_\infty \,dt + \Psi(x(T), y(T)) \right\} \\
& \qquad + (hK + r\nu K + C(p_l)N - J_\infty)t_\cA \\
& = \Psi(x(t_\cA), y(t_\cA)) + (hK + r\nu K + C(p_l)N - J_\infty)t_\cA < \infty.
\end{align}
We therefore have a finite upper bound independent of $T$, and taking $T\to\infty$ makes this an upper bound on the infinite-horizon problem.
\halmos \endproof



\proof{Proof of Proposition \ref{prop:Jinf}.}
It is clear that $J_\infty$ is an upper bound for the minimum long-run average cost, since it is attained by the equilibrium policy. Conversely, suppose for a contradiction that there exists a better policy $p(t)$ with long-run average cost of $J_\infty - \eps$ for some $\eps > 0$:
\begin{equation}
    \limsup_{T\to\infty} \frac1T \int_0^T h(x-N)^+ + r\nu y + \mu C(p(t)) (x \wedge N) \, dt = J_\infty - \eps.
\end{equation}
By definition, there exists a $T_0 > 0$ such that for all $T \geq T_0$:
\begin{align}
\frac1T \int_0^T h(x(t)-N)^+ + r\nu y(t) + \mu C(p(t)) (x(t) \wedge N) \, dt < J_\infty - \frac\eps2, 
\end{align}
implying that, 
\begin{equation}
    \int_0^T h(x(t)-N)^+ + r\nu y(t) + \mu C(p(t)) (x(t) \wedge N) - J_\infty \, dt < - \frac{\eps T}2.
\end{equation}
Taking $T\to\infty$ then yields divergence to $-\infty$.
\halmos\endproof

\proof{Proof of Proposition \ref{prop:linear_bangbang}.}
The claim follows from a simple application of the Minimum Principle. Under a linear intervention cost $C(p) = M \frac{p_u-p}{p_u-p_l}$, we have:
\begin{align}
    \frac{\partial}{\partial p} H(x(t), y(t), p, t) &= \mu (x(t) \wedge N)(C'(p) + \gamma_2(t) p) \\
    &= \mu(x(t) \wedge N) \left( -\frac{M}{p_u - p_l} + \gamma_2(t) \right).
\end{align}
Thus the Hamiltonian is increasing with respect to $p$ if $\gamma_2(t) > \frac{M}{p_u-p_l}$ and decreasing otherwise. Since $p^*(t)$ must minimize the Hamiltonian, it follows that we take $p^*(t) = p_u$ when $\gamma_2(t) > \frac{M}{p_u-p_l}$ and $p^*(t) = p_u$ otherwise. If $\gamma_2(t) = \frac{M}{p_u-p_l}$, then we are indifferent to all values of $[p_l, p_u]$, however, this can only occur momentarily and is hence inconsequential.
\halmos\endproof

\proof{Proof of Proposition \ref{policyformula}.}

The policy $p^*(t)$ is chosen to minimize the Hamiltonian function:
\begin{equation}h(x - N)^+ + r\nu y + C(p)\mu (x \wedge N) - J_\infty + (\lambda + \nu y - \mu(x \wedge N))\gamma_1(t) + (-\nu y + \mu p(x \wedge N))\gamma_2(t).
\end{equation}
Eliding the terms that do not depend on $p$, we are left with:
\begin{equation}
    \mu (x \wedge N) ( C(p) + \gamma_2(t) ).
\end{equation} 
It thus suffices to minimize the expression $\phi_t(p) = C(p) + \gamma_2(t) p$, which exactly represents the cost trade-off for a Needy patient finishing service at time $t$. We incur an immediate intervention cost of $C(p)$, and the patient has a probability $p$ of moving to the Content state (or, in the fluid context, the amount of fluid in the Content state increases by $p$ for every unit of fluid finishing service). Since the costate $\gamma_2$ represents the partial derivative of the value function in the $y$ direction (i.e. the marginal cost of one more Content patient), the state transition has an implicit cost of $\gamma_2(t) p$ which represents the additional future cost associated with the decision.

Because $C(\cdot)$ is continuous and convex, $\phi_t$ is continuous and convex as well. Similarly, since left and right derivatives of $C$ are non-decreasing, the left and right derivatives of $\phi_t$ (i.e., $\phi_{t,-}'(p) = C'_-(p) + \gamma_2(t)$ and $\phi_{t,+}'(p) = C'_+(p) + \gamma_2(t)$) are as well. Moreover, they are continuous and equal to each other except at the finitely many discontinuities of $C'(\cdot)$ in the interior of $[p_l, p_u]$, which must be upward jumps.

\begin{itemize}
    \item If $\gamma_2(t) < -C'_-(p_u)$, then $C'_-(p) + \gamma_2(t) \leq C'_+(p) + \gamma_2(t) < 0$ for all $p \in (p_l, p_u)$ (and $C'_+(p_l) + \gamma_2(t) < 0$) and hence $\phi_t(p)$ is minimized at $p_u$.
    \item If $\gamma_2(t) > -C_+'(p_l)$, then $C'_+(p) + \gamma_2(t) \geq C'_-(p) + \gamma_2(t) > 0$ for all $p \in (p_l, p_u)$ (and $C'_-(p_u) + \gamma_2(t) > 0$) and hence $\phi_t(p)$ is minimized at $p_l$.
    \item Otherwise, $\phi_t(p)$ is minimized at its critical points in $[p_l, p_u]$, i.e., those that satisfy $C'_- (q) \leq -\gamma_2(t) \leq C'_+(q)$.
\end{itemize}
Since $\gamma_2(t)$ is decreasing in $t$, we know that $C(p^*(t))$ must be increasing. Recalling that $C'(p)$ is non-increasing in $p$, we see that $p^*(t)$ must be non-decreasing in time. In the case where $-C'(p_u) \leq \gamma_2(t) \leq -C'(p_l)$, it is possible for $\phi_t(p)$ to have multiple minimizers, in which case we are indifferent between them at that instant $t$. However, $\phi_t(\cdot)$ is convex (for a given $t$), so its minimizers form a closed interval. Since $\gamma_2(t)$ is strictly decreasing, an interior value $p \in (p_l, p_u)$ cannot be optimal for multiple $t$, and these closed intervals must be disjoint. It follows that only countably many of these intervals can have positive length, i.e., there are at most countably many $t$ for which $\phi_t$ has multiple minimizers, so the optimal policy $p^*(t)$ is therefore unique almost everywhere on the interval $[0, \tau)$. \halmos \endproof

\proof{Proof of Proposition \ref{prop:noaggression}.}
We show the contrapositive: if we never enter $\cC$, then we must have never intervened. If the state never enters $\cC$, solving the differential equations \eqref{eq:Gd1}-\eqref{eq:G2} yields the constant solution $\gamma_1(t) = \frac{C(p_\infty) + rp_\infty}{1-p_\infty}, \gamma_2(t) = \frac{C(p_\infty) + r}{1-p_\infty}$. The costates remain constant at their equilibrium values, hence implying a constant policy of $p_\infty$ for all $t > 0$. \halmos\endproof

\proof{Proof of Proposition \ref{prop:twoswitches}.} From Proposition \ref{prop:noaggression} we know that if an optimal policy ever intervenes, it must enter region $\mathcal{C}$. We also know that the trajectory must eventually reach $\mathcal{A}$. If we start from $\mathcal{N}$, then either the trajectory proceeds directly from $\mathcal{N}$ to $\mathcal{A}$ and hence never intervenes, or it must exit $\mathcal{N}$ into $\mathcal{C}$, after which point we have a complete characterization of the optimal policy. Let $t_1$ be the time at which the trajectory enters $\mathcal{C}$ from $\mathcal{N}$. We focus our attention on the period $t < t_1$.

We show that if an optimal trajectory ever switches from intervention to non-intervention during this period, then it can never start intervening again. Suppose there is a time $t_0$ at which an optimal policy stops intervening (i.e., $p^*(t) = p_l$ for $t \uparrow t_0$ and $p^*(t) = p_h$ for $t \downarrow t_0$.) We know that $\gamma_2(t_0) = \frac{M}{p_h-p_l} > \frac{r + M}{1-p_h}$, and that $\dot\gamma_2(t_0) < 0$. Since we are in the non-congested region, the costates are governed by the system $\dot\gamma_1(t) = \mu(-M-\gamma_1(t) + p_l\gamma_2(t)), \dot\gamma_2(t) = \nu(-r+\gamma_1(t)-\gamma_2(t))$. This system has a general solution where $\gamma_2$ takes the form $\frac{r+M}{1-p_h} + c_1e^{a_1 t} + c_2 e^{a_2 t}$ with $a_1 > a_2 > 0$ (so it approaches the asymptote as $t \to -\infty$.) Even without full knowledge of the boundary value at $t_0$, the fact that $\gamma_2(t_0) = \frac{M}{p_h-p_l} > \frac{r+M}{1-p_h}$ with $\dot\gamma_2(t_0) < 0$ requires that $c_1 < 0, c_2 > 0$. It is straightforward to verify that $\gamma_2(t)$ continues to decrease for $t > t_0$, and hence we cannot ever get back to the intervention threshold $\frac{M}{p_h-p_l}$. It follows that there are at most two policy switches prior to time $t_1$. If there are two policy switches, then the first must be a switch from non-intervention to intervention and the second must be a switch from  intervention back to non-intervention (since we cannot start intervening again after we stop). \halmos \endproof


\section{Proof of Theorem \ref{policystructure}}

The proof proceeds in two parts. We first show that equation \eqref{lineequation} is a necessary condition for a point $(x,y) \in \cC$ to be associated with $\tau$. We do this again via Pontryagin's minimum principle. Specifically, we use the requirement that the Hamiltonian must equal zero under the optimal policy, and substitute expressions for the costates from Proposition \ref{policyformula} to obtain an equation expressed in terms of $x$, $y$, and $\tau$.

We then handle the converse direction, showing that equation \eqref{lineequation} is sufficient for a point $(x,y) \in \cC$ to be associated with $\tau$. We do this by showing that the lines of the form \eqref{lineequation} do not intersect in $\cC$. We first observe that the slope is increasing with respect to $\tau$, and then verify in Lemmas \ref{increasingintercept} and \ref{technicalthing} that the $y$-coordinate of the intersection with the line $x=N$ is also increasing in $\tau$. Since a point $(x,y)$ can lie on at most one such line, the result then follows.

We fix the congestion-clearing moment as the central reference point, and define the costates and policies going backwards in time from this moment. The \emph{backward costates} are,
\begin{align}
    \Gamma_1(s) &= hs + \frac{rp_\infty + C(p_\infty)}{1-p_\infty},\\
    \Gamma_2(s) &= \frac{h}\nu (e^{-\nu s} + \nu s - 1) + \frac{r + C(p_\infty)}{1-p_\infty}
\end{align}
and the optimal \emph{backward policy} $P^*(s)$ is given by $P^*(s) = \argmin_{p} \{ C(p) + \Gamma_2(s) p \}$.

For a given state $(x,y)$ and the associated congestion-clearing time $\tau > 0$, the costates at time 0 are $\gamma_1(0) = \Gamma_1(\tau)$ and $\gamma_2(0) = \Gamma_2(\tau)$. Substituting them into the equation $H(x,y,p^*(0),0) = 0$ yields the following relationship between $\tau$ and $(x,y)$:
\begin{align}
    h(x - N) + r\nu y + C(P^*(\tau))\mu N - J_\infty + (\lambda + \nu y - \mu N)\Gamma_1(\tau) + (-\nu y + \mu N P^*(\tau))\Gamma_2(\tau) & = \nonumber\\
    h(x-N) + \nu(r+\Gamma_1(\tau)-\Gamma_2(\tau))y + (\lambda - \mu N)\Gamma_1(\tau) + \mu N(C(P^*(\tau)) + P^*(\tau) \Gamma_2(\tau)) - J_\infty & = \nonumber\\
    h(x-N) + h(1-e^{-\nu\tau})y + (\lambda - \mu N)\Gamma_1(\tau) + \mu N(C(P^*(\tau)) + P^*(\tau) \Gamma_2(\tau)) - J_\infty &= 0.\nonumber
\end{align}

All optimal values of $p$ result in the same minimum value of $\Phi_s(p) := C(p) + \Gamma_2(\tau)p$, so this yields exactly one equation (even if we are indifferent between multiple values of $p$). All states $(x,y)$ associated with $\tau$ must lie on this one specific line, which intersects the line $x=N$ at 
\begin{equation}\label{lineintercept}
    y= \frac{(\mu N - \lambda)\Gamma_1(\tau) - \mu N(\Phi_s(P^*(s))) + J_\infty}{h(1-e^{-\nu \tau})},
\end{equation}
It is easy to note that the slope $-1/(1-e^{-\nu\tau})$ is an increasing function of $\tau$. It remains to show that the intercept \eqref{lineintercept} is also increasing in $\tau$.

\begin{lemma}\label{increasingintercept}
The $y$-coordinate at which the line \eqref{lineequation} intersects the line $x=N$, given by \eqref{lineintercept} is increasing with respect to $\tau$.
\end{lemma}

Since the intercept increases while the slope also increases, it follows that no lines of this family intersect in the congested region (the lines ``fan out'' from the line $x=N$). Every congested point $(x,y)$ lies on at most one such line. Conversely, every congested point lies on at least one line, and thus exactly one line. It follows that \eqref{lineequation} is exactly the set of points which the policy $p$ is optimal.

It remains to prove Lemma \ref{increasingintercept}. We first prove the following technical lemma regarding $\Phi_s(P^*(s)) := \min_p \{ C(P(s^*(t)) + \Gamma_2(\tau)P^*(\tau) \}$, the portion of the Hamiltonian whose minimization determines the optimal policy.

\begin{lemma}\label{technicalthing}
$\Phi_s(P^*(s))$ is increasing with respect to $s$, i.e., it increases moving backward in time, and is bounded below by $\frac{rp_\infty+C_\infty}{1-p_\infty} + h P^*(s) \left( s - \frac1\nu (1-e^{-\nu s}) \right)$. It is differentiable almost everywhere, and its derivative equals $h(1-e^{-\nu s})$ wherever it exists.
\end{lemma}
\proof{Proof of Lemma \ref{technicalthing}.}
Recall that $\gamma_2$ is decreasing in $t$, so $\Gamma_2$ is increasing in $s$. If $s_1 < s_2$, then:
\begin{align}
    \Phi_s(P^*(s_1)) &= C(P^*(s_1)) + P^*(s_1)\Gamma_2(s_1) \\
    &\leq C(P^*(s_2)) + P^*(s_2) \Gamma_2(s_1) \\
    &< C(P^*(s_2)) + P^*(s_2) \Gamma_2(s_2) \\
    &= \Phi_s(P^*(s_2)).
\end{align}
Danskin's theorem states that when $\Phi_s$ has a unique minimizer, $\Phi_s(P^*(s))$ is differentiable at $s$. Its derivative at that point is then given by $\Gamma_2'(s) P^*(s) = P^*(s) h(1-e^{-\nu s})$.

As in Lemma \ref{policystructure}, we know that $\{\argmin_{p} \Phi_s(P^*(s)); s \geq 0 \}$ is a family of disjoint closed intervals, and there are at most countably many $s$ for which $P^*(s)$ has more than one minimizer. Since $\Phi_s(P^*(s))$ is increasing, these are upward jump discontinuities. Integrating the derivative thus yields a lower bound: 
\begin{align}
    \Phi_s(P^*(s)) &\geq \Phi_0(P^*(0)) + \int_0^s P^*(u) h(1-e^{-\nu u}) du \\
    &\geq C(p_\infty) + \frac{r+C(p_\infty)}{1-p_\infty}p_\infty + h P^*(s) \int_0^s 1-e^{-\nu u}\, du \\
    &= \frac{rp_\infty+C_\infty}{1-p_\infty} + h P^*(s) \left( s - \frac1\nu (1-e^{-\nu s}) \right).
\end{align}
The proof is complete. \halmos \endproof

We are now ready to prove Lemma \ref{increasingintercept}.
\proof{Proof of Lemma \ref{increasingintercept}.}
The quantity of interest is
\begin{equation}\label{intercept}
    \frac{(\mu N-\lambda)\Gamma_1(\tau) - \mu N(C(P^*(\tau)) + P^*(\tau) \Gamma_2(\tau)) + J_\infty}{h(1-e^{-\nu\tau})}
.\end{equation}
The denominator is continuous, so any discontinuities are due to the (countably many) upward jump discontinuities in the numerator (which lead to increases). It suffices to show that the derivative of this expression is positive whenever it exists. To this end, we first establish an upper bound on the numerator:
\begin{align}
    (\mu N-\lambda)\Gamma_1(\tau) - \mu N(C(P^*(\tau)) + P^*(\tau) \Gamma_2(\tau)) + J_\infty &\leq \\
     (\mu N-\lambda)\Gamma_1(\tau) - \mu N\left(\frac{rp_\infty+C_\infty}{1-p_\infty} + h P^*(\tau) \left( \tau - \frac1\nu (1-e^{-\nu \tau}) \right)\right) + J_\infty &= \\
     (\mu N - \lambda)h\tau - \mu N h P^*(\tau)\left(\tau-\frac1\nu(1- e^{-\nu \tau}) \right).
\end{align}
Next, we differentiate (\ref{intercept}) and show that the numerator of its derivative (where the derivative exists) is positive:
\small
\begin{align*}
    \left[(\mu N - \lambda)h - \mu N P^*(\tau) h(1-e^{-\nu\tau})\right] h(1-e^{-\nu \tau}) - \\ 
    h\nu e^{-\nu \tau} \left[ (\mu N - \lambda)\Gamma_1(\tau) - \mu N(C(P^*(\tau)) + P^*(\tau) \Gamma_2(\tau)) + J_\infty \right] & \geq \\
    \left[(\mu N - \lambda)h - \mu N P^*(\tau) h(1-e^{-\nu\tau})\right] h(1-e^{-\nu \tau}) - h\nu e^{-\nu \tau} \left[ (\mu N - \lambda)h\tau - \mu N h P^*(\tau)\left(\tau-\frac1\nu(1-e^{-\nu \tau}) \right) \right] &= \\
    h^2 \left[ \left[(\mu N - \lambda) - \mu N P^*(\tau) (1-e^{-\nu\tau})\right] (1-e^{-\nu \tau}) - \nu e^{-\nu \tau} \left[ (\mu N - \lambda)\tau - \mu N P^*(\tau)\left(\tau-\frac1\nu (1-e^{-\nu \tau}) \right) \right] \right] &= \\
    h^2 \left[ \left( (\mu N - \lambda) - \mu N P(\tau) (1-e^{-\nu \tau}) \right)(1-e^{-\nu \tau}) - \nu e^{-\nu \tau} (\mu N - \lambda) \tau +  e^{-\nu\tau} \mu N P(\tau) \left( \nu\tau - (1- e^{-\nu \tau}) \right) \right] &=\\
     h^2 \left[ \left( (\mu N - \lambda) - \mu N P(\tau) \right)(1-e^{-\nu \tau}) - \nu e^{-\nu \tau} (\mu N - \lambda) \tau +  e^{-\nu\tau} \mu N P(\tau) \nu\tau \right] &=\\
     h^2 \left[ ( \mu N(1-P(\tau)) - \lambda) (1-e^{-\nu \tau}) - \nu\tau e^{-\nu \tau} (\mu N(1-P(\tau)) - \lambda) \right] &= \\
    h^2 ( \mu N(1-P(\tau)) - \lambda) (1-e^{-\nu \tau} - \nu\tau e^{-\nu\tau})
    &\geq 0
\end{align*}
\normalsize
This quantity is strictly positive for $\tau > 0$, so the intercept is always increasing. \halmos\endproof
\proof{Proof of Corollary \ref{cor:piecewise}.}
Theorem \ref{policystructure} gives us a monotone map from $\tau$ to $p^*$. In the proof of the theorem (see Lemma \ref{increasingintercept}) we show that the lines of the form \eqref{lineequation} are non-intersecting by virtue of their slope and $(x=N)$-intercept being both increasing with respect to $\tau$. This provides a monotone bijection between the lines and the values of $\tau$ (where monotone with respect to the lines means monotonicity with respect to their $(x=N)$-intercepts).

For a given $\tau$ and its associated line, we can determine the corresponding value of $p^*$ using proposition \ref{policyformula}. In the first case where $\gamma_2(t) < -C'_-(p_u)$, the policy is $p_{k+1} := p_u$. In the second case where $\gamma_2(t) > -C'_+(p_l)$, the policy is $p_0 := p_l$. Otherwise, we know that $p^*$ must be a critical point of $C(p) + \gamma_2(\tau) p$, and we require $C'_-(p^*) \leq -\gamma_2(\tau) \leq C'_+(p^*)$. Since $C(\cdot)$ is piecewise linear with $k$ pieces, $C'(\cdot)$ takes on $k$ values (increasingly, with jump discontinuities at the breakpoint $p_1, \ldots, p_{k-1}$). If $C'(p) = -\gamma_2(\tau)$ along the $i$th segment $(p_{i-1}, p_i)$, then we are indifferent between any value of $p$ in the interval $[p_{i-1}, p_i]$. However, we know that this can only occur momentarily at this value of $\tau$, and cannot occur for any other values of $\tau$ in a surrounding neighborhood. Otherwise, we can find a $p_i$ such that $C'(p) < -\gamma_2(\tau)$ for $p \in (p_{i-1}, p_i)$ and $C'(p) > -\gamma_2(\tau)$ for $p \in (p_i, p_{i+1})$. Then, $p_i$ is the unique value of $p$ satisfying $C'_-(p) \leq -\gamma_2(\tau) \leq C'_+(p)$, and hence the unique optimal value of the policy for any point along the line associated with $\tau$. It follows that almost everywhere in $\cC$, the optimal policy $p^*$ is unique and takes one of the values $p_0, \ldots p_{k+1}$. \halmos\endproof

\section{Filippov Solutions and Proof of Theorem \ref{thm:unique}}\label{ap:filippov}
For ease of exposition and without loss of generality, we focus on the case of a linear intervention cost with two regions. It will be apparent that the same argument can be applied to other boundaries in the case of multiple regions under a piecewise linear cost. Consider two disjoint regions $\Omega_+$ and $\Omega_-$ on the two sides of a boundary $\mathcal{S}=\partial \Omega_+=\partial \Omega_-$. The boundary in $\cC$ is represented by the level set $\{(x,y);h(x,y)=0\}$ where $h(x,y)=x+(1-e^{-\nu \tau})y-a$ for $a,\tau,\nu>0$. (Note that these may not necessarily be optimal parameters). In region $\Omega_+$ the return probability is $p_l$ and in $\Omega_-$ it is $p_h>p_l$. Consider the dynamics in each region and let,
\begin{equation}
    F_{+}(x,y) =
    \begin{pmatrix}
\lambda+\nu y-\mu (x\wedge N)  \\
-\nu y + p_l \mu (x\wedge N)  
\end{pmatrix},
\quad
    F_{-}(x,y) =
    \begin{pmatrix}
\lambda+\nu y-\mu (x\wedge N)  \\
-\nu y + p_h \mu (x\wedge N)  
\end{pmatrix}.
\end{equation}
To deal with the discontinuity at the boundary, we consider Filippov solutions. According to Filippov's definition, an absolutely continuous function is a solution of the switched system if it satisfies the \emph{differential inclusion}, 
\[
(\dot{x},\dot{y})^{T} \in F(x,y),
\]
where, 
\[
F(x,y) =    \begin{cases}
      F_+(x,y), & (x,y) \in \mathcal{D}_+\\
      F_-(x,y), & (x,y) \in \mathcal{D}_-\\
      \alpha F_+(x,y) + (1-\alpha) F_-(x,y), & (x,y) \in \mathcal{S},  
    \end{cases} 
\]
for $\alpha \in [0,1]$.

The existence of a solution is immediate, noting that the right-hand-side of the ODE is measurable and locally essentially bounded (see, e.g., Proposition 4 of \citealt{cortes2008discontinuous}). To establish uniqueness, we use the following sufficient conditions from  \cite{filippov2013differential}. Let level set $S=\{(x,y); h(x,y)=0\}$ where $h(x,y)=x+(1-e^{-\nu \tau})y-a$. At each boundary point $(x,y)\in \mathcal{S}$, $\nabla h(x,y)$ is a normal vector to $\mathcal{S}$. Without loss of generality, assume the normal vector points to $\Omega_+$.
\begin{theorem}\label{thm:filippov}
    (Theorem 2 (Chapter 2, Page 110) of \citealt{filippov2013differential}) Let the surface $\mathcal{S}$ be twice continuously differentiable, and the vector $F_+-F_-$ be continuously differentiable on $\mathcal{S}$. If for each $(x,y)\in \mathcal{S}$ either $(\nabla h)^T \cdot F_{+}<0$ or $(\nabla h)^T \cdot F_{-}>0$ holds then there exists a unique solution starting from each initial condition.
\end{theorem}
Roughly speaking, the conditions require that either $(\nabla h)^T \cdot F_{+}$ points in the direction of $\Omega_+$ or $(\nabla h)^T \cdot F_{-}$ points in direction of $\Omega_-$, otherwise multiple solutions may exist. Specifically, if $[(\nabla h)^T \cdot F_{+}][(\nabla h)^T \cdot F_{-}]>0$ then we have a \emph{Transversal Intersection} where trajectories that hit the boundary cross over to the other region, and if $(\nabla h)^T \cdot F_{+}<0$ and $(\nabla h)^T \cdot F_{-}>0$ then we have an \emph{Attracting Sliding Motion} where trajectories reaching $\mathcal{S}$ do not leave and \emph{slide} on the boundary; see \cite{dieci2009sliding}.
\proof{Proof of Theorem \ref{thm:unique}.} Consider the dynamics in the congested region $\mathcal{C}$. The first two conditions of Theorem \ref{thm:filippov} clearly hold. Therefore, to show uniqueness, we need to show that at least one of the conditions: $(\nabla h)^T \cdot F_{+}<0$ or $(\nabla h)^T \cdot F_{-}>0$ holds. To this end, we further divide $\cC$ into $\mathcal{C}_1$ and $\mathcal{C}_2$ depending on whether $y< (\mu N -\lambda) / \nu$ or not. Define,  
\begin{align}
    \mathcal{C}_2 &= (N,\infty) \times [0,(\mu N -\lambda) / \nu),\\
\mathcal{C}_1 &= (N,\infty) \times [(\mu N -\lambda) / \nu, \infty).
\end{align}
First consider $\mathcal{C}_2$. In this case, we have, 
\begin{equation}
    F_{+}(x,y) =
    \begin{pmatrix}
\lambda+\nu y-\mu N  \\
-\nu y + p_l \mu N  
\end{pmatrix},
\quad
    F_{-}(x,y) =
    \begin{pmatrix}
\lambda+\nu y-\mu N  \\
-\nu y + p_h \mu N  
\end{pmatrix}.
\end{equation}
Note that $\dot{x} = \lambda+\nu y-\mu N< \lambda + \mu N - \lambda - \mu N =0$, so on both sides of the boundary the Needy trajectory $x(t)$ is pointing to the negative side. Depending on the parameters, $\dot{y}=-\nu y + p_l \mu N$ could be positive or negative. 
To verify the condition, we check the sign of  $(\nabla h)^T \cdot F_{+} = \lambda+\nu y -\mu N + (1-e^{-\nu \tau})(-\nu y + p_l \mu N)$. Because  $(\lambda+\nu y -\mu N)<0$, if $(-\nu y + p_l \mu N)<0$ then the sign must be negative. If not, we show that in this case the absolute value of the negative part must be larger, making the sum negative. To see this, note that $|\dot{x}|=-\lambda-\nu y + \mu N > -\nu y +p_h \mu N$ because by Assumption \ref{assump:stability}, $p_h < 1- \lambda / \mu N $ or $p_h \mu N< \mu N - \lambda$. Similarly, and using the same argument, we have $(\nabla h)^T \cdot F_{-} = \lambda+\nu y -\mu N+(1-e^{-\nu y}) (-\nu y+p_h \mu N)<0$. This implies that we have a \emph{Transversal Intersection} where trajectories that hit the boundary cross over to the next region.

Next, consider $\mathcal{C}_1$. We have, $(\nabla h)^T \cdot F_{+} = \lambda+\nu y -\mu N + (1-e^{-\nu \tau})(-\nu y + p_l \mu N)$ and $(\nabla h)^T \cdot F_{-} = \lambda+\nu y -\mu N+(1-e^{-\nu y}) (-\nu y+p_h \mu N)$. Further, $\dot{x}=\lambda + \nu y - \mu N \geq 0$ by definition of $\mathcal{C}_1$ and $\dot{y}=-\nu y + \mu p_h N <0$ because $-y \nu <\lambda -\mu N$ by Assumption \ref{assump:stability}. Therefore, either both $(\nabla h)^T \cdot F_{-}$ and $(\nabla h)^T \cdot F_{+}$ are negative or positive, or $(\nabla h)^T \cdot F_{-}>0$ and $(\nabla h)^T \cdot F_{+}<0$. We can therefore either have Transversal Intersection in the latter case or Attracting Sliding Mode in the former case. In the case of a sliding motion, the motion is unstable, i.e., the trajectories leave the boundary after a finite time, eventually entering $\cA$. This completes the proof. \halmos \endproof
\subsection{Uniqueness for initial conditions in $\cN$} As discussed in Section \ref{subsec:empty}, we conjecture that the boundaries in $\cN$ are nonlinear extensions of the linear ones in $\cC$. Let us further assume that these boundaries are twice continuously differentiable. In $\cN$, we have $\dot{x}(t) > \lambda + \mu N -\lambda -\mu N =0$ and $\dot{y}(t) = -\nu y(t)+ \mu p_u x(t)< \lambda-\mu N + \mu N - \lambda =0$ because $-\nu y<\lambda - \mu N$ and $\mu p_u N < \mu N - \lambda$. Hence, around the boundary the vectors are pointing towards the boundary. It can then be shown, using a similar approach as above, that the system of ODEs must have a unique solution starting from any initial condition in $\cN$ and hence the entire state-space. We note that numerical experiments support the smoothness assumptions. For instance, we can fit a degree two polynomial to the numerically computed boundary with an $R^2>0.99$.
\section{Supplementary Numerical Experiments and Analysis}\label{ap:supnumerics}
The code to reproduce all numerical experiments in the paper is available from the following GitHub repository: \url{https://github.com/UofTSSL/Dynamic_Control_of_Service_Systems_with_Returns}.

\subsection{Optimal Fixed Return Probability for the Stochastic System}\label{ap:exact}
When restricted to fixed return probabilities, we can numerically compute the optimal probability using the exact analysis of the Erlang-R model; see, e.g., \cite{yom2014erlang}. Denote by $\mathbb{E}_p[Q_\infty]$ the expected steady-state queue-length under return probability $p\in [p_l,p_u]$. Then the steady-state total cost can be computed using, 
\[
\frac{\lambda}{1-p} h \mathbb{E}_p[Q_\infty] + \frac{\lambda}{1-p} rp + \frac{\lambda}{1-p} C(p),
\]
which after applying Little's Law and removing values that are independent of $p$ becomes,
\begin{equation} \label{eq:steadycost}
\mathbb{E}_p[W_\infty] + rp + C(p),    
\end{equation}
where $\mathbb{E}_p[W_\infty]$ is the steady-state expected waiting time under $p$. To find the optimal return probability, we use a grid search by varying $p\in[p_l,p_u]$. 
We next compare the steady-state cost obtained under the optimal $p$ computed numerically with that under the fluid policy which ignores holding costs.
\begin{figure}
    \centering
    \includegraphics[scale=0.7]{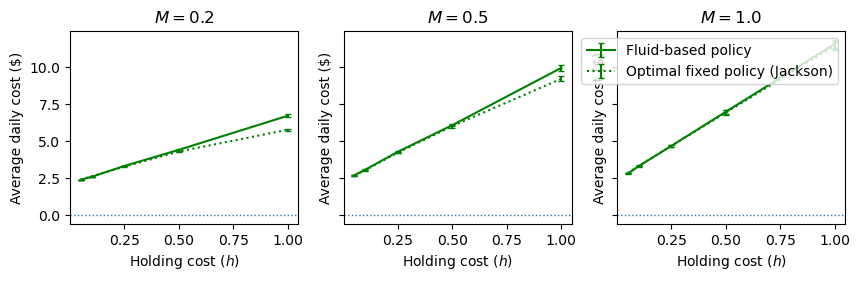}
    \includegraphics[scale=0.7]{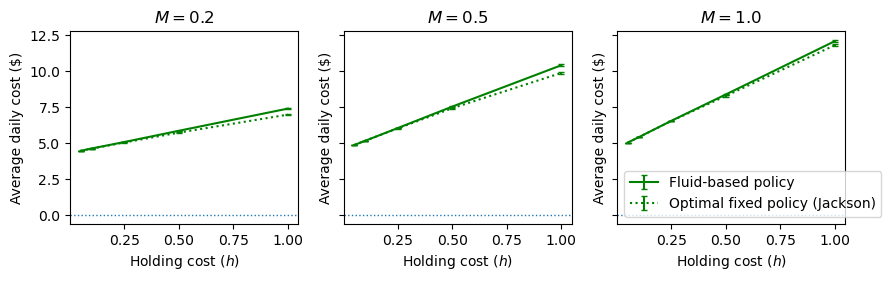}
    \caption{Comparing the long-run average cost under the optimal fixed return probability obtained using exact analysis and the optimal fixed fluid policy for varying maximum intervention $M$ and holding $h$ costs with $N=50$ (top) and $N=100$ servers (bottom). Other parameters are $\lambda=0.95$ (top), $\lambda = 1.9$ (bottom), $\mu=1/4,\nu=1/15, r=1$.}
    \label{fig:jackson}
\end{figure}
We present representative examples with the quadratic intervention cost and varying holding $h$ and intervention cost $M$ parameters. Figure \ref{fig:jackson} presents the results. With 50 servers, we find some level of suboptimality in cases where the incentive to intervene is high (i.e. the cost of intervention is low, and the holding cost is high). In these cases, the fluid policy underperforms because it is overly optimistic about holding costs, and foregoes the opportunity to make a cheap but significant reduction in long-term holding costs. However, the relative suboptimality is small, and decreases as we scale up the size of the system to 100 servers. 

\subsection{Supplement to Numerical Experiments of Sections \ref{sec:tv} and \ref{sec:numerics}}
\begin{figure}[hbt!]
    \centering
     \begin{subfigure}{0.4\textwidth}\centering
    \includegraphics[width=\linewidth]{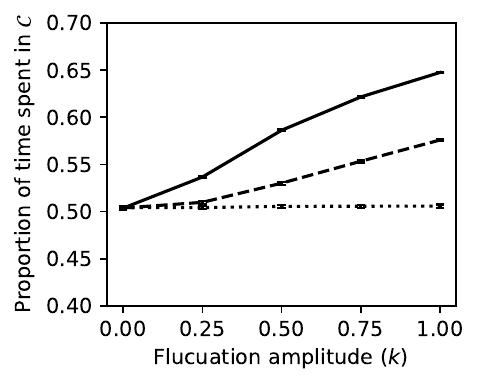}
    \caption{Linear Intervention Cost}
    \end{subfigure}
    \begin{subfigure}{0.4\textwidth}\centering
    \includegraphics[width=\linewidth]{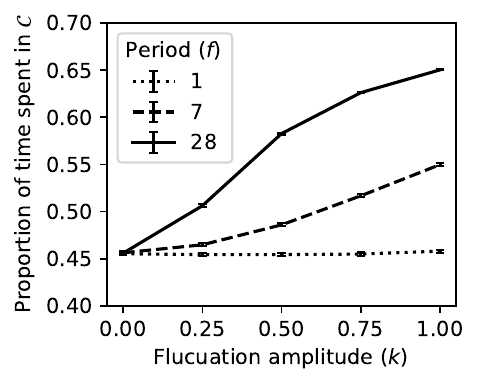}
    \caption{Quadratic Intervention Cost}
    \end{subfigure}
    \caption{Long-run fraction of time with all servers busy under the fluid policy and for three periods and varying amplitudes. Other system parameters are $N=50,\bar{\lambda}=0.95,\mu=1/4, \nu=1/15$; and cost parameters are fixed at $r=1, M=0.5, h=1/4$ with $C(p) = 5(0.2-p)$ (left) and quadratic $C(p) = 50(0.2-p)^2$ (right).}
\label{fig:varying_congestion}
\end{figure}

Figure \ref{fig:varying_congestion} presents supplementary results on the impact of time-varying arrivals on the performance of the fluid policy as discussed in Section \ref{sec:tv}.   

\begin{figure}[hbt!]
    \centering
    \begin{subfigure}{0.32\textwidth}\centering
    \includegraphics[width=\linewidth]{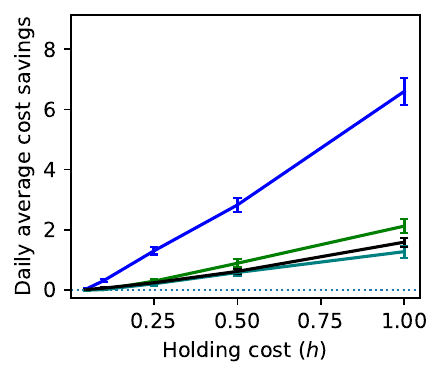}
    \caption{$M=0.2$}
    \end{subfigure}
    \begin{subfigure}{0.32\textwidth}\centering
    \includegraphics[width=\linewidth]{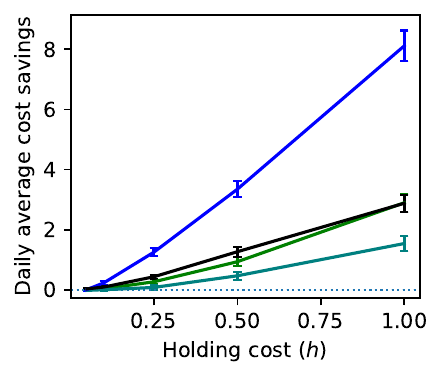}
    \caption{$M=0.5$}
    \end{subfigure}
    \begin{subfigure}{0.32\textwidth}\centering
    \includegraphics[width=\linewidth]{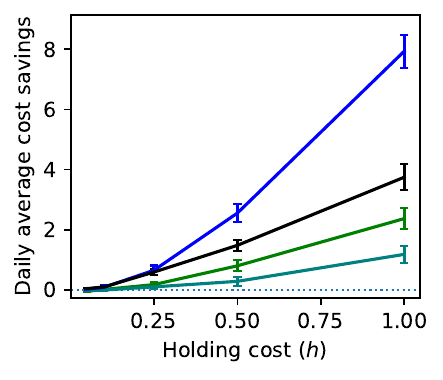}
    \caption{$M=1.0$}
    \end{subfigure}
    \begin{subfigure}{0.32\textwidth}\centering
    \includegraphics[width=\linewidth]{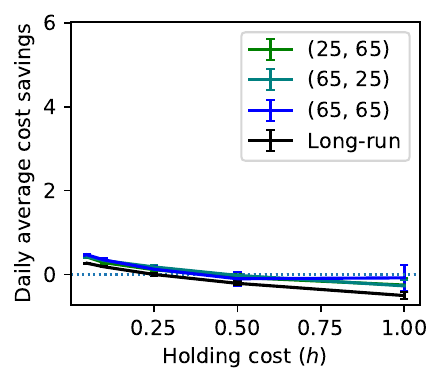}
    \caption{$M=0.2$}
    \end{subfigure}
    \begin{subfigure}{0.32\textwidth}\centering
    \includegraphics[width=\linewidth]{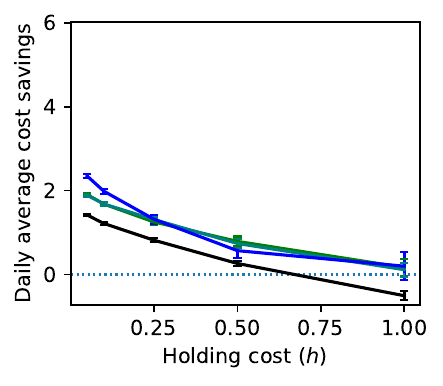}
    \caption{$M=0.5$}
    \end{subfigure}
    \begin{subfigure}{0.32\textwidth}\centering
    \includegraphics[width=\linewidth]{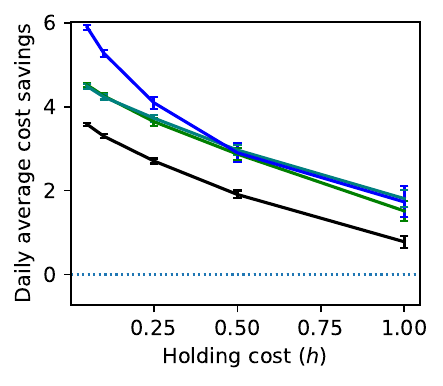}
    \caption{$M=1.0$}
    \end{subfigure}
    \caption{Absolute reduction in expected cost for both finite-horizon and long-run experiments under the fluid policy with respect to the equilibrium policy (top row) and simple policy (bottom row) and under different values of cost parameters and the quadratic intervention cost. System parameters are fixed at $N=50,\lambda=0.9,\mu=1/4,\nu=1/15$ and and the return cost is normalized at $r=1$.}
    \label{fig:cost_exp_quad_abs}
\end{figure}

\begin{figure}[hbt!]
    \centering
    \begin{subfigure}{0.32\textwidth}\centering
    \includegraphics[width=\linewidth]{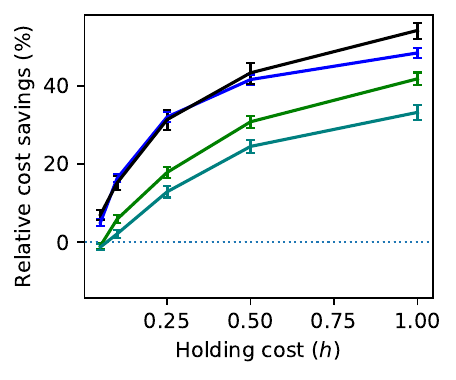}
    \caption{$M=0.2$}
    \end{subfigure}
    \begin{subfigure}{0.32\textwidth}\centering
    \includegraphics[width=\linewidth]{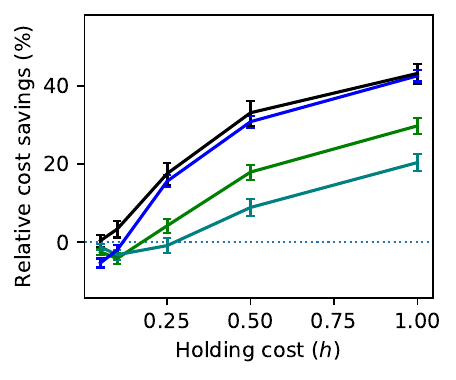}
    \caption{$M=0.5$}
    \end{subfigure}
    \begin{subfigure}{0.32\textwidth}\centering
    \includegraphics[width=\linewidth]{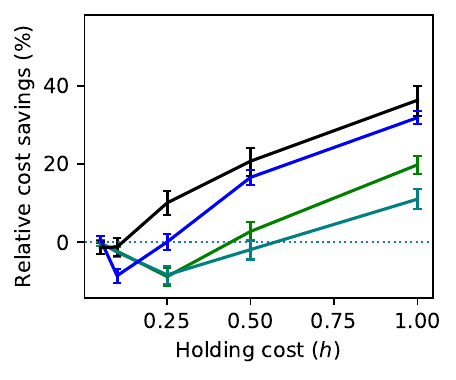}
    \caption{$M=1.0$}
    \end{subfigure}
    \begin{subfigure}{0.32\textwidth}\centering
    \includegraphics[width=\linewidth]{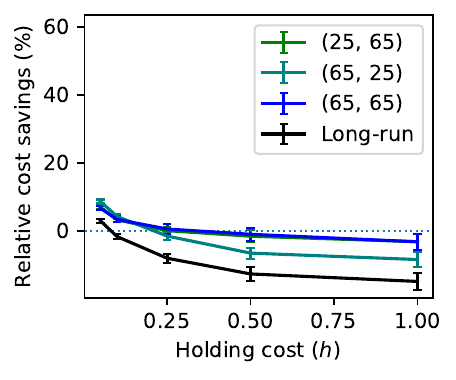}
    \caption{$M=0.2$}
    \end{subfigure}
    \begin{subfigure}{0.32\textwidth}\centering
    \includegraphics[width=\linewidth]{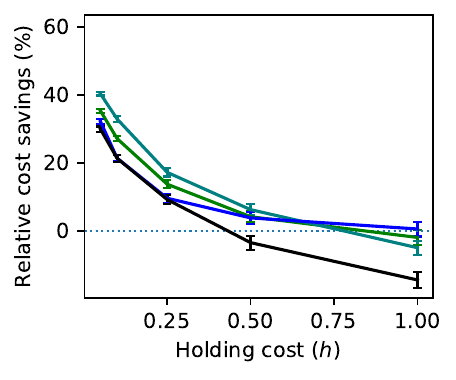}
    \caption{$M=0.5$}
    \end{subfigure}
    \begin{subfigure}{0.32\textwidth}\centering
    \includegraphics[width=\linewidth]{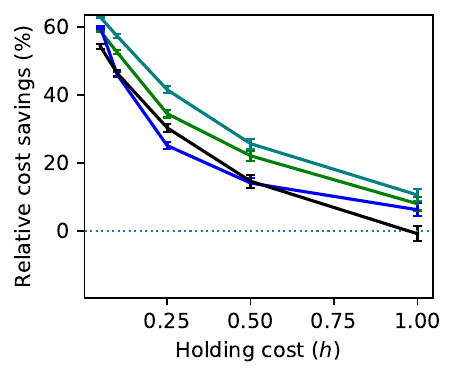}
    \caption{$M=1.0$}
    \end{subfigure}
    \caption{Relative reduction in expected cost for both finite-horizon and long-run experiments under the fluid policy with respect to the equilibrium policy (top row) and simple policy (bottom row) under the linear intervention cost and for different values of cost parameters. System parameters are fixed at $N=50,\lambda=0.95,\mu=1/4,\nu=1/15$ and the return cost is normalized at $r=1$.}
    \label{fig:cost_exp_linear}
\end{figure}

\begin{figure}[hbt!]
    \centering
    \begin{subfigure}{0.32\textwidth}\centering
    \includegraphics[width=\linewidth]{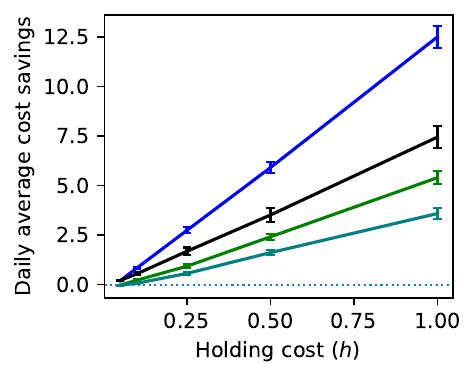}
    \caption{$M=0.2$}
    \end{subfigure}
    \begin{subfigure}{0.32\textwidth}\centering
    \includegraphics[width=\linewidth]{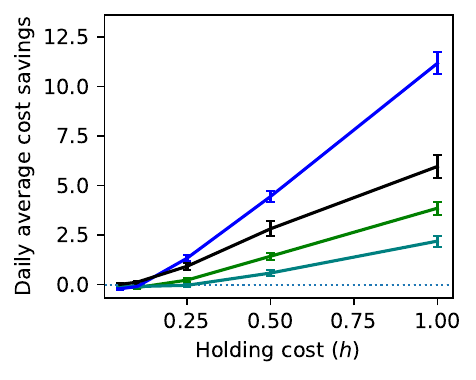}
    \caption{$M=0.5$}
    \end{subfigure}
    \begin{subfigure}{0.32\textwidth}\centering
    \includegraphics[width=\linewidth]{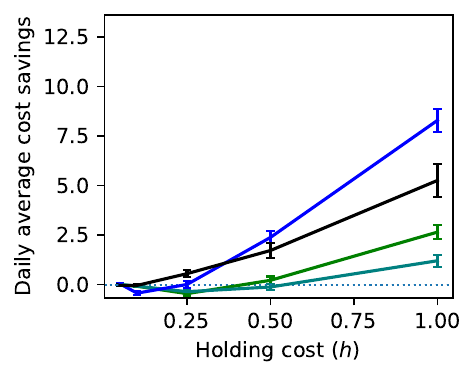}
    \caption{$M=1.0$}
    \end{subfigure}
    \begin{subfigure}{0.32\textwidth}\centering
    \includegraphics[width=\linewidth]{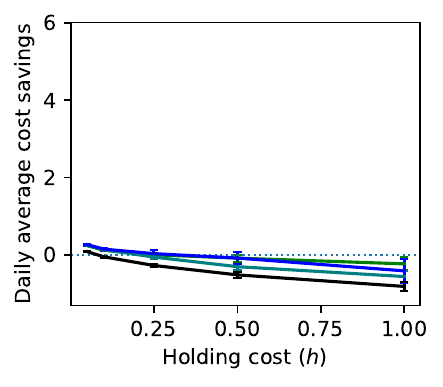}
    \caption{$M=0.2$}
    \end{subfigure}
    \begin{subfigure}{0.32\textwidth}\centering
    \includegraphics[width=\linewidth]{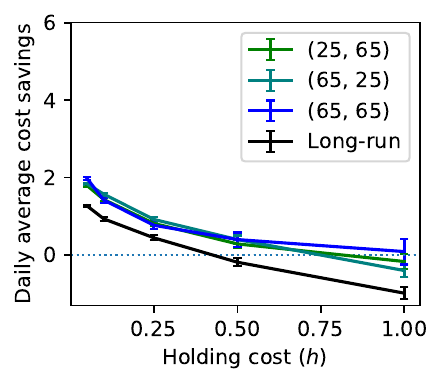}
    \caption{$M=0.5$}
    \end{subfigure}
    \begin{subfigure}{0.32\textwidth}\centering
    \includegraphics[width=\linewidth]{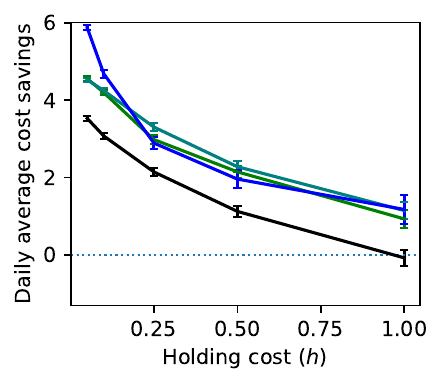}
    \caption{$M=1.0$}
    \end{subfigure}
    \caption{Relative reduction in expected cost for both finite-horizon and long-run experiments under the fluid policy with respect to the equilibrium policy (top row) and simple policy (bottom row) and under different values of cost parameters and the linear intervention cost. System parameters are fixed at $N=50,\lambda=0.95,\mu=1/4,\nu=1/15$ and the return cost is normalized at $r=1$.}
    \label{fig:cost_exp_lin_abs}
\end{figure}

\begin{figure}[hbt!]
    \centering
    \begin{subfigure}{0.32\textwidth}\centering
    \includegraphics[width=\linewidth]{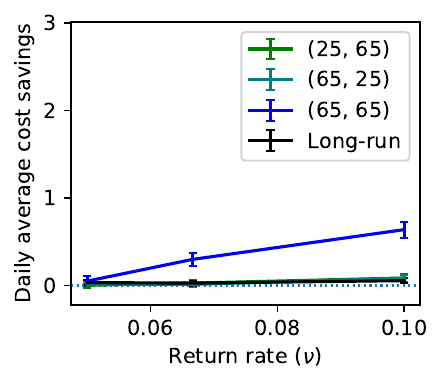}
    \caption{$\lambda=9.0$}
    \end{subfigure}
    \begin{subfigure}{0.32\textwidth}\centering
    \includegraphics[width=\linewidth]{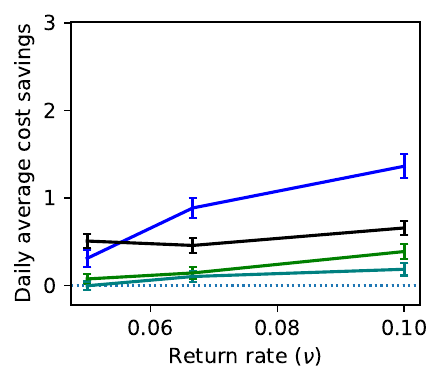}
    \caption{$\lambda=9.5$}
    \end{subfigure}
    \begin{subfigure}{0.32\textwidth}\centering
    \includegraphics[width=\linewidth]{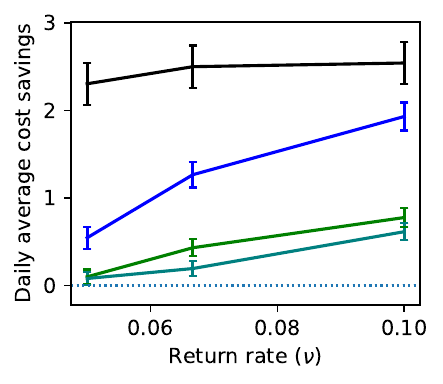}
    \caption{$\lambda=9.8$}
    \end{subfigure}
    \begin{subfigure}{0.32\textwidth}\centering
    \includegraphics[width=\linewidth]{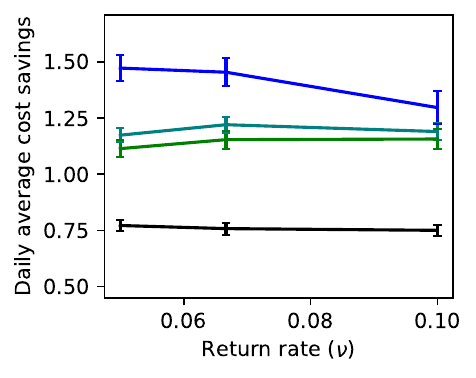}
    \caption{$\lambda=9.0$}
    \end{subfigure}
    \begin{subfigure}{0.32\textwidth}\centering
    \includegraphics[width=\linewidth]{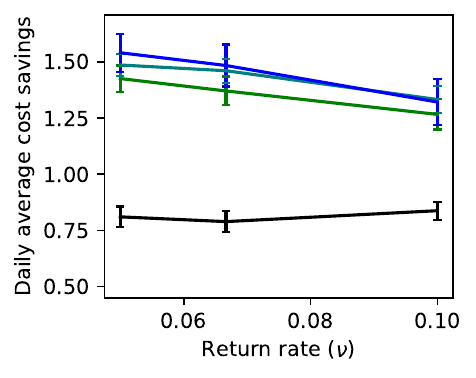}
    \caption{$\lambda=9.5$}
    \end{subfigure}
    \begin{subfigure}{0.32\textwidth}\centering
    \includegraphics[width=\linewidth]{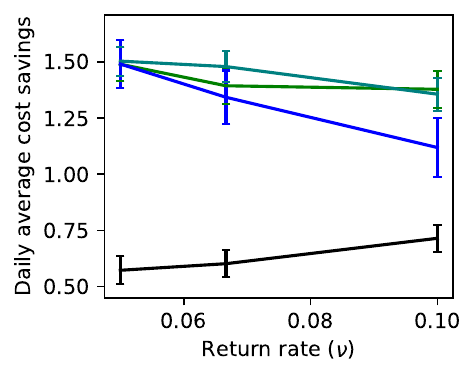}
    \caption{$\lambda=9.8$}
    \end{subfigure}
    \caption{Absolute reduction in expected cost for both finite-horizon and long-run experiments under the fluid policy with respect to the equilibrium policy (top row) and simple policy (bottom row) under the quadratic intervention cost and for different values of arrival $\lambda$ and return $\nu$ rates. Other cost and system parameters are fixed at $N=50,\lambda=0.95,\mu=1/4,\nu=1/15, r=1, M=0.5, h=1/4$.}
    \label{fig:rate_exp_quad_abs}
\end{figure}

\begin{figure}[hbt!]
    \centering
    \begin{subfigure}{0.32\textwidth}\centering
    \includegraphics[width=\linewidth]{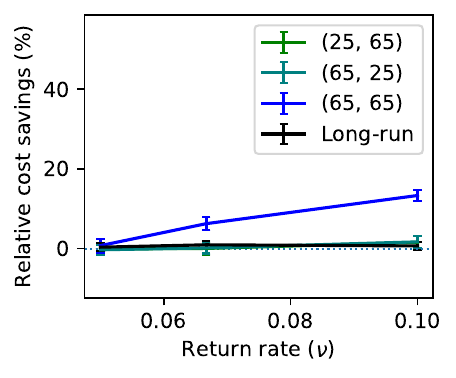}
    \caption{$\lambda=9.0$}
    \end{subfigure}
    \begin{subfigure}{0.32\textwidth}\centering
    \includegraphics[width=\linewidth]{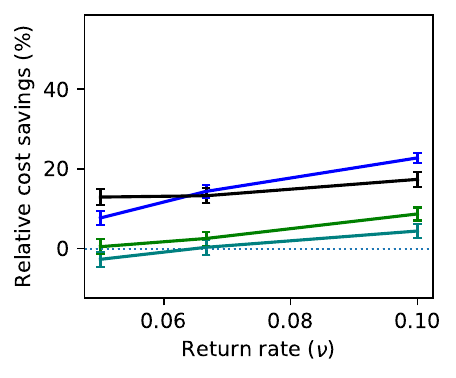}
    \caption{$\lambda=9.5$}
    \end{subfigure}
    \begin{subfigure}{0.32\textwidth}\centering
    \includegraphics[width=\linewidth]{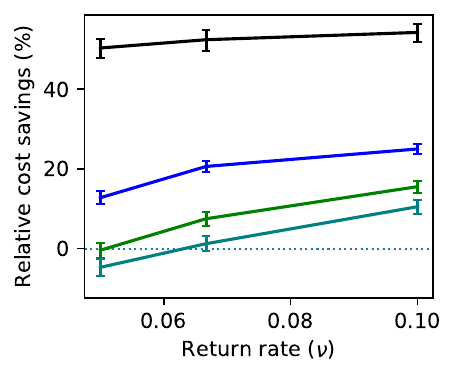}
    \caption{$\lambda=9.8$}
    \end{subfigure}
    \begin{subfigure}{0.32\textwidth}\centering
    \includegraphics[width=\linewidth]{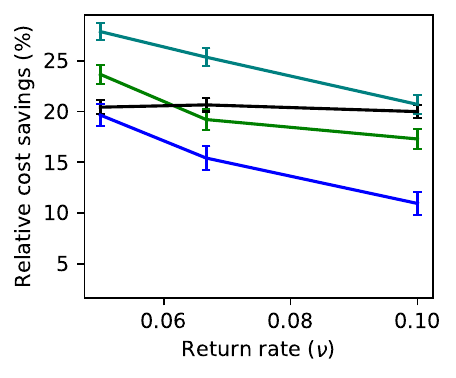}
    \caption{$\lambda=9.0$}
    \end{subfigure}
    \begin{subfigure}{0.32\textwidth}\centering
    \includegraphics[width=\linewidth]{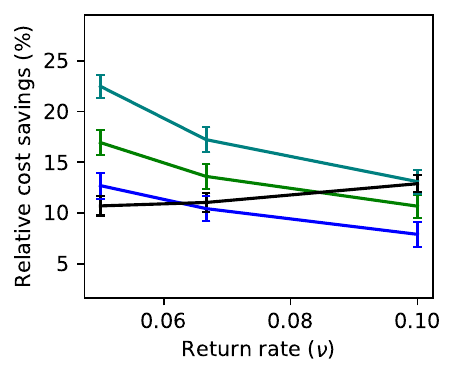}
    \caption{$\lambda=9.5$}
    \end{subfigure}
    \begin{subfigure}{0.32\textwidth}\centering
    \includegraphics[width=\linewidth]{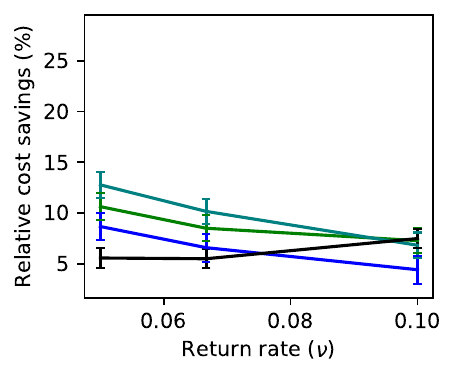}
    \caption{$\lambda=9.8$}
    \end{subfigure}
    \caption{Relative reduction in expected cost for both finite-horizon and long-run experiments under the fluid policy with respect to the equilibrium policy (top row) and simple policy (bottom row) under the linear intervention cost and for different values of arrival $\lambda$ and return $\nu$ rates. Other cost and system parameters are fixed at $N=50,\mu=1/4, r=1, M=0.5, h=1/4$.}
    \label{fig:rate_exp_lin}
\end{figure}

\begin{figure}[hbt!]
    \centering
    \begin{subfigure}{0.32\textwidth}\centering
    \includegraphics[width=\linewidth]{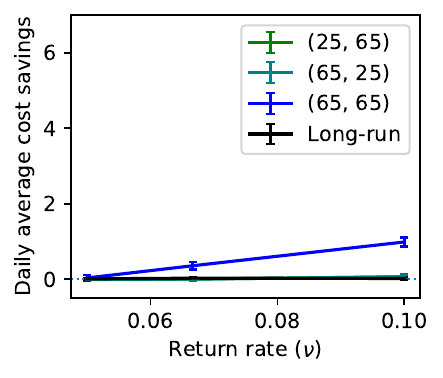}
    \caption{$\lambda=9.0$}
    \end{subfigure}
    \begin{subfigure}{0.32\textwidth}\centering
    \includegraphics[width=\linewidth]{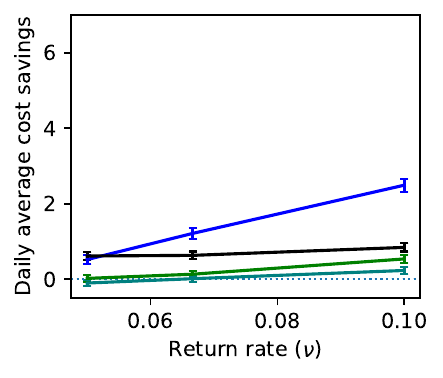}
    \caption{$\lambda=9.5$}
    \end{subfigure}
    \begin{subfigure}{0.32\textwidth}\centering
    \includegraphics[width=\linewidth]{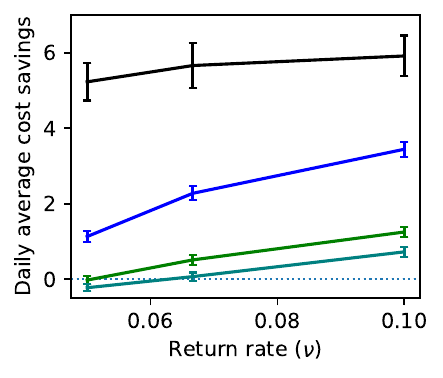}
    \caption{$\lambda=9.8$}
    \end{subfigure}
    \begin{subfigure}{0.32\textwidth}\centering
    \includegraphics[width=\linewidth]{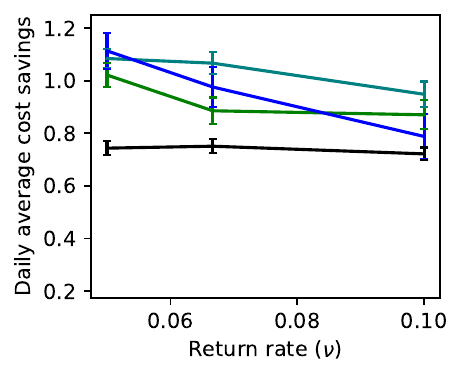}
    \caption{$\lambda=9.0$}
    \end{subfigure}
    \begin{subfigure}{0.32\textwidth}\centering
    \includegraphics[width=\linewidth]{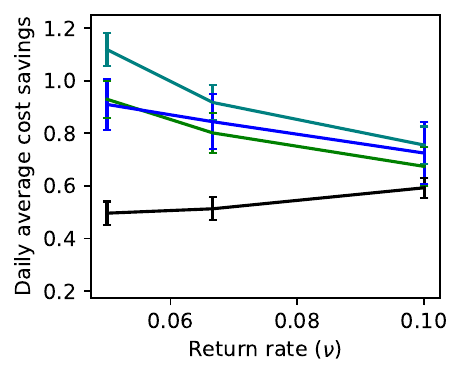}
    \caption{$\lambda=9.5$}
    \end{subfigure}
    \begin{subfigure}{0.32\textwidth}\centering
    \includegraphics[width=\linewidth]{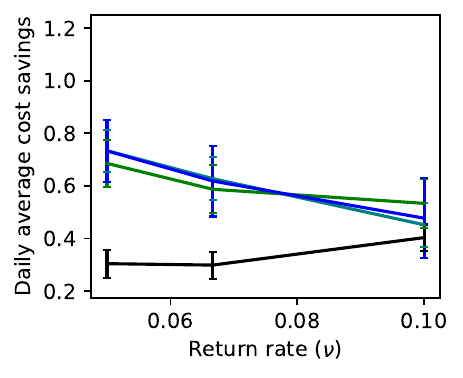}
    \caption{$\lambda=9.8$}
    \end{subfigure}
    \caption{Absolute reduction in expected cost for both finite-horizon and long-run experiments under the fluid policy with respect to the equilibrium policy (top row) and simple policy (bottom row) under the linear intervention cost and for different values of arrival $\lambda$ and return $\nu$ rates. Other cost and system parameters are fixed at $N=50,\mu=1/4, r=1, M=0.5, h=1/4$.}
    \label{fig:rate_exp_lin_abs}
\end{figure}

\begin{figure}[hbt!]
    \centering
    \begin{subfigure}{0.32\textwidth}\centering
    \includegraphics[width=\linewidth]{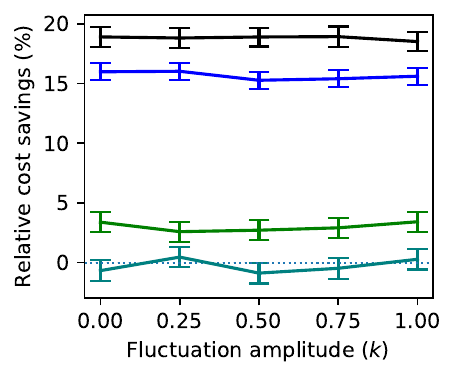}
    \caption{$f=1$}
    \end{subfigure}
    \begin{subfigure}{0.32\textwidth}\centering
    \includegraphics[width=\linewidth]{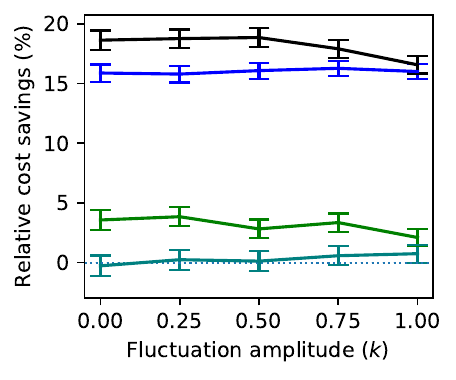}
    \caption{$f=7$}
    \end{subfigure}
    \begin{subfigure}{0.32\textwidth}\centering
    \includegraphics[width=\linewidth]{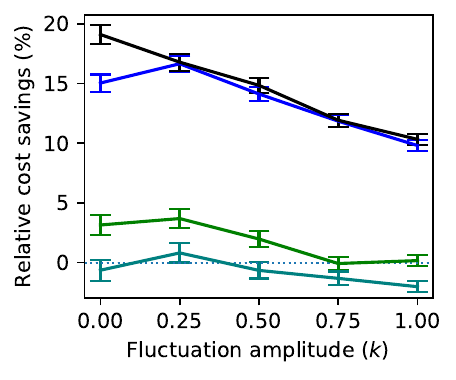}
    \caption{$f=28$}
    \end{subfigure}
    \begin{subfigure}{0.32\textwidth}\centering
    \includegraphics[width=\linewidth]{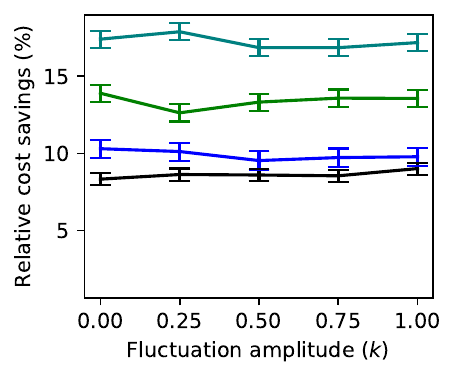}
    \caption{$f=1$}
    \end{subfigure}
    \begin{subfigure}{0.32\textwidth}\centering
    \includegraphics[width=\linewidth]{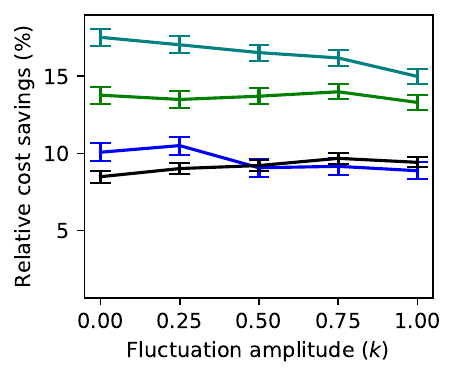}
    \caption{$f=7$}
    \end{subfigure}
    \begin{subfigure}{0.32\textwidth}\centering
    \includegraphics[width=\linewidth]{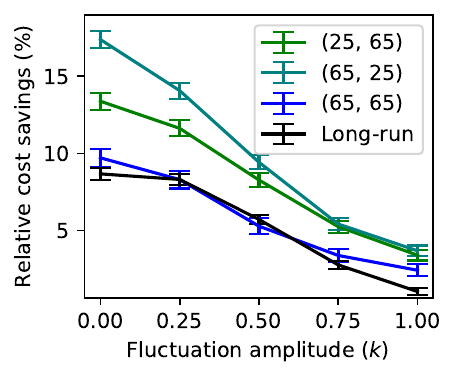}
    \caption{$f=28$}
    \end{subfigure}
    \caption{Relative reduction in expected cost under the linear intervention cost for the fluid policy with respect to the equilibrium policy (top row) and simple policy (bottom row) and under different time-varying arrival rate functions. System and cost parameters are fixed at $N=50,\bar\lambda=0.95,\mu=1/4,\nu=1/15, r=1, h=1/4, C(p)=5(0.2-p)$.}
    \label{fig:varying_exp_linear}
\end{figure}

\noindent Figures \ref{fig:cost_exp_quad_abs}--\ref{fig:cost_exp_lin_abs} present supplementary results for the impact of varying cost parameters on the performance of the fluid policy. 
\begin{itemize}
    \item Figure \ref{fig:cost_exp_quad_abs} presents the absolute cost reductions under the quadratic cost function.
    \item Figures \ref{fig:cost_exp_linear} and \ref{fig:cost_exp_lin_abs} respectively present the relative and absolute cost reductions for the linear cost function. 
\end{itemize}
Figures \ref{fig:rate_exp_quad_abs}--\ref{fig:rate_exp_lin_abs} present supplementary results for the impact of varying the system load and average time to return on the performance of the fluid policy. 
\begin{itemize}
    \item Figure \ref{fig:rate_exp_quad_abs} presents the absolute cost reductions under the quadratic cost function.
    \item Figures \ref{fig:rate_exp_lin} and \ref{fig:rate_exp_lin_abs} respectively present the relative and absolute cost reductions for the linear cost function. 
\end{itemize}
Figure \ref{fig:varying_exp_linear} presents the impact of time-varying arrivals on the performance of the fluid policy under the linear cost function and for varying period and amplitude parameters of the arrival rate.  

\newpage

\subsection{When Interventions Also Reduce Return Time: Preliminary Analysis}\label{ap:timetoreturn}
Here, we provide preliminary analysis of an extension of the problem to the case where interventions also reduce average return time $\nu$. In the original model, we allowed for a continuous intervention, allowing the readmission probability to take any value on an interval $[p_l, p_u]$. Regardless of the readmission probabilities chosen, the problem state remained two-dimensional. When we extend to multiple values of time-to-return, we (a) increase the dimension of $y(t)$ to allow for patients who received different interventions and will have different times-to-return; and (b) restrict ourselves to finitely many intervention options, in order for $y$ to be a finite-dimensional vector.
    
We assume that there is a ordering on the interventions: $(p_1, \nu_1) > (p_2, \nu_2) > \cdots > (p_k, \nu_k)$ (component-wise). In other words, when comparing any two intervention options, one must be ``better" (lower return probability \emph{and} slower return rate (i.e., longer time-to-return) than the other. We will not allow cases where one intervention is better along one dimension but worse along the other. We specify our policy as a vector $q$. Then the dynamics of the system can be easily presented in a ``vectorized" form:
    \begin{align}
        \dot x(t) &= \lambda + \nu^T \cdot y(t) - \mu (x(t) \wedge N),  \\
        \dot y(t) &= -\nu \odot y(t) + \mu (x(t) \wedge N) p \odot q,
    \end{align}
    where $\odot$ denotes component-wise multiplication.
    We incur cost at a rate of:
    \[
        h(x(t) - N)^+ + r\nu^T y(t) + \mu (x(t) \wedge N) C(p)^T \cdot q.
    \]
In the case of long-run average cost (i.e., in region $\cA$), the optimization remains the same as before, taking $p_\infty$ to be the value minimizing the expected lifetime cost per patient of $\frac{C(p) + rp}{1-p}$. In this region, the costates are $\gamma_1 = \frac{C(p_\infty) + rp_\infty}{1-p_\infty}$ and $\gamma_2 = \frac{C(p_\infty) + r}{1-p_\infty}$.
    
In the congested region, applying Pontryagin's Minimum principle yields the following results. We choose $q$ to minimize $(C(p) + (p \odot \gamma_2))^T \cdot q$. In other words, we choose the intervention $i$ which minimizes $C(p_i) + (\gamma_2)_i p_i$, analogous to what we did previously in the case of fixed $\nu$ (except in that case, $\gamma_2$ was a single scalar common across all intervention levels). The costates are governed by the system:
            \begin{align}
                \dot \gamma_1(t) &= -h, \\
                \dot \gamma_2(t) &= -r\nu - \gamma_1(t) \nu + \nu \odot \gamma_2,
            \end{align}
Note that $\gamma_1$ has a simple, linear solution $\gamma_1(t) = h(\tau - t) + \frac{rp_\infty + C(p_\infty)}{1-p_\infty}$. Each component of $\gamma_2$ (denoted by $(\gamma_2)_i$) is independent, and we only need to solve the equation $(\dot{\gamma}_2)_i = -r\nu_i -\gamma_1\nu_i + \nu_i(\gamma_2)_i$ for $i=1,2,\ldots,k$, which is one-dimensional once we substitute the expression for $\gamma_1$. Using our results from the simpler case with a single return probability, this equation has a closed-form solution:
\[(\gamma_2)_i(t) = \frac{h}{\nu_i} (e^{-\nu_i(\tau - t)} + \nu_i(\tau - t) - 1) + \frac{r+C(p_\infty)}{1-p_\infty}.
\]
Let $1 \leq i < j \leq k$. We assumed that $p_i > p_j$ and $\nu_i > \nu_j$ (i.e. intervention $j$ is better with respect to both return probability and return time). We can verify that $(\dot{\gamma_2})_i(t) < (\dot{\gamma_2})_j(t) < 0$ for all $t < \tau$. It then follows that $(\dot{\gamma_2})_i(t) p_i < (\dot{\gamma_2})_j(t) p_j < 0$, and hence $p_i(\gamma_2)_i - p_j(\gamma_2)_j$ is decreasing with respect to time (moving forward towards $\tau$). When we choose between two alternative intervention levels $i$ and $j$, we prefer $i$ over $j$ exactly when $C(p_i) - C(p_j) + p_i(\gamma_2)_i(t) - p_j(\gamma_2)_j(t) < 0$. It follows that we must reduce our intervention over time: it is impossible to prefer $i$ earlier in time than $j$, so if $i$ is the optimal policy at any time $t_0 < \tau$, then we can never choose intervention $j$ after that time $t_0$. This intuitively makes sense: when the system is very congested, it is optimal to apply high levels of intervention, and then gradually reduce the intensity as the congestion clears and the trajectories approach region $\cA$.

\subsection{Case Study Results with Noisy Prediction of the Content State}\label{ap:noisy}
\begin{figure}
    \centering
    \includegraphics[scale=0.7]{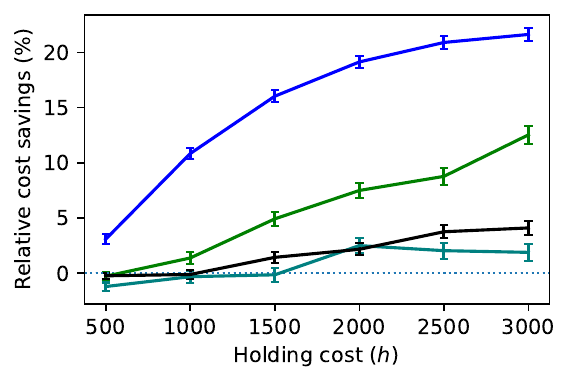}
\includegraphics[scale=0.7]{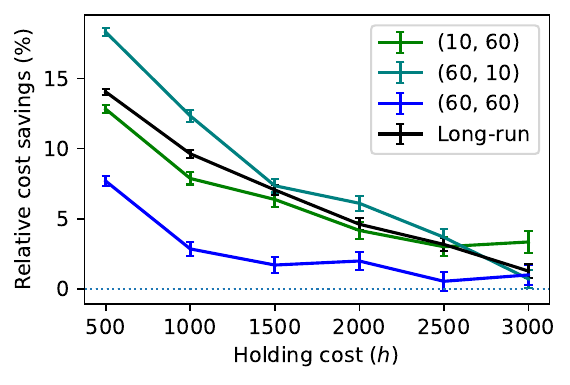}
    \caption{Case study results with noisy Content state. The plots present relative reduction in expected cost as measured against the equilibrium policy (left) and the simple policy (right) for different holding cost values.}
    \label{fig:noisy_case_opt}
\end{figure}

Here, we illustrate the robustness of our case study results when the Content state is not observable. We consider a simple estimation of the Content state: when a patient is discharged with return probability $p$, we increase our estimate by $p$, and when a patient returns, we decrease our estimate by the $p$ associated with that patient (note that we track each patient's $p$, since different patients may have different interventions and hence return probabilities depending on the decisions made at their discharge). If the patient does not readmit, then we wait until they ``fall off" 30 days after discharge to subtract their $p$ from the estimate. Note that this is a biased estimate of the Content state. For the transient experiments, we also need to determine an initial estimate. For simplicity we use the true initial value but as customers depart our estimate quickly deviates from the true value. Figure \ref{fig:noisy_case_opt} presents the relative cost reduction compared to the simple and equilibrium policies. We observe comparable reductions to those in Section \ref{sec:case} including for the long-run average case that is not sensitive to the estimate of the initial value at time zero. As discussed in Section \ref{sec:conc} we can attribute the robustness to the structure of the policy which only requires a correct classification of the Content state (above or below the boundary).

\subsection{Convergence of Scaled Sample Paths to the Fluid Trajectories}\label{ap:weak}
\begin{figure}[]
\begin{subfigure}{0.5\textwidth}
    \centering
    \includegraphics[width=0.9\textwidth]{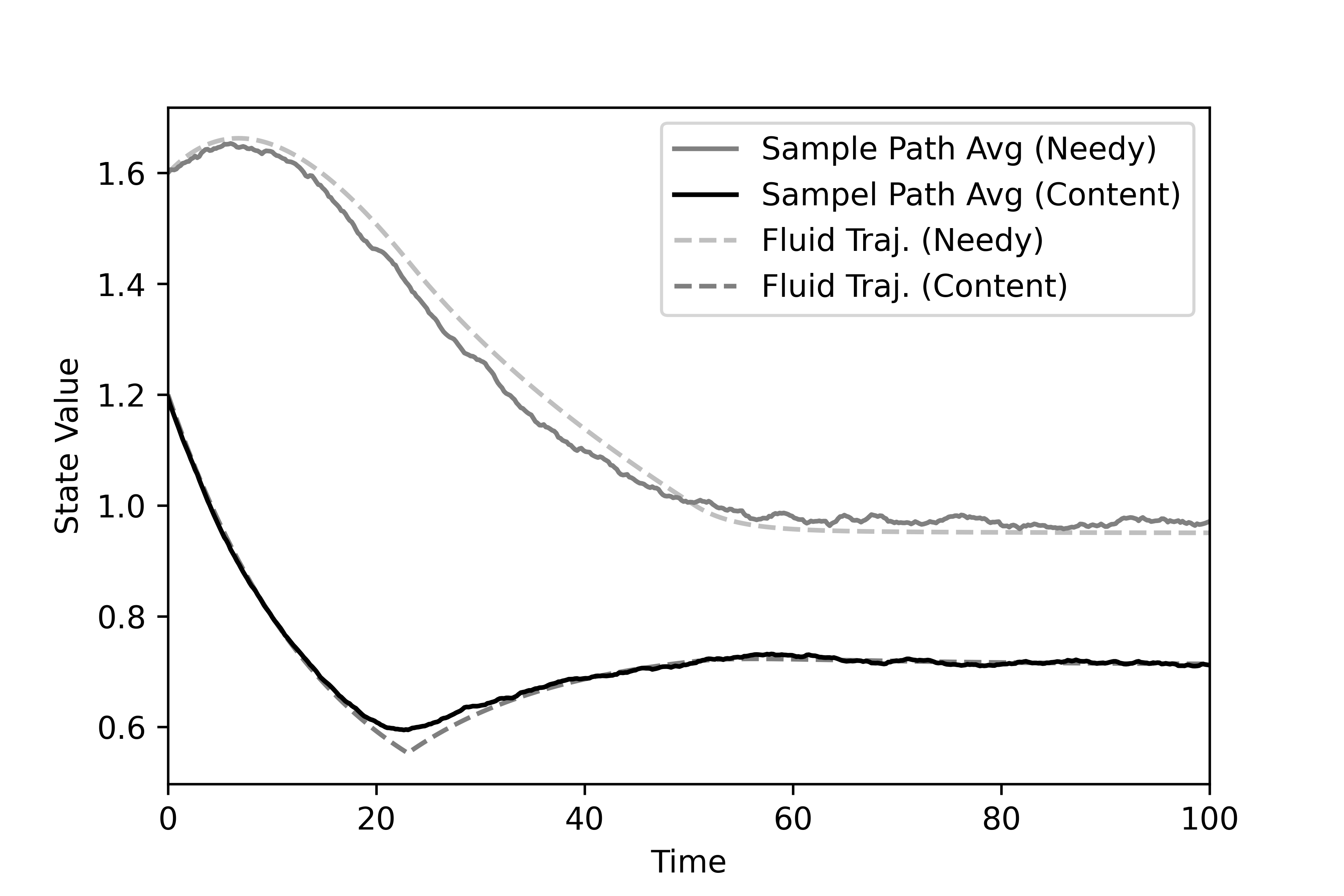}
    \caption{$(\bar{X}(0), \bar{Y}(0)) = (1.6, 1.2)$}
\end{subfigure}
\begin{subfigure}{0.5\textwidth}
    \centering
    \includegraphics[width=0.9\textwidth]{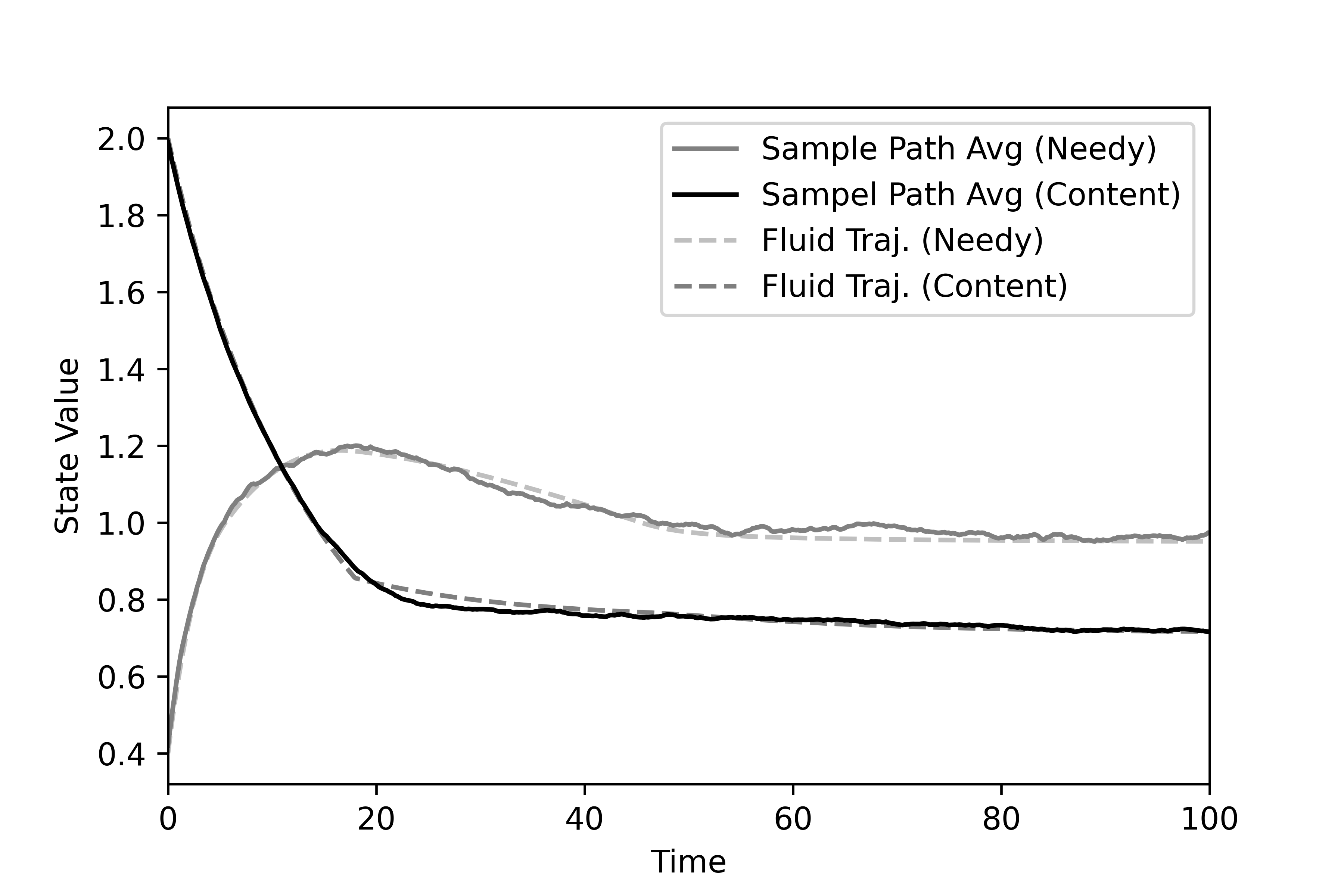}
    \caption{$(\bar{X}(0), \bar{Y}(0)) = (0.4, 2.0)$}
\end{subfigure}

\begin{subfigure}{0.5\textwidth}
    \centering
    \includegraphics[width=0.9\textwidth]{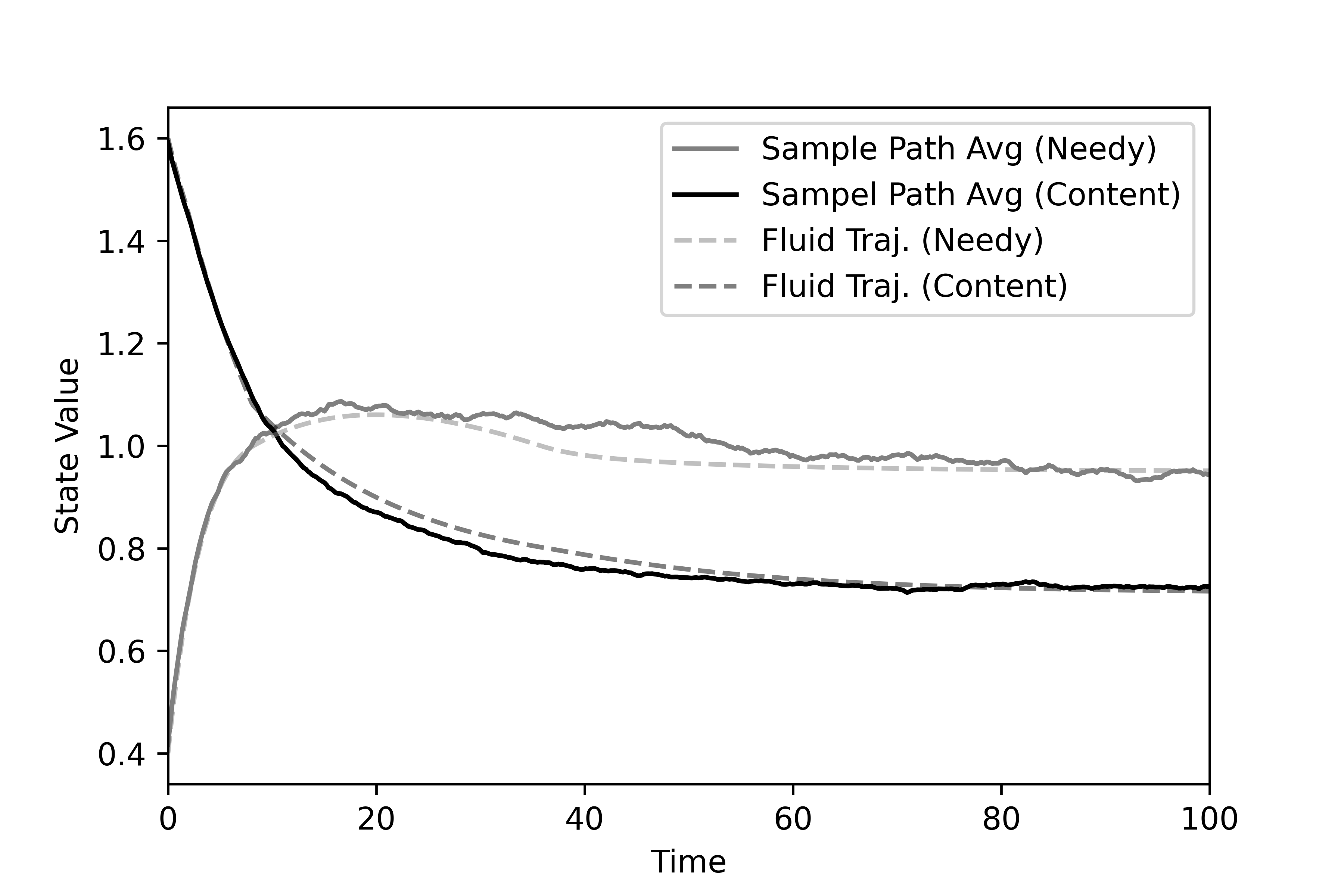}
    \caption{$(\bar{X}(0), \bar{Y}(0)) = (0.4, 1.6)$}
\end{subfigure}
\begin{subfigure}{0.5\textwidth}
    \centering
    \includegraphics[width=0.9\textwidth]{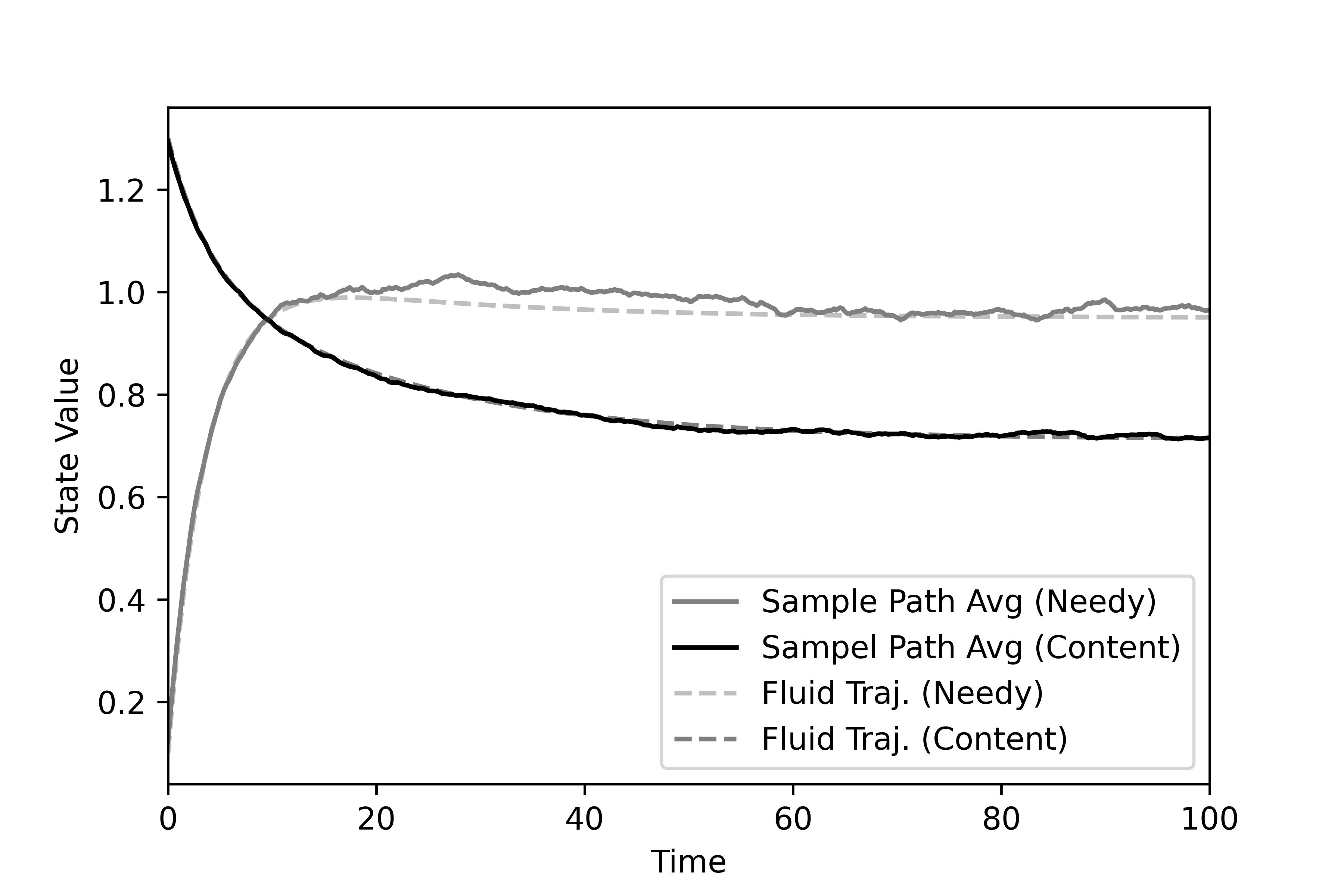}
    \caption{$(\bar{X}(0), \bar{Y}(0)) = (0.1, 1.3)$}
\end{subfigure}
\caption{Comparing fluid trajectories with estimates of expected scaled sample paths $(\bar{X}(t),\bar{Y}(t)) \equiv (X(t)/N, Y(t)/N)$ under discontinuous switching control for four initial conditions; $\lambda=0.19 N, \nu=1/15, \mu=1/4$, $N=500$. Cost parameters are the same as for the linear intervention cost example of Section \ref{sec:numerics}. }\label{fig:convergence}
\end{figure}

In Figure \ref{fig:convergence} we compare simulation estimates of the expected scaled queue-length (using 30 sample paths) under the optimal policy for the linear intervention cost example from Section \ref{sec:numerics} to the optimal fluid trajectories for four different initial conditions and a large system with $N=500$ servers. The results support our conjecture on the weak convergence of the fluid-scaled sample paths to the optimal fluid trajectories even under switching (discontinuous) control.  


\end{document}